\documentclass[twocolumn]{aastex631}
\usepackage{amsmath}
\usepackage{CJKutf8}

\definecolor{dblue}{rgb}{0.12, 0.56, 1.0}
\newcommand{\ch}[1]{#1}
\newcommand{\boxparam}{$\Omega_m$, $\sigma_8$, $A_{\rm SN1}$, $A_{\rm SN2}$, $A_{\rm AGN1}$, $A_{\rm AGN2}$}
\newcommand{\sna}{A$_{\rm SN1}$}
\newcommand{\snb}{A$_{\rm SN2}$}
\newcommand{\agna}{A$_{\rm AGN1}$}
\newcommand{\agnb}{A$_{\rm AGN2}$}
\newcommand{\massrange}[2]{$10^{#1} < $M$_*< 10^{#2} $M$_\odot$}

\newcommand{\mhalo}{M$_{\rm halo}$}
\newcommand{\fb}{$f_{\rm baryon}$}
\newcommand{\mbh}{M$_{\rm BH}$/M$_{\rm halo}$}

\definecolor{cosmocolor}{rgb}{0.2235, 0.5921, 0.0}
\newcommand{\colora}[1]{\textcolor{cosmocolor}{\bf #1}}
\definecolor{fbcolor}{rgb}{0.858, 0.188, 0.478}
\newcommand{\colorb}[1]{\textcolor{fbcolor}{\bf #1}}
\definecolor{halocolor}{rgb}{0.0, 0.4392, 0.8039}
\newcommand{\colorc}[1]{\textcolor{halocolor}{\bf #1}}

\graphicspath{{./}{new_figs/}}

\begin{document}

\title{How does feedback affect the star formation histories of galaxies?}

\author[0000-0001-9298-3523]{Kartheik G. Iyer}
\altaffiliation{Hubble Fellow}
\affiliation{Columbia Astrophysics Laboratory, Columbia University, 550 West 120th Street, New York, NY 10027, USA}
\affiliation{Center for Computational Astrophysics, Flatiron Institute, 162 Fifth Avenue, New York, NY 10010, USA}
\email{kgi2103@columbia.edu}

\author[0000-0003-2539-8206]{Tjitske K. Starkenburg}
\affiliation{Center for Interdisciplinary Exploration and Research in Astrophysics (CIERA), Northwestern University, 1800 Sherman Ave, Evanston, IL 60201, USA}
\affiliation{Department of Physics and Astronomy, Northwestern University, 2145 Sheridan Rd, Evanston IL 60208, USA}
\affiliation{NSF-Simons AI Institute for the Sky (SkAI), 172 E. Chestnut St., Chicago, IL 60611, USA}

\author[0000-0003-2630-9228]{Greg L. Bryan}
\affiliation{Columbia Astrophysics Laboratory, Columbia University, 550 West 120th Street, New York, NY 10027, USA}

\author[0000-0002-6748-6821]{Rachel S. Somerville}
\affiliation{Center for Computational Astrophysics, Flatiron Institute, 162 Fifth Avenue, New York, NY 10010, USA}

\author[0009-0007-3423-1332]{Juan Pablo Alfonzo}
\affiliation{Astronomical Institute, Tohoku University, 6-3, Aramaki, Aoba-ku, Sendai, Miyagi, 980-8578, Japan}

\author[0000-0001-5769-4945]{Daniel Angl\'es-Alc\'azar}
\affiliation{Department of Physics, University of Connecticut, 196 Auditorium Road, U-3046, Storrs, CT 06269-3046, USA}

\author[0000-0002-9217-1696]{Suchetha Cooray}
\affiliation{Kavli Institute for Particle Astrophysics and Cosmology, Stanford University, 452 Lomita Mall, Stanford, CA 94305, USA}
\affiliation{SLAC National Accelerator Laboratory, 2575 Sand Hill Road, Menlo Park, CA 94025, USA}

\author[0000-0003-2842-9434]{Romeel Dav\'e}
\affiliation{Institute for Astronomy, Royal Observatory, University of Edinburgh, Edinburgh EH9 3HJ, UK}
\affiliation{Department of Physics and Astronomy, University of the Western Cape, Bellville, Cape Town 7535, South Africa}
\affiliation{South African Astronomical Observatories, Observatory, Cape Town 7925, South Africa}

\author[0000-0003-4295-3793]{Austen Gabrielpillai}
\affiliation{Department of Astrophysics, The Graduate Center, City University of New York, 365 5th Ave, New York, NY 10016, USA}

\author[0000-0002-3185-1540]{Shy Genel}
\affiliation{Columbia Astrophysics Laboratory, Columbia University, 550 West 120th Street, New York, NY 10027, USA}
\affiliation{Center for Computational Astrophysics, Flatiron Institute, 162 Fifth Avenue, New York, NY 10010, USA}

\author[0000-0002-1050-7572]{Sultan Hassan}
\affiliation{Center for Cosmology and Particle Physics, Department of Physics, New York University, 726 Broadway, New York, NY 10003, USA}
\affiliation{Center for Computational Astrophysics, Flatiron Institute, 162 Fifth Avenue, New York, NY 10010, USA}

\author{Lars Hernquist}
\affiliation{Center for Astrophysics | Harvard \& Smithsonian, 60 Garden Street, Cambridge, MA 02138, USA}

\author[0000-0002-8896-6496]{Christian Kragh Jespersen}
\affiliation{Department of Astrophysical Sciences, Princeton University, Princeton, NJ 08544, USA}

\author[0000-0001-7964-5933]{Christopher C. Lovell}
\affiliation{Astronomy Centre, University of Sussex, Falmer, Brighton BN1 9QH, UK}
\affiliation{Institute of Cosmology and Gravitation, University of Portsmouth, Burnaby Road, Portsmouth, PO1 3FX, UK}

\author[0000-0003-4597-6739]{Boon Kiat Oh}
\affiliation{Department of Physics, University of Connecticut, Storrs, CT 06269, USA}

\author[0000-0003-4196-0617]{Camilla Pacifici}
\affiliation{Space Telescope Science Institute, 3700 San Martin Drive, Baltimore, MD 21218, USA}

\author[0000-0002-8449-1956]{Lucia A. Perez}
\affiliation{Center for Computational Astrophysics, Flatiron Institute, 162 Fifth Avenue, New York, NY 10010, USA}
\affiliation{Department of Astrophysical Sciences, Princeton University, Princeton, NJ 08544, USA}

\author[0000-0002-2906-2200]{Laura Sommovigo}
\affiliation{Center for Computational Astrophysics, Flatiron Institute, 162 Fifth Avenue, New York, NY 10010, USA}

\author[0000-0003-2573-9832]{Joshua S. Speagle (\begin{CJK*}{UTF8}{gbsn}沈佳士\ignorespacesafterend\end{CJK*})}
\affiliation{Department of Statistical Sciences, University of Toronto, 9th Floor, Ontario Power Building, 700 University Ave, Toronto, ON M5G 1Z5, CA}
\affiliation{David A. Dunlap Department of Astronomy \& Astrophysics, University of Toronto, 50 St George Street, Toronto ON M5S 3H4, CA}
\affiliation{Dunlap Institute for Astronomy \& Astrophysics, University of Toronto, 50 St George Street, Toronto, ON M5S 3H4, CA}
\affiliation{Data Sciences Institute, University of Toronto, 17th Floor, Ontario Power Building, 700 University Ave, Toronto, ON M5G 1Z5, CA}

\author[0000-0002-8224-4505]{Sandro Tacchella}
\affiliation{Kavli Institute for Cosmology, University of Cambridge, Madingley Road, Cambridge CB3 0HA, UK}
\affiliation{Cavendish Laboratory, University of Cambridge, 19 JJ Thomson Avenue, Cambridge CB3 0HE, UK}

\author[0000-0002-1185-4111]{Megan T. Tillman}
\affiliation{Department of Physics and Astronomy, Rutgers University, 136 Frelinghuysen Road, Piscataway, NJ 08854, USA}

\author[0000-0002-4816-0455]{Francisco Villaescusa-Navarro}
\affiliation{Center for Computational Astrophysics, Flatiron Institute, 162 Fifth Avenue, New York, NY 10010, USA}
\affiliation{Department of Astrophysical Sciences, Princeton University, Princeton NJ 08544, USA}

\author[0000-0002-5077-881X]{John F. Wu}
\affiliation{Department of Physics \& Astronomy, Johns Hopkins University, Baltimore, MD 21218, USA}
\affiliation{Space Telescope Science Institute, 3700 San Martin Drive, Baltimore, MD 21218, USA}
\affiliation{Department of Computer Science, Johns Hopkins University, Baltimore, MD 21218, USA}

\begin{abstract}
Star formation in galaxies is regulated by the interplay of a range of processes that shape the multiphase gas in the interstellar and circumgalactic media. Using the CAMELS suite of cosmological simulations, we study the effects of varying feedback and cosmology on the average star formation histories (SFHs) of galaxies at $z\sim0$ across the IllustrisTNG, SIMBA and ASTRID galaxy formation models. We find that galaxy SFHs in all three models are sensitive to changes in stellar feedback, which affects the efficiency of baryon cycling and the rates at which central black holes grow, while effects of varying AGN feedback depend on model-dependent implementations of black hole seeding, accretion and feedback. We also find strong interaction terms that couple stellar and AGN feedback, usually by regulating the amount of gas available for the central black hole to accrete. Using a double power-law to describe the average SFHs, we derive a general set of equations relating the shape of the SFHs to physical quantities like baryon fraction and black hole mass across all three models. We find that a single set of equations (albeit with different coefficients) can describe the SFHs across all three CAMELS models, with cosmology dominating the SFH at early times, followed by halo accretion, and feedback and baryon cycling at late times. Galaxy SFHs provide a novel, complementary probe to constrain cosmology and feedback, and can connect the observational constraints from current and upcoming galaxy surveys with the physical mechanisms responsible for regulating galaxy growth and quenching.
\end{abstract}


\section{Introduction} \label{sec:intro}

\begin{figure*}
    \centering
    \includegraphics[width=\textwidth]{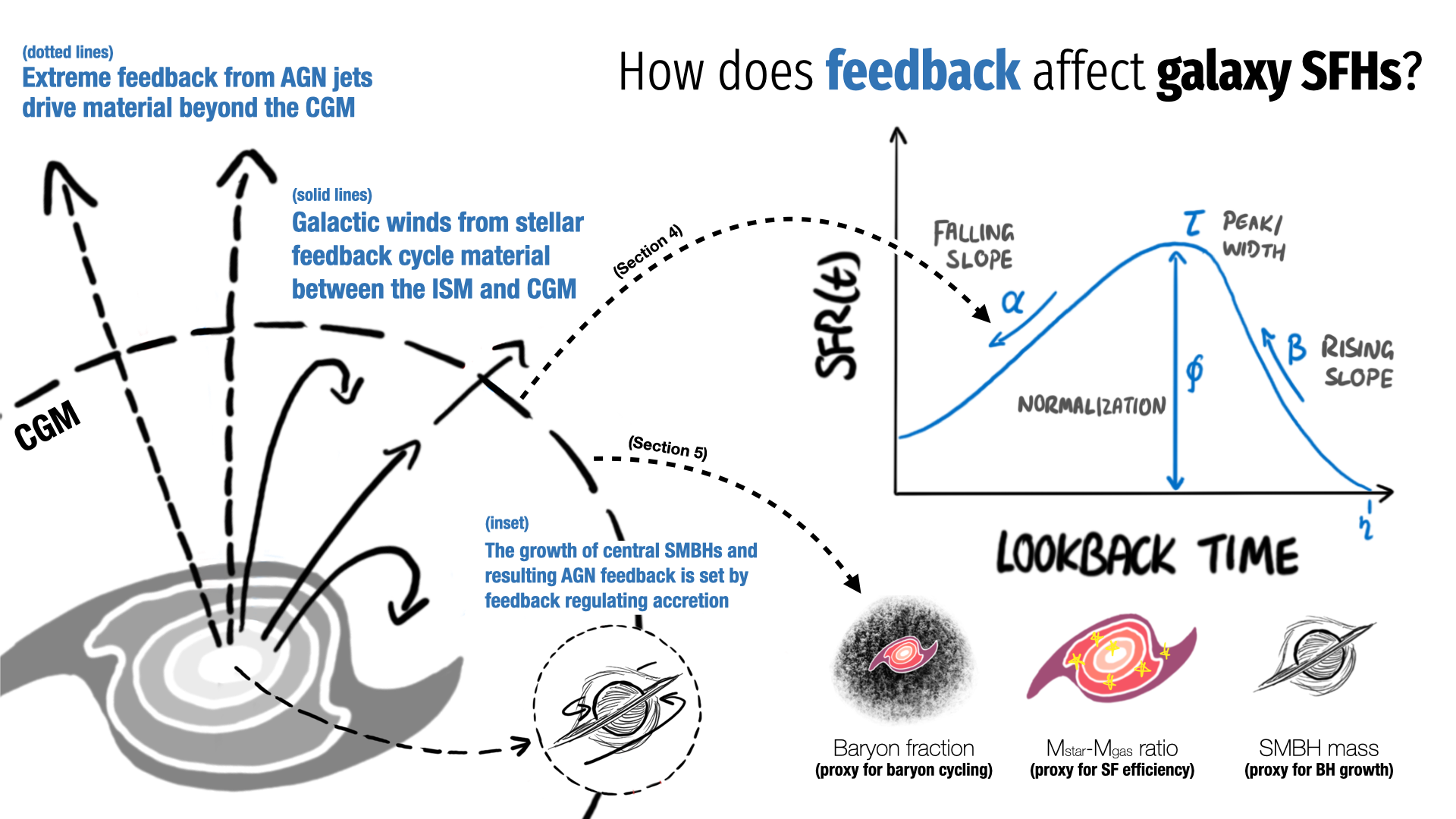}
    \caption{\textbf{Motivation:} The star formation histories (SFHs) of galaxies act as tracers of the overall gas regulation in the galaxy, which depend sensitively on the adopted prescription and strength of stellar and AGN feedback. By analyzing how the shape of the average SFH at a given mass \ch{(described using normalization ($\phi$), peak/width ($\tau$), and rising and falling slopes ($\alpha, \beta$)) depends} on the strength and nature of the feedback parameters in CAMELS, we aim to quantify their effects on galaxy evolution.}
    \label{fig:overall_schematic}
\end{figure*}

The formation and evolution of galaxies is a complex, multi-scale process, regulated by factors ranging from the growth of dark matter halos on cosmological scales to the accretion and feedback from supermassive black holes on \ch{sub-}parsec scales   \citep[for a more detailed description, see reviews by][]{2013ARA&A..51..511K, 2015ARA&A..53...51S, 2017ARA&A..55...59N, 2018ARA&A..56..435W}. 
On cosmological scales, the stochastic inflow of pristine gas into a dark matter halo follows the accretion history of the halo itself and provides both fuel for in-situ star formation as well as stellar populations from galaxy mergers \citep{2005MNRAS.363....2K, 2011MNRAS.414.2458V, 2015MNRAS.450.1514C, 2016MNRAS.455.2592R, 2017MNRAS.470.4698A}.
Preventative and ejective feedback regulate the availability of cold dense gas in the multiphase interstellar medium (ISM), and thus the rate at which galaxies form stars \citep{2017MNRAS.465.1682H, 2017ApJ...845..133S, 2019MNRAS.489.4233M, 2021MNRAS.508.2979P, 2022ARA&A..60..319S}. 
\ch{Stellar feedback works locally to disrupt and regulate star formation and drive galactic winds, and its effects are particularly studied in low-mass halos and early galaxies \citep{2014MNRAS.445..581H, 2015MNRAS.454.2691M, 2018Galax...6..114Z,2018Galax...6..138R, 2020ApJ...905....4P, 2023Galax..11...21H}.}
Supermassive black holes (SMBHs) in the centers of galaxies tend to co-evolve with the galaxy itself and eventually drive strong outflows that lift gas out of the halo and heat the gas in the CGM, preventing further inflows and eventually shutting down star formation in galaxies \citep{2003ApJ...582..559V, 2008ApJ...676...33D, 2012ARA&A..50..455F, 2013ARA&A..51..511K, 2017MNRAS.465.3291W, 2018ApJ...860...14B, 2018ApJ...866...91C, 2020MNRAS.499..768Z, 2021ApJ...917...53A, 2023ApJ...945L..17T, 2024OJAp....7E..18H, 2024arXiv240407252S}. 
As this often happens as a galaxy grows more massive, it is balanced against an increasingly \ch{deeper} gravitational potential well that can lead to baryon cycling, where gas that has been blown out of the galaxy spends time in the CGM before falling back in and fueling star formation at later times \citep{2015MNRAS.454.2691M, 2015ApJ...811...73L, 2017A&A...601A.143F, 2017MNRAS.470.4698A, 2020ApJ...905....4P, 2021MNRAS.508.2979P, 2024ApJ...976..151V, 2024ApJ...976..150V}.
Understanding the precise role that stellar and AGN feedback play in coupling the state of the galaxy's ISM and star formation to the state of its central SMBH and CGM thus lie at the heart of several current open questions in understanding how galaxies evolve, \ch{ranging from the core-cusp problem \citep{2012MNRAS.421.3464P, 2014MNRAS.437..415D, 2015MNRAS.454.2092O, 2016ASSL..418..317B} to quenching massive galaxies at low redshifts \citep{2008ApJS..175..390H, 2015MNRAS.449.4105C,  2017MNRAS.472..949B, 2020MNRAS.491.2939O, 2020MNRAS.493.1888T, 2021MNRAS.507..175S, 2021ApJ...915...53D, 2024MNRAS.tmp.1172L}}. 

This ongoing interplay between gas supply and feedback mechanisms leaves distinct imprints on a galaxy's star formation history (SFH), both on an individual and an ensemble level.
While variations across individual galaxies can be difficult to interpret due to the stochastic nature of gas inflows and mergers, broad trends observed in the SFHs of galaxy populations provide valuable clues about the strength and influence of the various feedback processes involved in galaxy evolution \citep{2020MNRAS.497..698T, 2020MNRAS.498..430I, 2024ApJ...961...53I}.
The goal of this paper is to examine how changes in stellar and AGN feedback strength in cosmological simulations affect overall trends in galaxy SFHs. This provides a basis for interpreting the growing body of observational data on galaxy SFHs across cosmic time in the context of physical processes regulating star formation.

The premise of constraining the nature and strength of feedback from observations has been explored in a variety of ways, including cosmological studies of the matter power spectrum \citep{2021MNRAS.502.6010H}, stellar mass functions \citep{2015MNRAS.450.1937C}, cosmic star formation rate density \citep{2010MNRAS.402.1536S}, IGM ionization using the Lyman alpha forest \citep{2014MNRAS.445.2313B}, CGM X-ray properties \citep{2024arXiv241016361M, 2024arXiv241204559L}, and other global variables, by studying the environments of extreme observables such as the properties of dwarf galaxies \citep{2019MNRAS.486..655D} or the state of the CGM in massive galaxy clusters \citep{2002ApJ...576..601V}. Methods that rely on large populations of galaxies tend to be extremely sensitive to modeling assumptions and selection effects, while direct measurements of outflows, X-ray measurements of hot gas, etc. tend to focus on small samples of nearby galaxies. 

\textbf{In this paper, we aim to understand the effects of feedback on the average SFHs of galaxies, and explore using them as a novel, complementary probe to constrain the strength of supernova and AGN feedback for populations of galaxies across a range of redshifts.} This is made possible by recent advances: (i) improvements in techniques and data quality for observationally inferring the SFHs of individual galaxies, (ii) suites of state-of-the-art simulations varying the strength of many different types of feedback, and (iii) the development of inference techniques to connect the two. This provides a complementary approach to constraining the strength of feedback across galaxy populations, and is especially timely considering the next generation of observational facilities and cosmological simulations to which this framework can be applied.

Recent advances -- including deeper observations, broader wavelength baselines, and much larger samples -- have accelerated the inference of SFHs from observed data on both an individual and ensemble level. These efforts have led to the development of a new set of tools to model the spectral energy distributions (SEDs) of galaxies with a range of different non-parametric SFH priors \citep[see e.g.,][]{2019ApJ...879..116I, 2024ApJ...961...53I, 2019ApJ...876....3L}. These new tools have enabled more robust measurements of galaxy stellar masses and star formation rates, as well as the scaling relations that depend on them. They have also driven measurements of typical SFHs across different populations, of quenching and rejuvenation timescales, and of the short-timescale burstiness. However, while these detailed measurements now exist for $\mathcal{O}(10^6)$ objects, it remains difficult to use these SFHs to identify signatures of feedback and connect them to the underlying physical processes.

In lockstep with observational advances, cosmological simulations - both hydrodynamical and semi-analytic, have advanced significantly over the last decade to include a variety of physical prescriptions for both stellar and AGN feedback. \ch{The effects of any single feedback channel in a given simulation are difficult to quantify given the wide range of spatiotemporal scales that these processes act over \citep[see][and references within]{2020MNRAS.498..430I}. As such, it is often easier to isolate the effects of a particular mechanism by turning it off and rerunning the simulation, or by studying a set of small parameter variations. This type of analysis has been done with some large hydrodynamical models including EAGLE \citep{2015MNRAS.450.1937C}, Illustris \citep{2015MNRAS.448...59N}, IllustrisTNG \citep{2018MNRAS.473.4077P}, and SIMBA \citep{2024arXiv240407252S}. However, these can not accurately account for the interactions between different types of feedback and the resulting covariances in observables, which required the development of large suites of simulations that simultaneously vary multiple types of feedback over a large parameter space.}

\ch{Moreover}, since the relative strength of these feedback processes remains difficult to constrain observationally (and galaxies with similar physical properties can be shown to exist in a diverse range of physical scenarios), it is imperative to consider previously unexplored domains to infer the relative strength of feedback parameters. While SFHs have been shown to vary sensitively with the implementation of feedback in galaxy simulations, it remains difficult to quantify the precise effects of varying stellar or AGN feedback on galaxy SFHs. The CAMELS (Cosmology and Astrophysics with MachinE Learning Simulations) suite \citep{2021ApJ...915...71V, 2023ApJS..265...54V} provides an opportunity to fill this missing gap through sets of simulations with multiple models that vary the strength of different types of feedback (summarized in Table \ref{tab:datasets}). 
By studying the SFHs of galaxies in these simulations, we can quantify the effect of feedback on the average SFH for a sample of galaxies. 

Finally, it is also important to stress the difficulties associated with framing this as an inference problem. \ch{Up until recently}, it has been computationally intractable to produce a large set of simulations varying the strengths of a diverse set of feedback types. It has also been extremely difficult to run inference with large populations of galaxies, since there are a number of other parameters (dust, stellar population properties, and so on) that need to be marginalized over. Even with the large dataset provided by the CAMELS simulations, it would have been difficult to cover the full parameter space (and even judge coverage) of possible SFHs, let alone do inference to constrain the feedback strength. Modern inference techniques bridge this gap by training a neural network to learn a joint probability distribution using an `emulator' that maps between parameters of interest (i.e. the CAMELS box parameters controlling feedback strength in this case) and observables (the SFHs in this work and forward modeled photometry in a companion paper \citep{2024arXiv241113960L}). This allows for a more thorough exploration of parameter space and enables inference that is essential for practical applications of this kind of analysis. 

In this paper, we compile the SFHs for galaxies from the CAMELS suite of simulations, and study the effects of varying feedback and cosmology on average trends of star formation over time. To quantify the effect of varying frameworks and sub-grid recipes, we analyze the CAMELS/TNG, CAMELS/SIMBA and CAMELS/ASTRID sets of simulations that have varying implementations of stellar and AGN feedback. To facilitate comparison with observations and account for the limited resolution and box size of the CAMELS boxes, we only consider galaxies in a mass range that are not strongly affected by resolution or cosmic variance. In this range, we 
build a mapping between the CAMELS feedback parameters for that box and the SFHs of the galaxies in it. Using these results, for the first time, we can link aspects of the SFH shape and variability to the underlying parameters governing cosmology and effective feedback physics. While this study does have limitations, we take care to predict general trends that can be tested across other cosmological simulations and observational datasets.

This paper is structured as follows: we describe the CAMELS suite of cosmological hydrodynamical simulations in Section \ref{sec:dataset}, and describe the methods used to compile and analyze the SFHs in Section \ref{sec:methods}. In Section \ref{sec:result_sfh} we present our results detailing the effects of varying stellar and AGN feedback on star formation in the three simulation suites. In Section \ref{sec:result_galstate}, we correlate signatures in the SFHs with the baryon content and black hole masses of galaxies and discuss their implications in Section \ref{sec:discussion}, along with the caveats of the current analysis and its implications in context with the current literature. We summarize our conclusions in Section \ref{sec:conclusions} and describe plans for future work.

\section{Dataset: The CAMELS suite of simulations} \label{sec:dataset}

An `experimental' setup for studying the effects of varying feedback should involve varying the strength of the various feedback prescriptions in a cosmological simulation (i.e. perturbing the system) and measuring their effects on galaxy SFHs (studying the response). However, since the effects of stellar and AGN feedback act across a wide range of scales in both space and time, it is difficult to generalize varying a particular kind of feedback in isolation in a single model \citep[e.g., the TNG variations;][]{2018MNRAS.473.4077P}) to understand the full effects and interactions of that parameter in other regimes. We also need to be careful when comparing simulations, as it is difficult to disentangle the effects of numerical methods and input assumptions from varying physics \citep{2020MNRAS.498..430I}. 

We therefore use the CAMELS suite of magnetohydrodynamic (MHD) simulations \citep{2021ApJ...915...71V, 2023ApJS..265...54V, 2023ApJ...959..136N}, which provides a rich baseline of three different physical models (IllustrisTNG, SIMBA and ASTRID, and soon Swift-EAGLE\footnote{and more, see https://camels.readthedocs.io/en/latest/index.html}, \citealt{2024arXiv241113960L}), with a set of different realizations for each model \ch{that vary the cosmology and strengths of SNe and AGN feedback}. For the purposes of this paper, we use the CAMELS CV\footnote{Cosmic Variance} (27 runs), 1P\footnote{1 Parameter} (61 runs) and LH\footnote{Latin Hypercube} (1000 runs) datasets. The CV dataset includes multiple runs with the same fiducial cosmology and feedback parameters but different random seeds. The 1P set includes runs varying a single parameter at a time, holding all the others constant at the fiducial values, and are mostly used for validation purposes. The LH datasets contain 1000 runs per model, employing a Latin hypercube sampling of the 6-dimensional parameter space varying cosmology and feedback and are used to train the machinery used in the rest of this work. All runs have the same initial conditions and are evolved over a volume of $(25 h^{-1}\rm{Mpc})^3$, with further details found in \cite{2021ApJ...915...71V, 2023ApJS..265...54V, 2023ApJ...959..136N}. 

\subsection{Feedback Across the Three Models} \label{sec:model_desc}

\begin{figure*}
    \centering
    \includegraphics[width=\textwidth]{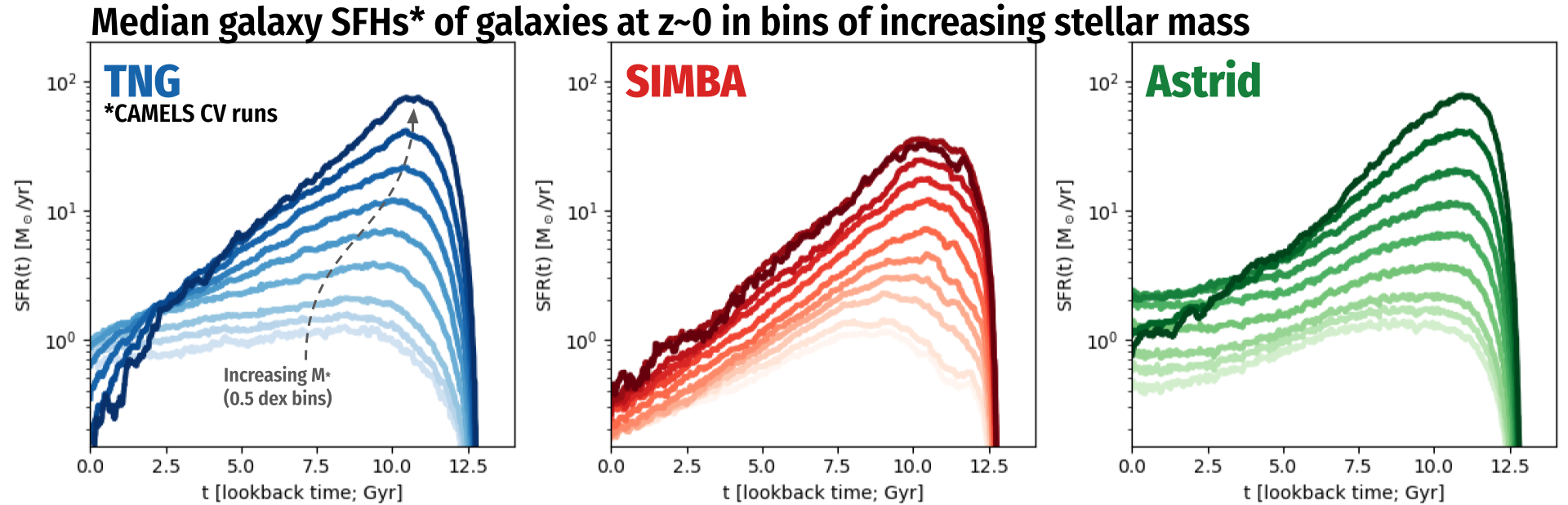}
    \caption{Median SFHs in different stellar mass bins from the three CAMELS models we use in this work, compiled using the CV datasets. The plot shows the SFHs of galaxies starting at $10^{9.25}$M$_\odot$ in increments of 0.2 dex up to $10^{10.85}$M$_\odot$, with darker colors indicating higher mass bins.}
    \label{fig:fig1_cv_sfhs}
\end{figure*}

We briefly describe the three galaxy formation and evolution models implemented in CAMELS that we use in the present analysis and the various feedback parameters that they use. In all three models the scalings for galactic winds driven by stellar feedback are described using the \sna{} and \snb{} parameters, which are meant to scale the energy from stellar feedback and wind speed respectively. AGN feedback is described using the \agna{} and \agnb{} parameters, which are roughly equivalent to their stellar counterparts but tend to have distinct implementations based on the model considered due to the way SMBH seeding, accretion and feedback are implemented. While a brief summary of the various models and their implementations of the feedback parameters is given below, we recommend that the readers refer to \citealt{2021ApJ...915...71V} for the TNG and SIMBA models, and \citealt{2023ApJ...959..136N} for the ASTRID model specificiations. A summary of the relevant properties for the three models is given in Table \ref{tab:datasets}.  We describe the relevant quantities below for completeness.

\begin{table*}
    \centering
    \begin{tabular}{c | c c c}
        \hline
       Feedback type & TNG & SIMBA & ASTRID \\
\hline
\sna{}  & \textbf{Energy per unit SFR} & \textbf{Mass loading factor} & \textbf{Energy per unit SFR} \\
 &  (Galactic Winds; range: (0.25,4)) &  (Galactic Winds; range: (0.25,4)) &  (Galactic Winds; range: (0.25,4)) \\
\snb{}  & \textbf{Wind speed} & \textbf{Wind speed} & \textbf{Wind speed} \\
&  (Galactic Winds; range: (0.5,2)) &  (Galactic Winds; range: (0.5,2)) &  (Galactic Winds; range: (0.5,2)) \\
\agna{}  & \textbf{Energy per unit accretion \ch{(Kinetic)}} & \textbf{Momentum flux} & \textbf{Energy per unit accretion (Kinetic)} \\
&  (AGN feedback; range: (0.25,4)) &  (AGN feedback; range: (0.25,4)) &  (AGN feedback; range: (0.25,4)) \\
\agnb{}  & \textbf{Ejection speed / burstiness} & \textbf{Jet speed} & \textbf{Energy per unit accretion (Thermal)} \\
&  (AGN feedback; range: (0.5,2)) &  (AGN feedback; range: (0.5,2)) &  (AGN feedback; range: (0.25,4)) \\
\hline
    \end{tabular}
    \caption{Summary of the various feedback implementations across the three models, adapted from Table 1 in \citealt{2023ApJ...959..136N}. \ch{The \sna{} and \snb{} parameters scale the energy per unit star formation rate (or mass loading factor in SIMBA) and wind speed for stellar feedback, respectively. The \agna{} and \agnb{} parameters scale the energy per unit accretion or momentum flux of AGN feedback and determine which specific feedback modes are affected in each model, as detailed in Section \ref{sec:model_desc}. The scenario \sna$ = $\snb$ = $\agna$ = $\agnb$ = 1$ corresponds to the fiducial runs of each simulation.}}
    \label{tab:datasets}
\end{table*}

\subsubsection{TNG}

The Next Generation of the Illustris simulation (IllustrisTNG; \citealt{2017MNRAS.465.3291W, 2018MNRAS.473.4077P}) is the successor to the AREPO-based Illustris simulation that implements MHD equations to simulate galaxy formation and evolution. 

Star formation follows the \citet{2003MNRAS.339..289S} formalism in which stars form in gas above a threshold of $n_H \sim 0.13$ cm$^{-3}$ on timescales that depend on the density. 
Galactic winds driven by stellar feedback are implemented by particles that are hydrodynamically decoupled and stochastically ejected from star forming gas with an additional thermal component. The wind speed $v_w$ and total energy injection rate per unit star formation $e_w$ depend on local gas conditions, metallicity and dark-matter velocity dispersion, along with redshift and two global normalization parameters, \sna{} and \snb{}, described in more detail in \cite{2021ApJ...915...71V}. \ch{For completeness, we also include the equations here. The energy per unit SFR is given by,} 
\begin{equation}
    e_w = A_{\rm SN1} \bar{e}_w \left[ f_{w,Z} + \frac{ 1-f_{w,Z} }{(1+Z/Z_{\rm ref})^{\gamma_{w,z}}} \right] N_{\rm SNII} E_{\rm SNII,51} 
\end{equation} 
\ch{and the wind speed by,}
\begin{equation}
    v_w = A_{\rm SN2} \times {\rm max} \left[ \kappa_w \sigma_{\rm DM} \left(\frac{H_0}{H(z)}\right)^{1/3}, v_{\rm w,min} \right]
\end{equation}
\ch{where $E_{\rm SNII,51}$ is the energy per type-II SNe in units of $10^{51} {\rm erg /M} _\odot$, Z is the gas metallicity, $\sigma_{\rm DM}$ is the local dark matter velocity dispersion calculated by SUBFIND, H is the Hubble constant, and values for the parameters $\bar{e}_w, f_{w,Z}, \gamma_{w,z}, N_{\rm SNII}, E_{\rm SNII,51}, Z_{\rm ref},\gamma_{w,z}, \kappa_w, v_{\rm w,min}$ can be found in \citet{2018MNRAS.473.4077P}. The resulting mass loading factor is $\eta_{TNG} = \dot{M}_{w} /SFR = 1.8 v_w^2e_w$.} 

Halos with M$_{FoF} > 5\times 10^{10}$M$_\odot h^{-1}$ are seeded with supermassive black holes (SMBHs) with an initial mass of M$_{BH} = 8\times 10^{5}$M$_\odot h^{-1}$, with instantaneous merging of SMBH particles that are within each other's `feedback spheres'. SMBHs grow by \citet{1952MNRAS.112..195B} accretion, and AGN feedback is implemented in three modes: thermal, kinetic and radiative. Although minor, the radiative mode is always active and modifies the heating and cooling of the host halo of the SMBH by adding radiative flux to the cosmic ionizing background. The thermal and kinetic modes are coupled to the Eddington ratio of the instantaneous accretion rate onto the SMBH, with kinetic feedback operating at low accretion rates \ch{(and generally when $M_{\rm BH} \gtrsim 10^8$M$_\odot$; see \citealt{2023AJ....166..228T})} scaled by the \agna{} and \agnb{} parameters, and thermal feedback operating at high accretion rates. \agna{} scales the energy per unit BH accretion, \ch{given by,} 
\begin{equation}
    \dot{E}_{low} = A_{\rm AGN1} \times \dot{M}_{\rm BH} c^2 {\rm min} \left[ \frac{\rho}{0.05\rho_{\rm SFthresh}}, 0.2 \right]
\end{equation}
\ch{where $\rho$ is the gas density around the SMBH and $\rho_{\rm SFthresh}$ is the density threshold for star formation. This energy builds up over time and is released into the `feedback sphere' as discrete bursts in a random direction, when the energy accumulated since the last feedback event exceeds a threshold $E_{inj, min}$.} \agnb{} controls the ejection speed or burstiness of how gas is ejected into the `feedback sphere', \ch{given by,} 
\begin{equation}
    E_{\rm inj, min} = A_{\rm AGN2} \times f_{\rm re} \frac{1/2}{\sigma_{\rm DM}^2 m_{\rm enc}}
\end{equation}
\ch{where $\sigma_{\rm DM}^2$ is the one-dimensional dark matter velocity dispersion around the SMBH, $m_{\rm enc}$ is the gas mass in the feedback sphere, and  $f_{re} = 20$ is a constant of the fiducial TNG model.}
A lower \agnb{} value results in more feedback events, but with each event having lower energy, while higher \agnb{} results in fewer, higher energy events.  Since the Eddington ratio depends on SMBH mass, the transition from high to low mode occurs around $M_{\rm BH}\sim 10^8$M$_\odot$, corresponding to stellar masses of roughly $M_* \sim 10^{10.5} - 10^{10.8}$M$_\odot$, which means that our TNG sample of galaxies unfortunately does not have many galaxies that show dramatic quenching due to AGN feedback, or display significant effects from varying the strengths of AGN feedback \citep{2023AJ....166..228T, 2023ApJ...945L..17T}, because of the relatively small volumes of the simulations we analyze.

\subsubsection{SIMBA}

SIMBA \citep{2019MNRAS.486.2827D} is the successor to the MUFASA \citep{2016MNRAS.462.3265D} simulation that implemented a GIZMO-based hydrodynamics model for modeling galaxy evolution. 

Star formation occurs as molecular gas is probabilistically turned into stars on a dynamical timescale $SFR = \epsilon_* \rho_{\rm H_2} / t_{\rm dyn}$ where $\epsilon_* = 0.02$ \citep{1998ApJ...498..541K}.
Galactic winds from stellar feedback are implemented by hydrodynamically decoupled two-phase winds. Star forming gas particles are stochastically ejected according to the mass factor $\eta = \dot{M}_{\rm wind} / SFR$ and wind velocity $v_w$. The mass loading factor scales with stellar mass using scalings from the FIRE zoom-in simulations as implemented in \cite{2017MNRAS.470.4698A}, scaled by an overall factor given by \sna{}, so that the total mass ejected goes up with higher \sna{}. \ch{This is given by,}
\begin{equation}
    \eta (M_*) = A_{\rm SN1} \times 9 \left( \frac{M_*}{M_0} \right)^{\gamma_{M_*}}
\end{equation}
\ch{where $\gamma_{M_*} = -0.317$ if $M_* < M_0$ and $\gamma_{M_*} = -0.761$ if $M_* > M_0$ for $M_0 = 5.2(10)^9$M$_\odot$.}
The wind velocity is also based on FIRE and scales with circular velocity following \cite{2015MNRAS.454.2691M}, scaled overall by \snb{}, \ch{given by,}
\begin{equation}
    v_w = A_{\rm SN2} \times 1.6 \left( \frac{v_{\rm circ}}{200km/s} \right)^{0.12}v_{circ} + \Delta v (0.25 R_{\rm vir})
\end{equation}
\ch{where $\Delta v (0.25 R_{\rm vir})$ is the velocity corresponding to the potential difference between the launch point and $0.25R_{\rm vir}$ where \citet{2015MNRAS.454.2691M} is calibrated.}
While SIMBA limits the wind kinetic energy to the available SN energy CAMELS leaves this free to scale with the feedback strength parameters. Stellar feedback from Type 1a SNe and AGN wind heating remains unchanged. 

SMBHs with $M_{\rm seed} = 10^4$M$_\odot$ are seeded in galaxies with M$_* \sim 10^{9.5}$M$_\odot$, since FIRE shows that stellar feedback suppresses the growth of SMBHs in lower-mass galaxies. SMBH growth follows a two-phase model with cold gas accreting using the gravitation torque accretion model presented in \cite{2017MNRAS.464.2840A} and hot gas accretion proceeds through the \cite{1952MNRAS.112..195B} model. AGN feedback is implemented in a two-mode approach motivated by observations, with high mass loading outflows in the radiative quasar mode (higher Eddington ratios) and faster outflows with lower mass loading in the jet mode (lower Eddington ratios, \ch{triggered for black holes with M$_{\rm BH} > 10^{7.5}$M$_\odot$ and $\lambda_{Edd} < 0.2$}). The low Eddington ratio mode also has an X-ray mode that turns on when jet speeds reach the maximum allowed velocity, which helps quench galaxies at lower redshift \citep{2024arXiv240407252S}. The jet mode decouples hydrodynamically for a time that scales with $t_H$, can travel up to 10 kpc before recoupling, and heats the ejected gas to the virial temperature of the host. The total momentum flux of the ejected gas in both modes scales with \agna{}, \ch{given by,}
\begin{equation}
    \dot{P}_{\rm out} = \dot{M}_{\rm out} v_{\rm out} = A_{\rm AGN1} \times 20 L_{\rm bol}/c
\end{equation}
\ch{where $L_{\rm bol} = \epsilon_r \dot{M}_{\rm BH} c^2$ is the bolometric luminosity}.
and the jet velocity scales with \agnb{}, \ch{given by,} 
\begin{equation}
    v_{\rm out} = v_{\rm rad} + A_{\rm AGN2} \times v_{\rm jet} 
\end{equation}
\ch{only in the case where $M_{\rm BH} > 10^{7.5}$M$_\odot$ and $\lambda_{\rm Edd} < 0.2$, and $v_{\rm jet} = 7000 \times$min$[1,log_{10}(0.2/\lambda_{\rm Edd})] $km/s, and $\lambda_{\rm Edd} = \dot{M}_{\rm BH}/\dot{M}_{\rm Edd}$.}
In contrast with the random direction in TNG and ASTRID along which gas is ejected during kinetic feedback, the jet mode bipolar injections in SIMBA are parallel to the angular momentum vector of the disk, which makes it easier for the feedback to consistently eject gas out to large distances \citep{2023ApJ...945L..17T, 2023AJ....166..228T}. 

\subsubsection{ASTRID}

The ASTRID simulation \citep{2022MNRAS.512.3703B} is \ch{based on} an MP-Gadget based MHD code designed to be extremely scalable. Star formation is implemented using the \cite{2003MNRAS.339..289S} model based on gas fraction and local column density. Galactic winds are driven by stellar feedback and implemented with hydrodynamically decoupled wind particles. A wind particle remains decoupled for a minimum of 60 Myr or $20$ kpc/$v_w$, and recouples when its density drops by a factor of 10. Similar to TNG, \sna{} scales the total energy injected per unit SFR, and \snb{} controls the wind speed, which is proportional to the local 1D dark matter velocity dispersion, \ch{given by,}
\begin{equation}
    v_w = A_{\rm SN2} \times \kappa_w \sigma_{\rm DM}
\end{equation}
\ch{where $\kappa_w=3.7$} following the Illustris model. The SN feedback model in ASTRID is purely energy driven, so the mass loading asymptotically scales with wind speed as $\eta \propto v_w^{-2}$. \ch{In the fiducial ASTRID model, $\eta_{\rm w,fid} = (v_w/\sigma_{\rm 0,fid})^{-2}$ with $\sigma_{\rm 0,fid} = 353 $km/s \citep{2022MNRAS.512.3703B, 2023ApJ...959..136N}. This results in a mass loading for the ASTRID suite given by,}
\begin{equation}
    \eta_w = \frac{\dot{M}_w}{\dot{M}_{\rm SFR}} = A_{\rm SN1} \times \eta_{\rm w, fid} = A_{\rm SN1} \left( \frac{\sigma_{\rm 0,fid}}{v_w}\right)^2
\end{equation}

\ch{In contrast to the full-resolution ASTRID simulation, CAMELS/ASTRID follows the BH seeding and dynamics prescription from the BLUETIDES simulation due to its lower resolution.} SMBHs with $M_{seed} = 5\times 10^5 h^{-1}$M$_\odot$ are seeded in galaxies with $M_h \sim 5 \times 10^{10}h^{-1}$M$_\odot$. The SMBH models in CAMELS/ASTRID are described in \cite{2023ApJ...959..136N} and have a thermal and kinetic component. SMBH accretion follows the \citet{1952MNRAS.112..195B} prescription with short periods of super-Eddington accretion allowed, \ch{with the Eddington ratio capped at $\chi_{thr,max} = 0.05$}. SMBH feedback follows a two-mode approach delineated by the Eddington ratio of the instantaneous accretion rate. The kinetic mode is only turned on for massive SMBHs with $M_{\rm BH} \gtrsim 5\times 10^8 h^{-1}$M$_\odot$. \agna{} scales the kinetic feedback strength in the same way as CAMELS/TNG, \ch{given by,}
\begin{equation}
    \Delta \dot{E}_{\rm low} = A_{\rm AGN1} \times \epsilon_{\rm f,kin}\dot{M}_{\rm BH} c^2 ~~~(\lambda_{\rm Edd} < \chi_{\rm thr})
\end{equation}
\ch{where $\epsilon_{\rm f,kin}$ scales with the local BH gas density and has a maximum value of 0.05. This energy is accumulated over time and released in a burst once it exceeds a threshold $E_{\rm inj, min} = f_{re}\frac{1}{2}\sigma_{\rm DM}^2m_{\rm enc}$, where $\sigma_{\rm DM}$ is the 1D DM velocity dispersion around the BH, $m_{\rm enc}$ is the gas mass in the feedback sphere, and $f_{re} = 5$ for the fiducial case.}
In contrast to SIMBA and TNG, \agnb{} scales the amount of energy deposited through thermal feedback, \ch{given by,}
\begin{equation}
    \Delta \dot{E}_{\rm high} = A_{\rm AGN2} \times \epsilon_{\rm f, th} \epsilon_r \dot{M}_{\rm BH}c^2 ~~~(\lambda_{\rm Edd} > \chi_{\rm thr})
\end{equation}
\ch{where the fiducial value for the mass-to-light conversion is $\epsilon_{r} = 0.1$ and $\epsilon_{\rm f, th} =0.05$ assumes that 5\% of the energy is injected thermally into the the surrounding gas.}
Given the more stringent conditions for turning on kinetic AGN feedback in ASTRID \citep{2023ApJ...959..136N}, the effects of this mode are expected to be milder compared to TNG. 

\subsection{Sample selection}

The CAMELS hydro boxes have relatively low resolution, with an initial star particle mass of $1.27\times 10^7h^{-1}$M$_\odot$ in a periodic box of volume of $(25 h^{-1}\rm{Mpc})^3$. To avoid contamination from `shot noise' contributions to the SFH arising from the probabilistic way star particles form from gas particles that satisfy star formation criteria, we restrict ourselves to subhaloes with stellar masses above M$_* \gtrsim 10^{9.5}$M$_\odot$ \ch{(in order to have at least $250$ star particles per subhalo to robustly calculate the SFH, using the robustness metrics calculated in \citet{2020MNRAS.498..430I})}. Similarly, to reduce the impact of stochasticity at high masses due to the small volume, we impose an upper threshold on stellar mass of M$_* \lesssim 10^{11.5}$M$_\odot$. To maintain comparisons with observational data, we use both centrals and satellites in the current analysis. We also repeat the analysis with only centrals and report wherever this makes a significant difference. 

\subsection{Computing galaxy SFHs}

To calculate the SFHs of individual galaxies, we follow a procedure similar to \citet{2020MNRAS.498..430I} considering the set of all star particles in a given subhalo and computing a mass-weighted histogram of their ages, correcting for an age and metallicity dependent mass loss in cases where the initial masses of star particles are not stored. For simplicity and comparison with results from previous work, we restrict the current analysis to the $z\sim 0$ snapshot. Since the observationally determined galaxy SFHs contain information about the (lookback time) ages of all the stars in the galaxy at the time of observation (i.e, they do not differentiate between in-situ and ex-situ star formation and only contain information \ch{about the stars present in a galaxy at the time of observation}), we match this by computing the lookback time SFHs using all the star particles in the subhalo at the time of `observation'; i.e., $z\sim 0$.

Figure \ref{fig:fig1_cv_sfhs} shows the median star formation histories of galaxies in stellar mass bins of 0.2 dex from $10^{9.25}$M$_\odot$ to $10^{10.85}$M$_\odot$ in the three models we consider (ASTRID, TNG, SIMBA). While the overall trends in the three simulations with increasing mass are similar, the times at which star formation peaks and the median quenching timescale differ across the simulations. 

\subsection{Considerations when comparing against observations}

\ch{To understand how feedback influences star formation across cosmic time, previous studies have mapped the relationship between feedback strength and a few key observables including the cosmic star formation rate density (SFRD), the stellar mass function (SMF) and galaxy color distributions} \citep{2021ApJ...915...71V, 2023ApJ...944...67J, 2024arXiv241113960L}.
However, given the relatively small box size and low resolution of the galaxies in the CAMELS multiverse, the mass range that we are confident \ch{is robust for our analysis} is quite limited. Since we are therefore incomplete for significant populations of both low mass dwarfs and massive cluster galaxies, instead of looking at individual galaxies or the cosmic SFRD, we instead consider \textbf{the average SFHs of 100 randomly selected galaxies} from any given box, acting as a rough proxy of the `mean' SFH of a galaxy in that mass range, and study the effects of varying feedback on this quantity instead. 

The strength of this approach is that we are sampling the SFHs of generic massive galaxies that are abundant in surveys like CANDELS/3D-HST\footnote{https://www.ipac.caltech.edu/project/candels} \citep{2011ApJS..197...35G, 2011ApJS..197...36K}, COSMOS\footnote{https://cosmos.astro.caltech.edu/page/hst} \citep{2007ApJS..172....1S, 2022ApJS..258...11W}, PanSTARRS \citep{2002SPIE.4836..154K, 2016arXiv161205560C}, GAMA\footnote{https://www.gama-survey.org/} 
\citep{2011MNRAS.413..971D}, DES\footnote{https://www.darkenergysurvey.org/} \citep{2005astro.ph.10346T}, and DESI \citep{2016arXiv161100036D, 2024arXiv240919066S} at redshifts up to cosmic noon, 
and JWST surveys like CEERS \citep{2023ApJ...946L..13F}, PRIMER \citep{2024MNRAS.533.3222D}, NGDEEP \citep{2024ApJ...965L...6B}, JADES \citep{2023arXiv230602465E} and COSMOS-Web \citep{2023ApJ...954...31C} at even higher redshifts, \textit{which makes it \ch{more straightforward} to compare against observations}. Another important caveat is that the varying depth and geometry of each survey impose complicated selection functions on the observed galaxies (although we expect most of these surveys to be mass complete at the range we \ch{consider here}, $M_* > 10^{9.5}$M$_\odot$).

\subsection{Other physical properties}

In addition to the galaxy SFHs, we also calculate the masses of each component (dark matter, stellar, gas and BH) for each galaxy in the sample, along with stellar and gas phase metallicity. These quantities are used to estimate the baryon fraction, stellar-to-gas mass ratio, and black hole masses at a given halo mass, which act as `state variables' that describe how galaxy properties change as feedback varies across the CAMELS boxes.

We also compute the specific star formation rate (${\rm sSFR} \equiv {\rm SFR}/$M$_*$), quenched fraction \citep[$f_Q$; defined as the fraction of galaxies with $sSFR < 0.2/t_H$ following e.g.,][and other related works]{2016ApJ...832...79P}, and the average duty cycle of the SFHs (the fraction of time a galaxy spends actively forming stars, defined as $f_{\rm duty} = \Delta t_{\rm sf}/ (t_{\rm obs}-t_{\rm form})$ where $t_{\rm sf}$ is the duration the galaxy spends actively forming stars and $t_{\rm form}, t_{\rm obs}$ are the times at which the galaxy formed and when it was observed, following \citealt{2023ApJ...954L..11G}) for the ensemble of galaxies in a given box, which helps to better connect to observable quantities that depend on the SFHs.  

\section{Methods} \label{sec:methods}

In this section, we briefly describe the methods used to train normalizing flows used to sample the average SFH (SFH$|\Theta$) space (Section \ref{sec:method_flows}), where $\Theta$ are a set of CAMELS box parameters and galaxy properties, to either predict the SFHs given a set of CAMELS parameters or perform inference to predict the CAMELS parameters given a distribution of SFHs. Followed by this, in Section \ref{sec:method_sfhshape} we compute summary statistics that describe the overall shape of the SFH on long timescales.  
In sections \ref{sec:method_gam} and \ref{sec:method_pysr}, we describe how we use this parameterization with generalized additive models (GAMs) and symbolic regression to determine feature importance in order to construct general equations that describe the dependence of the SFHs on $\Theta$ across the three CAMELS models.

\subsection{Sampling the average SFHs with normalizing flows} \label{sec:method_flows}

The LH set of simulations provides a way to sparsely sample a large parameter space, and are an ideal starting point to train machine learning based samplers that allow us to smoothly explore different parts of the CAMELS parameter space. 

We train two implicit likelihood inference (ILI, also sometimes called likelihood-free inference (LFI) or simulation-based inference (SBI) \citealt{2014MNRAS.443.1252L, 2023ApJ...944...67J, 2023ApJ...952...69D}) models using the LtU-ILI package \citep{2024arXiv240205137H} to bijectively learn the distributions $P(\Theta(\{ SFH_i \}) | \Theta_{\rm sim})$ and $P(\Theta_{\rm sim} | \Theta(\{ SFH_i \}) )$, where $\Theta_{\rm sim}$ are the CAMELS box parameters and halo mass ($\Omega_m$, $\sigma_8$, $A_{\rm SN1}$, $A_{\rm SN2}$, $A_{\rm AGN1}$, $A_{\rm AGN2}$, $M_{\rm halo}$) and the $\Theta(\{ SFH_i \})$ are samples of the average star formation histories of 100 galaxies in a given box at a fixed set of lookback times, or other state variables of interest ($M_*$, $M_{\rm gas}$, $M_{\rm BH}$, $f_{\rm baryon}$ etc.). We calculate the average SFH by sampling SFHs within the mass range from within a box, repeating the procedure 10 times to augment the training set size. We train the bijective neural networks for the dual purposes of this paper: (i) understanding the effect of varying cosmology and feedback on the average SFHs of galaxies, and (ii) to quantify the constraints on feedback and cosmology given observational measurements of galaxy SFHs in the context of the different CAMELS models. 

Practically, we use the SBI package within LtU-ILI, training a neural posterior estimator \citep[NPE;][]{greenberg2019automatic} consisting of an ensemble of a Masked Autoregressive Flow (MAF) with 50 hidden features and five neural transformations and a Mixture Density Network (MDN) with 50 hidden features and six mixture components, similar to the example in Section 5.5 of \citet{2024arXiv240205137H}. Similar to \cite{2024arXiv241113960L}, we use a batch size of 32 and a learning rate of $10^{-4}$, and an improvement-based stopping criterion during training, along with priors based on the parameter limits described in Table \ref{tab:datasets}. While other packages exist for training normalizing flows and SBI-based methods, we use LtU-ILI for its built-in validation and coverage metrics. 
\textbf{This allows us to sample the full distribution of SFHs at any point in the CAMELS parameter space, and thus run subsequent analyses to study the dependence of the SFH parameters on specific parameters while keeping the others fixed}. It also allows us to account for the fact that, given the fixed box size, there is an inherent variability in the random sample of 100 SFHs used to calculate the average which implicitly is incorporated as uncertainty in the predicted distribution. Finally, it allows us to invert the trained model and instead predict the box parameters given a set of observed SFHs. Appendix \ref{app:1p_vs_nn} show a comparison between the 1P set and samples from the trained model. More details are available in a companion paper by \cite{2024arXiv241113960L}. Unless otherwise mentioned, all the SFHs in the following sections are generated by sampling the trained normalizing flows. 

\subsection{Parametrizing SFH shapes} \label{sec:method_sfhshape}

To describe the effects of varying feedback strength on a galaxy's SFH, it is helpful to quantify a few key summary statistics that encode information about the SFH's shape. Some ways of doing this include fitting the SFH with a parametric form like a lognormal (which has been shown to accurately describe ensembles of galaxy SFHs; \citealt{2016ApJ...832....7A, 2017ApJ...839...26D}) and using the lognormal parameters ($\mu, \sigma$) as summary statistics. Another way is to use a non-parametric description, e.g. using a set of numbers $\{ t_X \}$ that correspond to the lookback times at which the galaxy formed X\% of its mass \citep{2013ApJ...762L..15P, 2019ApJ...879..116I}.

While these are all viable approaches, we adopt a double power-law form for the SFH motivated by a combination of simplicity and interpretability (following \citealt{2019ApJ...873...44C}), summarized in Figure \ref{fig:overall_schematic}, and described as follows:
\begin{equation}
    SFH(t) = \phi  \left( \left(\frac{t-\eta}{\tau}\right)^{\alpha} +  \left(\frac{t-\eta}{\tau}\right)^{-\beta} \right)^{-1}
\label{eqn:dbplaw}
\end{equation}

\begin{figure*}
    \centering
    \includegraphics[width=0.99\textwidth]{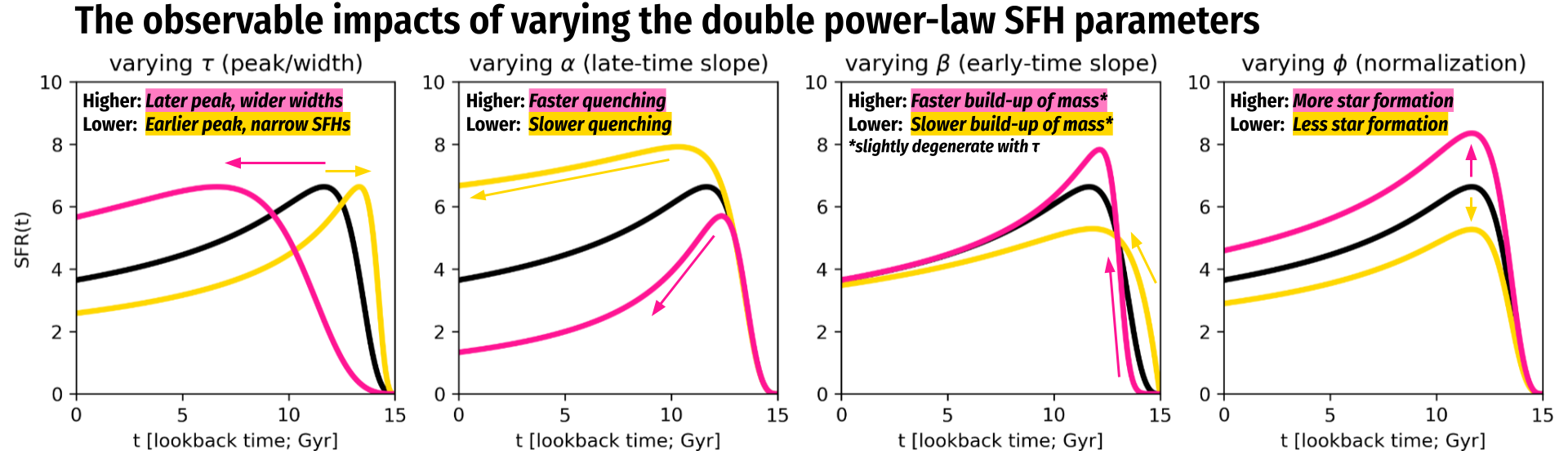}
    \caption{Changes in the average SFHs of galaxies with variations in the double power-law parameters $\tau$ (peak/width), $\alpha$ (falling slope), $\beta$ (rising slope), and $\phi$ (overall normalization), as described in Eqn. \ref{eqn:dbplaw}.}
    \label{fig:sfh_param_explain}
\end{figure*}

This effectively parametrizes the SFHs in terms of $\phi$ - an overall normalization, $\alpha, \beta$ - the slopes for the fall and rise of the SFH respectively, $\tau$ - a parameter that corresponds to the peak and width of the SFH, collectively denoted $\Theta_{SFH} \equiv \{ \alpha, \beta, \tau, \phi, \eta \}$. We also include an additional parameter for when star formation first begins ($\eta$) to account for changes in the age of the universe with varying $\Omega_m$. While these might not be optimal descriptors for the SFHs of individual galaxies because of multiple epochs of SFR, complex SFH trajectories or rejuvenation, they are useful for describing the broad rise and fall of the \textit{average} SFHs that we consider in this work (except for a small fraction of ASTRID SFHs that have a long sustained tail of  low-level star formation at late times that are not well-described by a power-law slope). The double power-law form also shows generally better convergence in fitting galaxy SFHs compared to other forms we tested (lognormal, power-law rise followed by exponential decline, and a six-parameter model that decouples $\tau$ into separate peak and width parameters), and relates to theoretical considerations for galaxy growth based on DM accretion in an extended Press-Schecter formalism \citep{2018ApJ...868...92T}. It is also related to other parametrizations such as the lognormal SFHs studied in \citet{2017ApJ...839...26D}  (where $\tau$ (peak/width) correlates with the mode of the lognormal $\exp (\mu - \sigma^2)$) and $t_{50}$ in non-parametric SFHs from \citet{2019ApJ...876....3L, 2019ApJ...879..116I}.

In addition to this, these parameters also lend themselves to physical interpretation - $\beta$ provides information about the epoch of most rapid star formation and mass buildup, $\alpha$ describes the behavior of the SFH during the epoch of quenching, and $\tau$ provides information about whether the star formation was concentrated at a single epoch or sustained over time, which \ch{is connected to the feedback strength}. 
\ch{Galaxies that have strong star early star formation followed by rapid quenching tend to have small $\tau$ values, in comparison to galaxies with large $\tau$ that have sustained star formation over a long period of time.}
Finally, the overall normalization ($\phi$) corresponds to both the relative enhancement or suppression of star formation in galaxies. Since we are sampling galaxies from a fixed range in stellar or halo mass, this tells us whether the average galaxy is more or less massive given a set of feedback conditions. 

A caveat in defining the parameters corresponding to these average SFHs is that for populations with significant dispersion in a certain parameter (e.g. quenching timescales), computing the average SFH and then estimating $\alpha$ might not accurately represent the behavior of individual galaxies. While we check in the current analysis that with $0.5$ dex bins in halo mass this is not a significant issue, it is something to be mindful of while extending the analysis across other samples and redshifts. 

\subsection{Quantifying the drivers of SFH change}

The effects of varying cosmology and feedback on galaxy SFHs are complex, and building a model to explain them needs to account for both the effects of individual parameter variations as well as interactions. In this section, we briefly describe the three methods we use to examine the relation between the average galaxy SFHs at a given halo mass and the corresponding feedback and state variable terms, using (i) feature importance with random forests, (ii) a linear functional decomposition with generalized additive models, and (iii) model construction with symbolic regression. While none of the methods are perfect in understanding the data and offering a causal interpretation, they offer complementary insights that we use to construct our model. 

\subsubsection{Feature importance with Random Forests}

Random forests have offered interpretable ways to approach feature importances by quantifying the mean decrease in node impurity or model accuracy for different features in the dataset \citep[see e.g.][]{2022A&A...659A.160B, 2023ApJ...944..108B}. We follow a similar procedure, training a random forest to predict the average SFR of a sample of galaxies at a given lookback time as a function of both CAMELS box parameters (\boxparam) and average galaxy state variables (\mhalo, \fb, \mbh). We repeat this across a range of lookback times to estimate the feature importances over a range of epochs from the early universe to present day. We then use the feature importances over cosmic time as an independent sanity check to determine whether the equations we derive in Section \ref{sec:disc_univmodel} encode physical meaning - i.e. do the rising slopes ($\beta$) of SFHs depend on similar features to what the random forest finds important at early times, and do the falling slopes ($\alpha$) depend on similar features as what the random forest finds important at late times? 

While random forests provide an informative estimate of the feature importance, they do not account for nonlinear interactions (e.g. when a feature is only important in combination with other variables) and parameter covariances (e.g. if two parameters are heavily correlated the feature importance might be spread out among them) very well, which can play a significant role in the current scenario. Moreover, our goal is to build a model rather than simply identify important features in the dataset. We therefore use it more as a sanity check to compare against our understanding from the following two methods. 

\subsubsection{Understanding nonlinearities and interactions using GAMs} \label{sec:method_gam}

Generalized additive models (GAMs; \citealt{hastie1990generalized}) are models of the form $f(x) = \sum_a f_a (x_b) + c$ that decompose the regression problem of predicting a response variable $f(x)$ as a sum of additive contributions from each input/dependent variable $(x_b)$. This forms a specialized subset of the Kolmogorov-Arnold representation theorem \citep{kolmogorov1957representation}, and is equivalent to a single-layer version of modern implementations like Kolmogorov-Arnold networks \citep{liu2024kan}. The individual variables can have arbitrary (and often nonlinear) dependencies, and are often estimated using a spline or other non-parametric form. In this work, we use the pyGAM package \citep{daniel_serven_2018_1476122} for our analysis, performing a regression of the form 
\begin{equation}
    \Theta_{\rm SFH} = \sum s_a(q_a) + \sum {s_{ab}}(q_a,q_b) + c
    \label{eqn:gam_params}
\end{equation}
where $q_a \in (\Omega_m$, $\sigma_8$, \sna{}, \snb{}, \agna{}, \agnb{}, \mhalo, \fb, \mbh$)$, $\Theta_{\rm SFH} \in (\tau, \alpha, \beta, \phi, \eta)$ and the functions $s_a$ and $s_{ab}$ are spline functions that account for the individual nonlinear dependence for each feedback parameter and interaction terms between the different feedback parameters respectively. In this case, we determine the $\Theta_{\rm SFH}$ parameters using a simple MCMC fit to each SFH in our sampled dataset. We initially ran the GAM analysis with no interaction terms (i.e. $s_{a,b} = 0$, and the SFH parameters are purely decomposed into a set of additive functions of the galaxy state variables), followed by a re-analysis adding each individual interaction term (i.e. a term of the form $s_{ab}(q_a,q_b)$) and measuring the corresponding decrease in the loss. This allows us to quantify the interaction terms that are responsible for $\Theta_{\rm SFH}$ values that are difficult for the model to predict without accounting for interactions. We show the effects of these interaction terms in Section \ref{sec:result_sfh_interactions}. Because of their inherent flexibility from not assuming a functional form for the relation between a parametric form for the relation between $q_a$ and $\Theta_{\rm SFH}$, they generally provide a lower limit of how well each $\Theta_{\rm SFH}$ parameter can be predicted with a maximally flexible model. We therefore use the GAM loss metrics as a proxy of the Bayes risk (i.e., the lowest error possible with a given model) while building and fine-tuning our equations for the different SFH terms. 

\begin{figure*}
    \centering
    \includegraphics[width=0.95\textwidth, trim={0 1.4cm 0 0},clip]{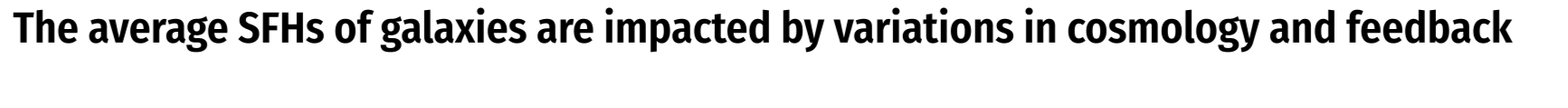}
    \includegraphics[width=0.95\textwidth, trim={0 0 0 1cm}, clip]{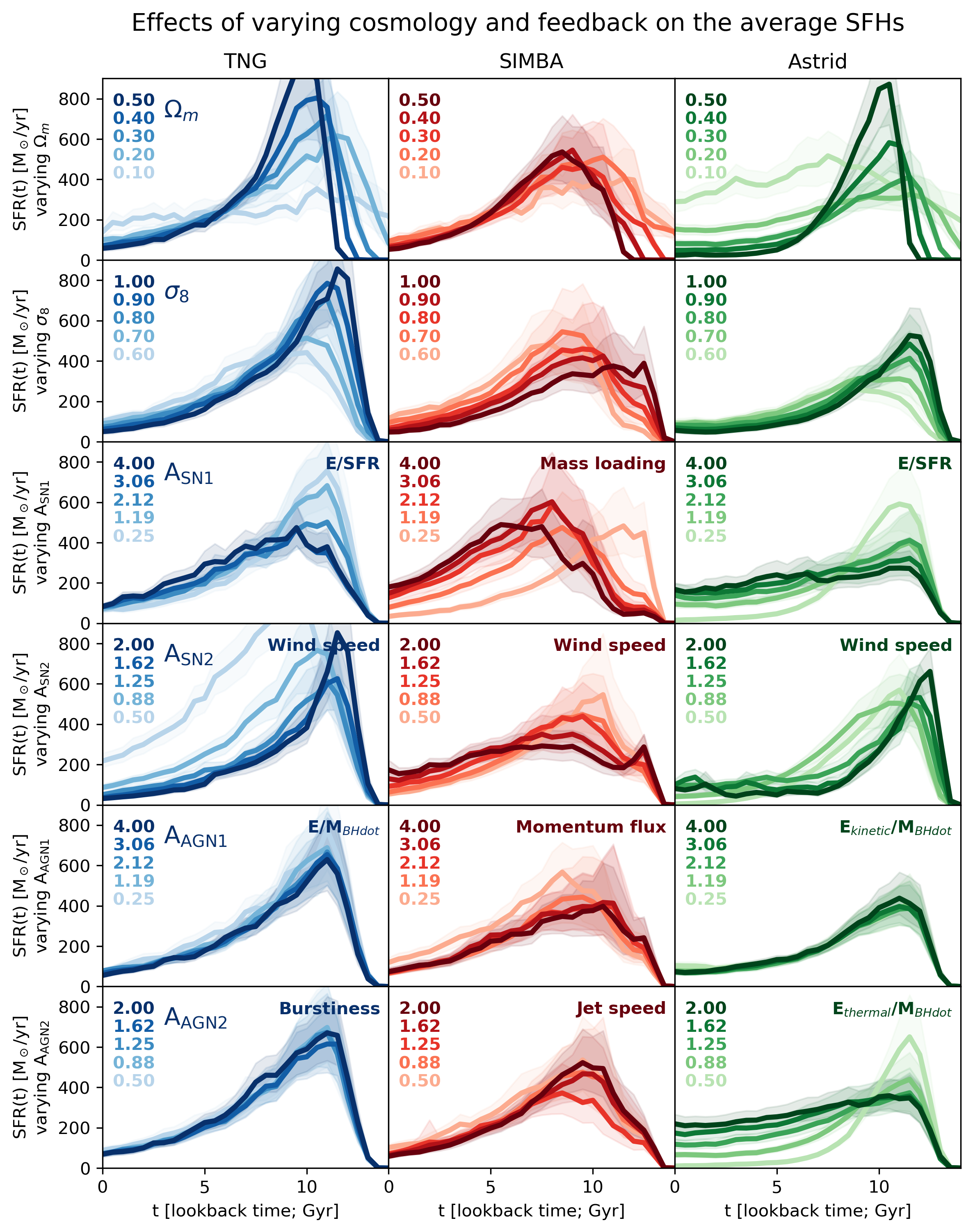}
    \caption{Variation in the average SFHs of 100 randomly selected galaxies with \textbf{stellar masses} \massrange{9.5}{11.5} caused by varying the cosmology and feedback strengths across the three CAMELS models (ASTRID, TNG, SIMBA), predicted using the trained normalizing flows. Solid lines and shaded regions show the median SFH and corresponding $1\sigma$ uncertainties from the bootstrap sampling used to train the normalizing flow. While some parameters can cause large variations (e.g. \snb{} in TNG), others do not affect the SFHs nearly as much (e.g. \agna{} in ASTRID).}
    \label{fig:sfh_1param_exploration}
\end{figure*}

\subsubsection{Feature and term importance using Symbolic Regression} \label{sec:method_pysr}

Similar to the GAM analysis, we use symbolic regression using the pySR package \citep{2024ascl.soft09018C} to help us determine functional forms that describe the relations between the SFH parameters ($\Theta_{\rm SFH}$) and the galaxy state variables ($\Theta_{\rm gal}$). In doing so, we found that the resulting equations relating these quantities are not very stable (i.e. the form of the equation is not preserved when changing the random initialization, when derived with bootstrap subsamples of the data, or when changing the sample selection criteria). Thus, instead of running symbolic regression a single time, we rerun the symbolic regression across multiple sub-samples and collate the resulting families of equations. We use these families of equations to determine the frequency of certain terms (see Figure \ref{fig:sfh_interaction_terms}) and interactions. These terms are subsequently used to construct equations that generalize across the different CAMELS models. This analysis uses the symbolic regression framework more as a feature importance metric, but in the space of equations describing the mapping between SFHs and the galaxy state variables. 

The other reason for not using the symbolic regression to directly derive equations is that the equations are likely to be different for each of the three CAMELS models, due to the differing implementations of sub-grid physics and parametrizations of the \sna{}, \snb{}, \agna{}, \agnb{} parameters. In addition to this, there are heavy covariances between the CAMELS box parameters and the galaxy state variables, due to how black hole mass or baryon fractions are affected by changing the strength of feedback. Rather than finding an optimal mapping between these parameters and the SFHs in a single model, our goal is to determine whether it is possible to empirically create a system of equations that can relate the shape of an SFH to the present-day state of the galaxy in a way that is independent of the choice of model and strength of feedback. 

In order to construct a generalizable system of equations, we use the terms from the symbolic regression as a starting point in constructing an equation for each SFH parameter, given by,
\begin{equation}
    \Theta_{\rm SFH} = \sum_i c_{i} f(\Theta_{{\rm sfh},i})
\end{equation}
where $\Theta_{\rm SFH} \in \{ \alpha, \beta, \tau, \phi, \eta \}$ and $f(\Theta_{\rm sfh})$ consists of terms from the symbolic regression that include both the CAMELS box parameters and average galaxy state variables like halo mass and baryon fraction, this time leaving the coefficients free in each model and minimizing the net loss across the models, i.e. 
\begin{equation}
    \mathcal{L} =  \sum_{\rm model} (\Theta_{\rm SFH} -  \sum_i c_{i,{\rm model}} f(\Theta_{{\rm sfh},i}))^2
\end{equation}
where $f(\Theta_{sfh,i})$ is the same term in the equation describing how to predict $\Theta_{\rm SFH}$ across the different models, but with a different coefficient. Thus the physical model describing the shape of the SFH based on the galaxy parameters does not change across TNG, SIMBA and ASTRID, but the coefficients can (and should) change depending on the extent to which different parameters affect the SFH. For example, \sna{} and \snb{}, play a strong role in suppressing black hole growth and increasing baryon cycling, with distinct imprints across the three models. Now, if cases where this effect can be fully described in terms of the resulting effect on the relative black hole mass and baryon fraction, then our equations are constructed to only include the latter, more general, terms. However, if the effect of \sna{} or \snb{} on the SFHs goes beyond what the summary variables alone can describe, then they are also included in the equations and are likely to have different coefficients depending on \ch{how strongly they impact} the SFHs across the three models.

\section{Results: How are SFHs Affected by Feedback?} \label{sec:result_sfh}

Armed with the machinery to sample and characterize SFHs across the CAMELS parameter space, we explore the effect of varying the cosmology and feedback scaling parameters on the average galaxy SFHs across the three CAMELS models in this section, analyzing SFHs sampled from a broad range of stellar masses (to match observational samples; Section \ref{sec:result_sfh_1P}) and in bins of halo mass (Section \ref{sec:result_sfh_mhalo}). While these focus on the effects of varying individual parameters one at a time, we also consider the effects of interactions in Section \ref{sec:result_sfh_interactions}. We connect variations \ch{of the CAMELS parameters} to variations in physical properties in Section \ref{sec:result_galstate}, and discuss the implications in Section \ref{sec:discussion}. 

\subsection{Single Parameter Variations}
\label{sec:result_sfh_1P}

\begin{figure*}[ht]
    \centering
    \includegraphics[width=0.99\textwidth, trim={0.9cm 3.6cm 3cm 0},clip]{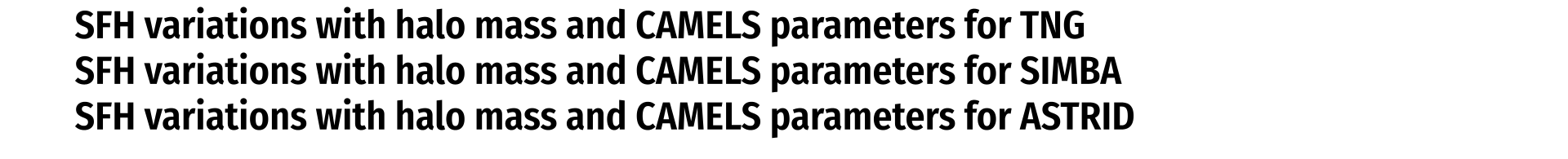}
    \includegraphics[width=0.99\textwidth]{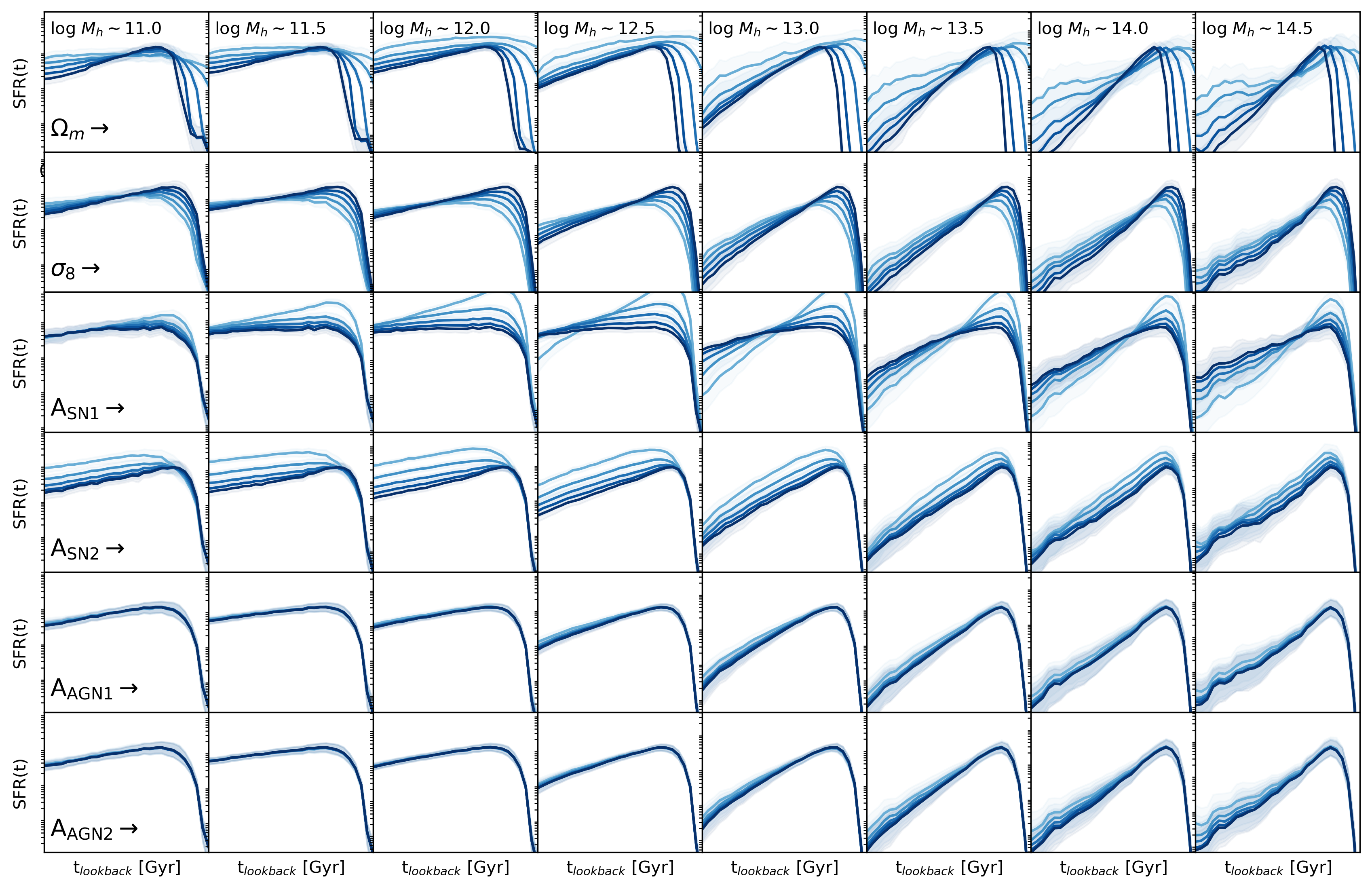}
    \caption{Similar to Figure \ref{fig:sfh_1param_exploration}, but breaking down the variations in the average SFHs for 100 galaxies within 0.5 dex \textbf{halo mass} bins predicted with the trained normalizing flows for parameter variations across \textbf{CAMELS/TNG}. Note that in contrast to Figure \ref{fig:sfh_1param_exploration},  the y-scale in these plots is logarithmic to account for the increased dynamic range in the SFHs. The various curves in each panel show the average SFHs for different values of each parameter, with darker colors indicating higher values. There is increased shot noise in the higher mass bins where the galaxies tend to be poorly sampled due to the limited volume of the CAMELS boxes. }
    \label{fig:sfh_1param_massbins_tng}
\end{figure*}

We first study the effects of varying feedback on the broad, overall shape of the SFH. Figure \ref{fig:sfh_1param_exploration} shows the overall effects of varying each of the CAMELS parameters on the resulting average SFHs of galaxies in \massrange{9.5}{11.5} across the three simulation suites (TNG, SIMBA and ASTRID). Each row presents the effect of varying a single parameter while holding others constant, allowing us to isolate their individual impacts on star formation over time. 
We use this stellar mass selected sample to facilitate comparisons with observations, using a range in stellar mass that is accessible to (and highly complete in) most modern galaxy surveys.
The SFHs shown are sampled from the normalizing flows trained on the LH datasets for the three models, though this particular figure can also be computed using the CAMELS 1P runs. The main advantage of using the normalizing flows are that we can also estimate the uncertainties in the SFHs (shown as shaded regions in Figure \ref{fig:sfh_1param_exploration}), which account for the variance due to randomly choosing 100 galaxies to compute the mean SFH as well as the epistemic uncertainty learned by the normalizing flow. When comparing to observations, this is crucial as it allows us to formulate likelihoods that take into account the modeling uncertainties in SFH space. 

\begin{figure*}
    \centering
    \includegraphics[width=0.99\textwidth, trim={0.9cm 2.1cm 3cm 1.5cm},clip]{sfh_var_mhalo_title.png}
    \includegraphics[width=0.99\textwidth]{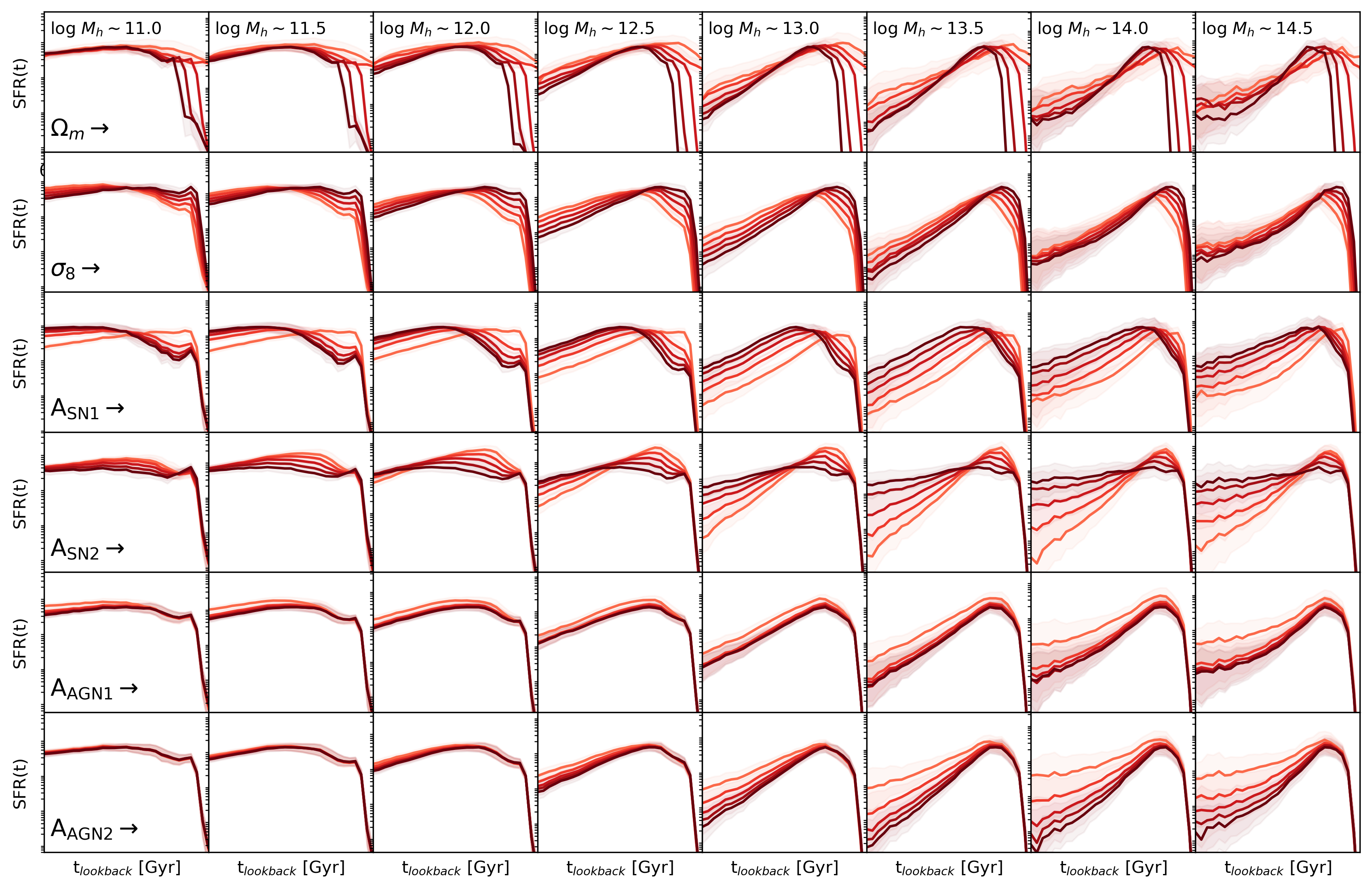}
    \caption{Similar to Figure \ref{fig:sfh_1param_massbins_tng}, but for \textbf{CAMELS/SIMBA}.}
    \label{fig:sfh_1param_massbins_simba}
\end{figure*}

\begin{figure*}[ht]
    \centering
    \includegraphics[width=0.99\textwidth, trim={0.9cm 0.6cm 3cm 3.0cm},clip]{sfh_var_mhalo_title.png}
    \includegraphics[width=0.99\textwidth]{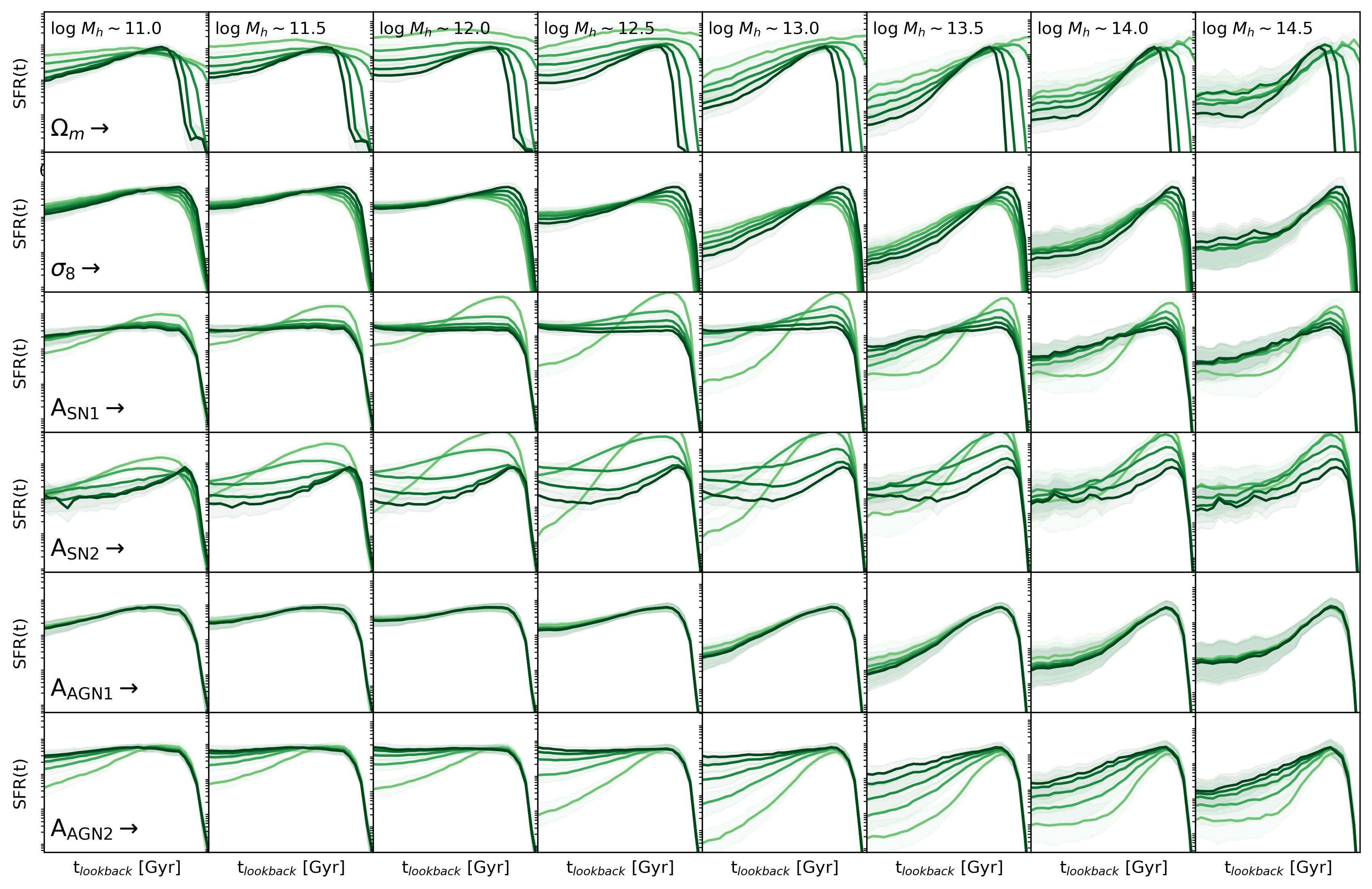}
    \caption{Similar to Figure \ref{fig:sfh_1param_massbins_tng}, but for \textbf{CAMELS/ASTRID}.}
    \label{fig:sfh_1param_massbins_astrid}
\end{figure*}

\begin{figure*}
    \centering
    \includegraphics[width=0.81\textwidth, page=1]{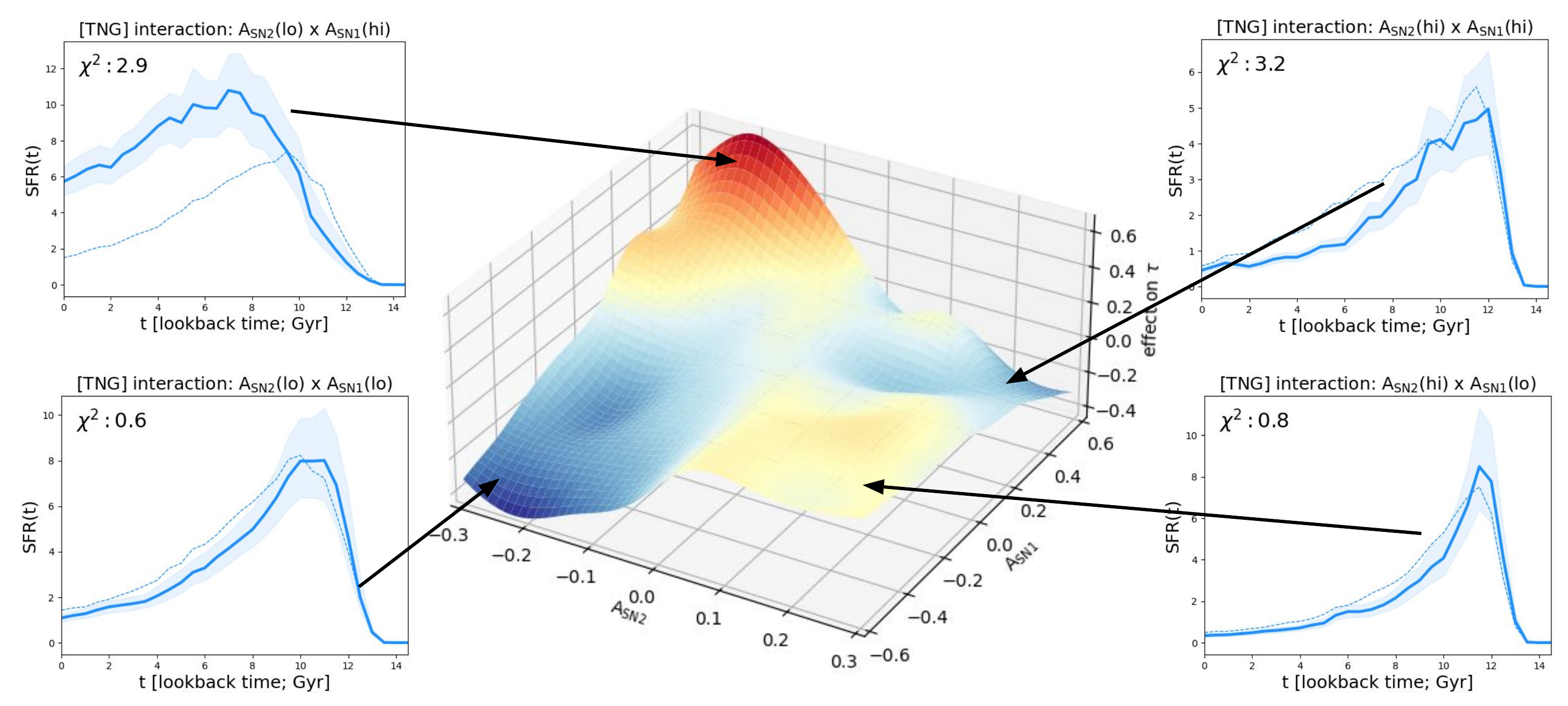}
    \includegraphics[width=0.81\textwidth, page=2]{sfh_interactions.pdf}
    \includegraphics[width=0.81\textwidth, page=3]{sfh_interactions.pdf}
    \caption{Interactions between different feedback and cosmology channels (thicker lines) can sometimes affect the SFHs beyond the \ch{linear} additive effects of variations in the individual parameters (thin dashed lines). The three examples \ch{illustrate the} effects of the interactions between the  galactic wind parameters \ch{(\sna{} and \snb{})} in TNG (top), between cosmology and feedback \ch{($\Omega_m$ and \sna{})} in SIMBA (middle) and between different modes of feedback \ch{(\snb{} and \agnb{})} in ASTRID (bottom). These interactions show \ch{one example from each model of} the nonlinear effects, indicating the need to go beyond simply studying 1-parameter variations to understand the full effects of feedback. \ch{A fuller treatment of this for all combinations of parameters can be found in Appendix \ref{app:all_interactions}.}}
    \label{fig:sfh_interaction_terms}
\end{figure*}

\textbf{Cosmology:} Varying the cosmology changes the total matter content of the universe, \ch{which impacts} halo clustering and formation time. These effects as seen in Figure \ref{fig:sfh_1param_exploration} are more rapid growth at early times with increasing $\Omega_m$ or $\sigma_8$, with all three models showing similar trends. This indicates that the change in galaxy properties (at least the SFH) with cosmology generalizes in a reasonably model-agnostic manner. The number of galaxies at all masses scales with $\Omega_m$, with fewer galaxies populating halos as the amount of matter in the universe decreases. TNG and ASTRID exhibit the strongest sensitivities to $\Omega_m$, with higher values leading to substantially enhanced star formation rates at early times, while SIMBA shows qualitatively similar but more moderate responses. On the other hand, varying $\sigma_8$ mostly affects the formation of high-mass galaxies, with a strong correlation between the number of galaxies in high mass ($M_h\gtrsim 10^{13.5}$M$_\odot$) halos and clustering. There is also a resulting difference in the SFHs at late times, especially for ASTRID. 

\textbf{Feedback:} Since the three CAMELS models have varying implementations of the feedback parameters, we consider these individually: 

\textbf{Varying feedback in CAMELS/TNG:} Overall, the average SFHs of galaxies in the TNG model are primarily affected by the strength of galactic winds (\sna{}, \snb{}), with AGN feedback only playing a role in specific regimes or for high-mass galaxies. This is not to say that the AGN feedback doesn't play a role in influencing galaxy growth and quenching, but rather that varying the \agna{} and \agnb{} parameters does not have a significant impact on the average SFHs of galaxies randomply sampled from our mass range. This is partly because the kinetic mode of AGN feedback that is controlled by the \agna{} and \agnb{} parameters \ch{is only active in} galaxies with $M_{\rm BH} > 10^{8.5}M_\odot$, which are limited due to the volume of the CAMELS boxes.

While an increase in either \sna{} (energy per unit SFR) or \snb{} (increased wind speeds) tends to suppress the amount of stars formed in galaxies at all masses, \sna{} also reduces the number of galaxies at all stellar masses while \snb{} preferentially affects lower-mass galaxies in TNG. At low values of \snb{}, \ch{slower} wind speeds result in supernova feedback that does not drive gas \ch{as far. This leads to more gas available for star formation in the ISM, which allows} 
galaxies to rapidly grow in mass and \ch{effectively} convert their gas into stars at higher efficiencies. 

\textbf{Varying feedback in CAMELS/SIMBA:} Galaxies in the SIMBA model are strongly affected by changes in \sna{} (mass loading factor) and \snb{} (wind speed), and also show sensitivity to AGN feedback, especially at high masses. Unlike TNG and ASTRID, where \sna{} regulates the energy per unit SFR, varying \sna{} in SIMBA regulates the mass loading factor. This has some notable differences in the SFHs when comparing with the other models, since in TNG and ASTRID, the wind speed is varied while keeping the energy per unit SFR constant, while in SIMBA it is varied while keeping the mass loading factor constant. Since these quantities are related by $\dot{E} \propto \eta v_w^2$, this means that varying \snb{} in SIMBA sometimes shows the inverse effect to TNG and ASTRID. This is explored further in Section \ref{sec:disc_cgm_juggling}.

Similar to TNG, \ch{increasing} \sna{} (mass loading) pushes star formation to later times, with a much more dramatic change in the peak time instead of the SFH width (i.e. \ch{the timescale over which the galaxy forms stars}), while not significantly affecting the normalization. Changing the wind speed tends to change the SFH width and normalization, with galaxies forming stars at a roughly constant rate for a much longer duration when at the highest wind speeds (again, note that in this case wind speed is changed at constant mass loading, which means that the total energy also scales as as a result).  

Increasing the momentum flux from AGN feedback (\agna{}) or jet speed (\agnb{}) suppresses star formation, especially at later times and for more massive galaxies. This also manifests as changes in the average SFHs, but they tend to be more subtle given the relatively small number of massive galaxies in our sample. 

\textbf{Varying feedback in CAMELS/ASTRID:} Galaxies in ASTRID are affected by galactic winds in a very similar manner to TNG, but unlike TNG, they are also strongly affected by the strength of the thermal AGN feedback term (\agnb{}). 

Galaxies in ASTRID tend to generally quench earlier than TNG or SIMBA, so effects on later SFR are more difficult to discern unless we restrict our analysis to lower-mass galaxies (as in the next section). Overall, increasing \sna{} tends to reduce SFR \ch{(and thus M$_*$ at a given halo mass)} but spread it out over a longer period of time. The effect of varying \snb{} is similar to TNG, but the effects are more pronounced, with a high SFR followed by rapid quenching at both the high and low end \ch{of the parameter values} (at low \snb{} due to AGN feedback becoming important more quickly and at high \snb{} due to gas being removed from the shallow potential wells of low-mass halos). Thermal feedback from the AGN (\agnb{}) also plays a major role in regulating star formation even in low-mass galaxies (for the default parameter choice), with increasing energy reducing the overall normalization but prolonging SFR to much later times \ch{compared to the fiducial ASTRID model}. This makes it more challenging to tease apart the relative roles of the SN parameters in ASTRID.

\subsection{Dependence on Halo Mass} \label{sec:result_sfh_mhalo}

We now shift away from observationally accessible quantities to better understand the impact of varying feedback and cosmology on subsets of galaxy populations characterized by halo mass. Instead of training a normalizing flow to learn the joint distributions of galaxies in a wide ($\sim 2$ dex) bin of stellar mass, we instead train the flow to learn the distribution of SFHs in narrower ($\sim 0.5$ dex) bins of halo mass, with the average halo mass as an additional parameter. Figures \ref{fig:sfh_1param_massbins_tng}, \ref{fig:sfh_1param_massbins_simba} and \ref{fig:sfh_1param_massbins_astrid} show the results of this analysis, showing the average SFHs at the fiducial, minimum and maximum values for each parameter varied. This allows us to see if a change in the overall SFHs in the previous Section is a result of a similar change across the entire population of galaxies, or if a particular mass range preferentially contributes to the shape of the average SFHs. Note that in contrast to Figure \ref{fig:sfh_1param_exploration}, these show the SFHs in log-SFR space, mainly to capture the wide range of scales across different halo masses in a single plot. The three sets of plots show the gradual onset of quenching with increasing halo mass, and also highlight the mass dependence of certain modes of feedback (e.g. varying AGN feedback primarily affects high-mass galaxies that host massive black holes). These plots are also designed to be read in conjunction with the analysis in Section \ref{sec:result_galstate} where we also study the changes in the baryon fraction and black hole mass. 

The \ch{effects of varying the cosmological parameters in all three of the baryonic physics models} tend not to be strongly mass-dependent, while for feedback (particularly in the case of AGN feedback), it is common to see the effects become more prominent at high halo masses where the galaxies also tend to host more massive black holes. In TNG, we see a mass dependence for \sna{}, where lower values lead to \ch{more rapid} quenching at masses above $M_h\gtrsim 10^{12.5}$, and the effects of varying the strength of AGN feedback starts to be visible above $M_h\gtrsim 10^{13}$, mostly as a shallower slope for quenching with lower feedback strength. In SIMBA, \snb{} has a strong effect on galaxy quenching at high halo masses, while galaxies in lower halo masses are relatively unaffected. A similar effect is seen for \agnb{}, where lower values of the jet speed lead to more sustained star formation over a longer period of time. In ASTRID, \snb{} shows a large effect over a range of $10^{12} < M_h < 10^{13}$, with the effects diminishing at higher and lower masses. Perhaps counterintuitively, scaling the thermal feedback from AGN affects quenching across all halo masses, with a decreased strength of \agnb{} leading to faster quenching. The sustained tail of low-level star formation at late times seen in some ASTRID panels in Figure \ref{fig:sfh_1param_exploration} can be attributed to the high mass halos ($M_h \gtrsim 10^{13.5}$) in ASTRID, which continue to form a small \ch{amount} of stars despite \ch{(partially)} quenching. 

\subsection{Where are Interaction Terms Needed?} \label{sec:result_sfh_interactions}

In addition to the single-parameter interactions, in this section we study the effects of interactions between different feedback and cosmology scalings. This is important to determine whether the interactions between the different parameters significantly change the effects they have on galaxy properties, since for example the effect of varying the strength of supernova feedback might be different depending on the underlying cosmology, which can either amplify or suppress the effect. To study this, we sample the average SFHs from the trained normalizing flow varying two parameters simultaneously and compare it to the linear average of the SFHs while varying both parameters individually (i.e., is SFH($\theta_1, \theta_2$) significantly different from ((SFH($\theta_1$) + SFH($\theta_2$))/2)?). We compute this difference using a standard $\chi^2$ metric described as, 
\begin{equation}
    \chi^2_{\rm SFH_{1P,2P}} = \frac{1}{t_{\rm univ}}\int_{t=0}^{t_{\rm univ}} \frac{(\rm SFH_{1P}(t) -  SFH_{2P}(t))^2}{(\rm \sigma_{\rm SFH_{1P}}(t)^2 + \sigma_{\rm SFH_{2P}}(t)^2)} dt
\end{equation}
where ${\rm SFH}_{2P} \equiv {\rm SFH}(\theta_1, \theta_2)$ is the SFH sampled while simultaneously varying two parameters, and ${\rm SFH}_{1P} \equiv {\rm SFH}(\theta_1)$ is the linear average of the SFHs sampled while varying the parameters individually, with the corresponding uncertainties added in quadrature.

The three rows in Figure \ref{fig:sfh_interaction_terms} show three examples of such interactions, which we find broadly fall into three classes: (i) between parameters describing the same kind of feedback (top row), (ii) between cosmology and feedback terms (middle row), and (iii) between different kinds of feedback (bottom row). We quantify these changes by rerunning the GAM model to learn the mappings between SFH parameters and the CAMELS parameters with only one parameter terms, compared to a GAM model with the one parameter terms plus each individual interaction terms included, in turn. We then compute the accuracy of the model in predicting the SFHs with and without the interaction terms, to quantify the impact and statistical significance of each interaction term. An example of each type of interaction term is shown in the Figure, highlighting the regimes in which the SFHs are strongly affected (in ways that are not captured by the 1P variations, shown as thin dotted lines). A full list of all the 2P interactions is provided in Appendix \ref{app:all_interactions}. 

In TNG, there is a strong interaction between the galactic wind parameters (\sna{} and \snb{}) where the effects of changing the energy per unit SFR (\sna{}) are amplified when the wind speeds (\snb{}) are low. At low wind speeds, material is not ejected from the galaxy's ISM. In combination with the increased \sna{}, which inhibits the early growth of the central SMBH, this leads to a scenario where the overall star formation is significantly enhanced. 

In SIMBA, we see a strong interaction between $\Omega_m$ and \sna{}, where the effects of changing the mass loading (\sna{}) are amplified at lower values of $\Omega_m$ - when the universe has less matter, star formation rates are suppressed at high mass loading, but only at early times. At late times the star formation rates are enhanced compared to the combined 1P prediction. The inverse is also true at low mass loading, where galaxies have higher early star formation and lower late-time SFRs on average.

In ASTRID, we highlight a strong interaction between the wind speed (\snb{}) and the scaling for thermal AGN feedback (\agnb{}). At low values of the thermal AGN feedback, we see faster quenching, with the timescale depending on the precise value of the wind speed. Conversely, at high values of thermal feedback and high wind speeds, we see evidence for late time rejuvenation of star formation, although this is not statistically significant and requires further exploration.

\section{Results: Galaxy state and Feedback} \label{sec:result_galstate}

In this section, we turn our attention to studying the effect of varying feedback and cosmology on three galaxy properties that act as `state variables', which help  to contextualize trends in the average SFHs studied in Section \ref{sec:result_sfh}. While they carry information about both the galaxy's present state, they are also affected by the way the galaxy evolved over time. Similar to Section \ref{sec:result_sfh_mhalo}, we study the change in each of these properties with varying feedback as a function of halo mass.
\begin{enumerate}
	\item The baryon fraction $f_b$, defined as $f_b = (M_* + M_{\rm gas}) / {M_h f_{\rm b,univ}}$ where $f_{\rm b,univ}$ is the universal baryon fraction for a given cosmology. The baryon fraction defined here is for the entire subhalo, which includes gas in both the galaxy's ISM and CGM. A decrease in this number generally indicates more material blown out of the subhalo potential well due to feedback \ch{or preventative feedback making accretion less efficient}, which then provides less gas for baryon cycling. 
	\item The stellar-to-gas mass ratio $M_*/M_{\rm gas}$. Rather than the absolute value of this ratio, relative changes with varying feedback indicate higher or lower star formation efficiencies due to varying feedback that affect how the gas in the subhalo is turned into stars.
	\item The SMBH-to-halo mass ratio $M_{\rm BH}/M_h$, which serves as a proxy for the strength of AGN feedback and the growth of the SMBH over time \citep{2023AJ....166..228T}. Lower values of $M_{\rm BH}/M_h$ correlate with scenarios where feedback inhibits the accretion of gas onto the SMBH, which in turn affects the amount of AGN feedback. 
\end{enumerate}

\begin{figure*}
    \centering
    \includegraphics[width=0.99\textwidth]{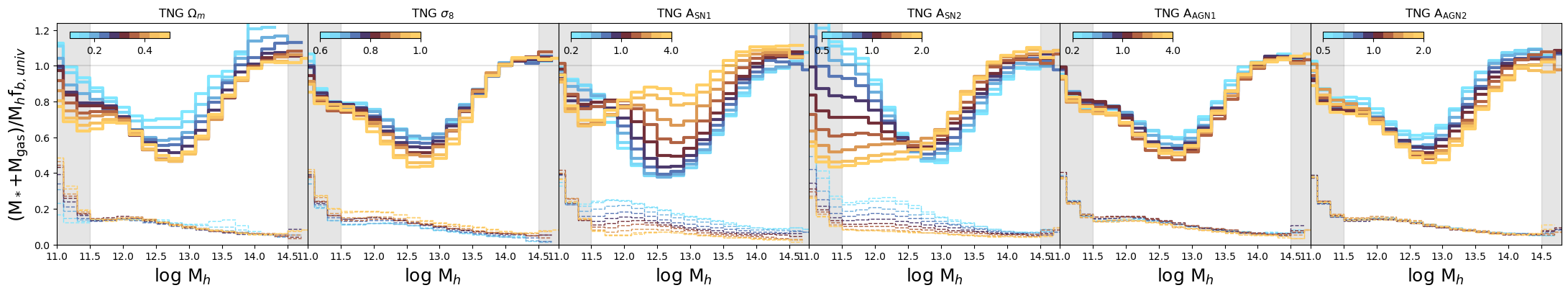}
    \includegraphics[width=0.99\textwidth]{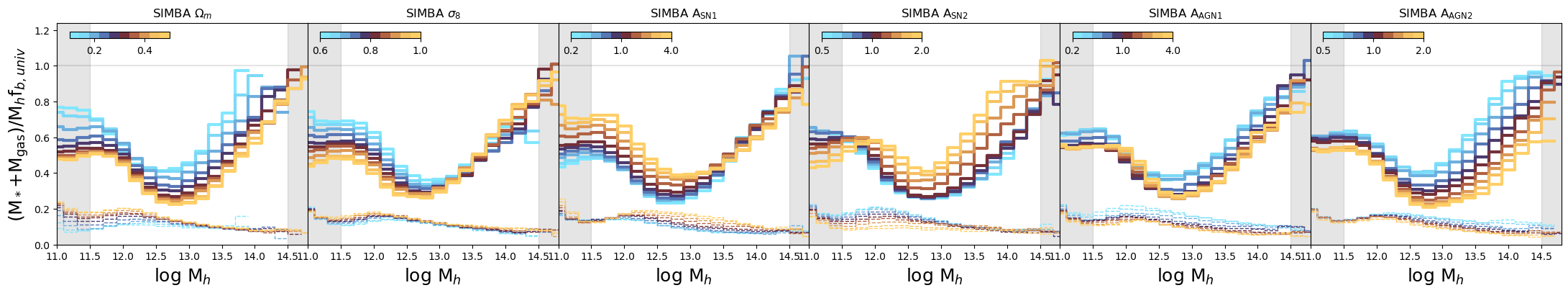}
    \includegraphics[width=0.99\textwidth]{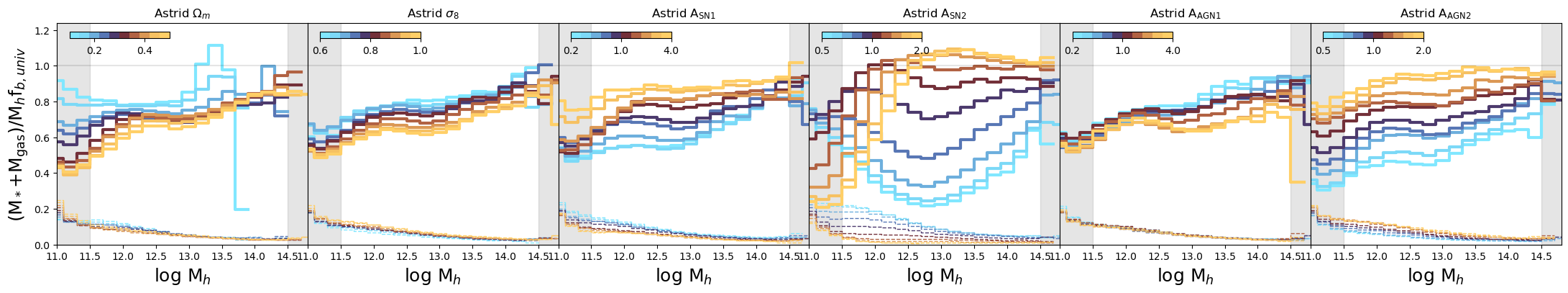}
    \caption{\textbf{Baryon content as a function of halo mass across the three CAMELS models}. Each row focuses on a specific model (TNG, SIMBA, ASTRID respectively), while each column focuses on the effect of varying a single feedback parameter (\sna{}, \snb{}, \agna{}, and \agnb{} respectively). Thick solid lines show a running mean of the total baryon content as a function of halo mass across the CAMELS LH runs, and thin dashed lines show the fraction of baryons in stars. The baryon content shows how the amount of gas in the ISM and CGM of the halo is affected by varying feedback, and higher $f_{\rm baryon}$ generally correlates with larger SFH widths and more efficient baryon cycling. 
    }
    \label{fig:baryon_content}
\end{figure*}

The key results showing the effect of varying feedback on the baryon fraction, stellar-to-gas mass ratio and SMBH-to-halo mass ratio for the three models are shown in Figures \ref{fig:baryon_content}, \ref{fig:bh_content} and \ref{fig:gas_content}. Lines of different colors in each panel trace how these quantities change as a function of the different CAMELS parameters, e.g. showing the effects of varying $\Omega_m$ on the baryon fraction in TNG at a given halo mass in the top-left panel of Figure \ref{fig:baryon_content}. In contrast to the previous section, the plots here use measurements of these quantities directly from the LH dataset and are not sampled from the normalizing flow because the plots use smaller bins than what is used to train the normalizing flows. In our checks where we compare these against samples drawn from the normalizing flow we do not find a significant difference between the two.

\subsection{Baryon Content and feedback} \label{sec:result_galstate_fbaryon}

The analysis in \cite{2024arXiv240208408W} described how studying the baryon fraction as a function of halo mass could be used to determine the regimes in which stellar and AGN feedback dominate the growth and baryon cycling of the galaxy as a function of the halo potential well depth. While their analysis uses the fiducial EAGLE, SIMBA and TNG volumes, here we extend the analysis to the entire CAMELS multiverse in Figure \ref{fig:baryon_content}. Similar to that paper, the fiducial models for TNG and SIMBA show three distinct `regimes', (i) a stellar feedback dominated regime at low halo masses where $f_{\rm baryon}$ increases with increasing halo mass, with the impact declining as the potential well becomes deeper with increasing mass (seen as an increase in the baryon fraction). In comparison to \ch{the fiducial TNG/SIMBA models studied} in \cite{2024arXiv240208408W}, \ch{the three regimes are not as prominent, nor their transitions distinctly separated for large portions} of the CAMELS parameter space. This is succeeded by (ii) an intermediate-mass regime where both stellar and AGN feedback act in concert to determine the baryon fraction and stellar-to-gas mass ratios where $f_{\rm baryon}$ is negatively correlated or uncorrelated with halo mass, and finally (iii) a high-mass regime where AGN feedback is the dominant effect and $f_{\rm baryon}$ is positively correlated with halo mass. Due to the low resolution of the CAMELS boxes, we do not trust the trends at low halo masses (M$_h < 10^{11.5}$M$_\odot$) due to stochasticity introduced by the particle mass resolution, or very high halo masses (M$_h > 10^{13.5}$M$_\odot$). In the following sections these regions are simply referred to as regions (i, ii and iii). In all three models, the increase in $f_{\rm baryon}$ with halo mass is simply tied to the increasing depth of the gravitational potential well, which means that feedback can not as easily blow material out of the halo (or prevent it from entering) as it grows more massive. The thin dashed lines in the plot show the corresponding stellar mass in each halo mass bin, primarily to contrast whether the stellar mass also increases when the baryon fraction decreases. In situations where that is not the case (for example, when varying \sna{} in TNG), it indicates that even though the total baryon content is increasing, the fraction of those baryons that are converted into stars is not. In comparison to TNG and SIMBA, the dependence on halo mass for ASTRID is generally shallower, \ch{extremely dependent on feedback strength, and generally monotonic (i.e. it does not always show three distinct regimes)}.

\subsubsection{Baryon fraction and cosmology} 

The response of the baryon fractions to variations in cosmological parameters ($\Omega_m$ and $\sigma_8$), \ch{seen in the left two panels of Figure \ref{fig:baryon_content}}, show distinct behaviors across the three simulation suites. In TNG and SIMBA, increasing either parameter generally leads to lower baryon fractions across all halo masses, though the effect is more pronounced at intermediate masses (10$^{12}$-10$^{13}$ M$_\odot$). Though the change in response to $\Omega_m$ is qualitatively similar, ASTRID shows a weaker overall dependence on halo mass and more uniform enhancement of baryon fractions across the entire mass range. The relative insensitivity to cosmological variations of $\sigma_8$ at both low and high mass ends in TNG and at the high mass end in SIMBA suggests that feedback processes, rather than cosmology, dominate the baryon cycling in these regimes. This is particularly evident in the consistent spread of the curves at intermediate masses (region ii) where both stellar and AGN feedback are important.

\subsubsection{Baryon fraction and stellar feedback} 

Variations in the strength of stellar feedback, \ch{seen in the middle two panels of Figure \ref{fig:baryon_content}}, produce qualitatively similar responses across the three models with some notable distinctions. In all three models, increasing \sna{} or \snb{} generally leads to an increase in $f_{\rm baryon}$, except in certain notable regimes that correspond to region (i), where $f_{\rm baryon}$ is suppressed when the halo potential well is not deep enough to hold on to baryons ejected by feedback. In TNG, increasing the energy per unit SFR (\sna{}) leads to an overall increase in $f_{\rm baryon}$, and changes the boundary between regions (ii) and (iii), extending the regime where stellar feedback dominates the baryon cycling of low-mass halos. Increasing the wind speed (\snb{}) primarily affects halos below $M_h < 10^{12.5} M_\odot$, reducing their baryon fractions while slightly increasing the baryon fractions of more massive systems. SIMBA exhibits a similar but more extended effect, with mass loading (\sna{}) predominantly affecting low-mass halos, but with wind speed variations influencing baryon fractions up to higher halo masses. \ch{ASTRID shows a flatter profile in baryon fraction as a function of halo mass compared to the other models, with very subtle transitions between stellar- to AGN-dominated regimes. ASTRID shows similar trends to TNG when varying stellar feedback, albeit with a much larger effect on $f_{\rm baryon}$ when varying \snb{}, with increasing wind speed driving material out at low halo masses (and thus reducing $f_{\rm baryon}$) to a large increase in baryon content at the highest wind speeds, which is with consistent increased baryon cycling, also seen in the prolonged duration of late-time star formation at high \snb{} in Figure \ref{fig:sfh_1param_massbins_astrid}.}

Notably, the stellar masses (thin dashed lines in Figure \ref{fig:baryon_content}) show a systematic decrease with increasing feedback across almost all the models (with the notable exception of \sna{} in SIMBA), indicating an inverse relationship between feedback strength \ch{or wind speed} and star formation efficiency. 
\ch{In SIMBA, where \sna{} explicitly varies the mass loading factor, we see that the effect on stellar masses (at a given $M_h$) is lower, but does result in lower CGM baryon fractions, in agreement with other studies using analytic and semi-analytic approaches \citep{2023ApJ...949...21C, 2023ApJ...956..118P, 2024ApJ...976..151V, 2024ApJ...976..150V}.}

\subsubsection{Baryon fraction and AGN feedback} 

The impact of AGN feedback variations (\agna{}, \agnb{}) on $f_{\rm baryon}$, \ch{seen in the two right panels of Figure \ref{fig:baryon_content}}, is more pronounced than the changes in the corresponding SFHs, and to some extent is dependent on fundamental differences in how the three models implement black hole growth and feedback. It is also notable that for TNG (and \agna{} in ASTRID), changing AGN feedback affects the baryon fraction but not the stellar mass. In TNG, $f_{\rm baryon}$  is affected in intermediate mass halos ($M_h \sim 10^{12.5} M_\odot$) when the energy per unit accretion (\agna{}) is varied, and in high mass halos ($M_h \gtrsim 10^{12.5} M_\odot$) when the stochasticity of the AGN feedback (\agnb{}) is varied. SIMBA shows a similar trend but with a more gradual transition between stellar and AGN feedback dominated regimes, suggesting a more continuous coupling between these feedback mechanisms. ASTRID's response to AGN feedback variations is more uniform across the mass range.

\subsection{SMBHs and feedback} \label{sec:result_galstate_mbh}

\begin{figure*}
    \centering
    \includegraphics[width=0.99\textwidth]{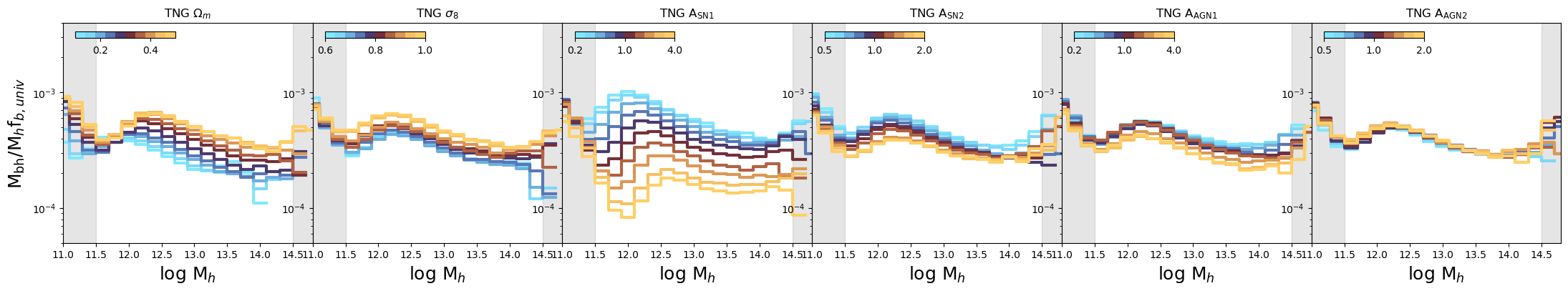}
    \includegraphics[width=0.99\textwidth]{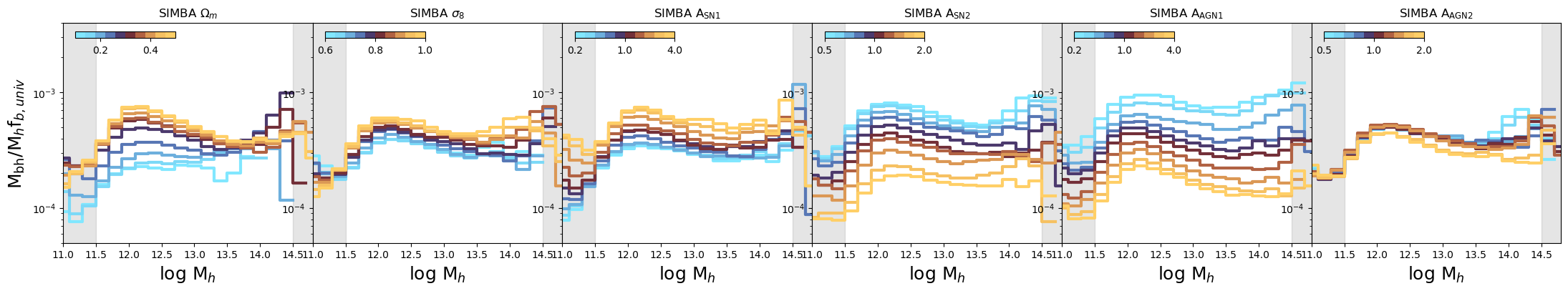}
    \includegraphics[width=0.99\textwidth]{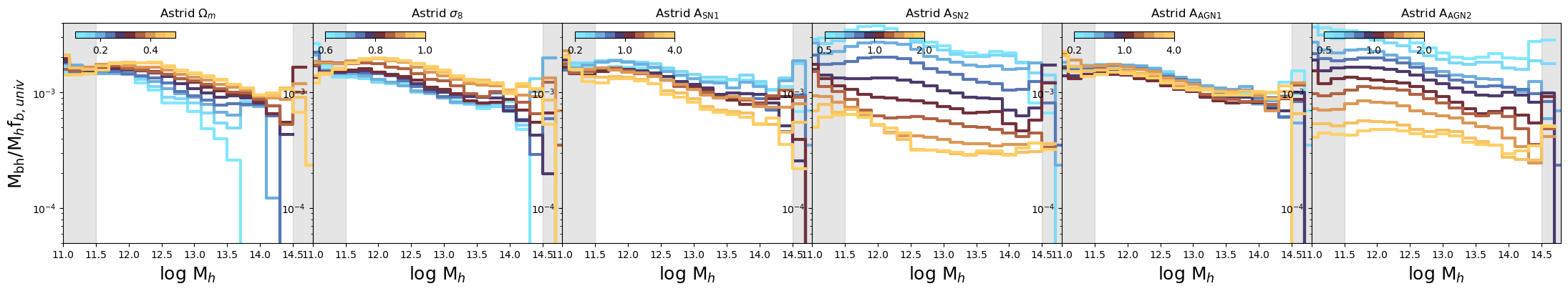}
    \caption{Similar to Figure \ref{fig:baryon_content}, but for the relative \textbf{mass of the central SMBH as a function of halo mass} across the three simulations while varying feedback. Black holes in ASTRID are generally more massive at $z\sim 0$ compared to the other models, and are correlated with the earlier quenching observed in their SFHs.
    }
    \label{fig:bh_content}
\end{figure*}

While \cite{2023AJ....166..228T} and other CAMELS papers provide a much more detailed exploration of varying the CAMELS parameters on the growth and characteristics of their constituent supermassive black holes, here we simply look at the relative mass of the black hole across the three models, which (to an extent) traces the growth of the black hole, since SMBHs that grow faster also tend to quench their host galaxies earlier, making it difficult to continue growing except by mergers \citep{2017MNRAS.465.3291W, 2020MNRAS.499..768Z, 2023ApJ...944..108B}. On the other hand, stronger supernova feedback has also been shown to inhibit accretion onto SMBHs, inhibiting their growth \citep{2017MNRAS.472L.109A, 2017MNRAS.468.3935H, 2023MNRAS.520..722B}. Our main results are summarized in Figure \ref{fig:bh_content}, which shows the ratio of black hole to halo mass as a function of halo mass in the different models while varying the CAMELS parameters. Overall, mass in black holes scales almost linearly with halo mass, with a slope of slightly less than unity, peaking around $M_h \sim 10^{12 - 12.5} M_\odot$. Black holes in ASTRID are generally a factor of $\sim 3\times$ more massive than those in TNG or SIMBA. 

\subsubsection{SMBHs and cosmology} 

In general, higher $\Omega_m$ or $\sigma_8$ leads to earlier galaxy growth (as seen in Section \ref{sec:result_sfh} \ch{and the two left panels of Figure \ref{fig:bh_content}}), and thus an overall increase in the final specific black hole mass at all halo masses. Compared to $\sigma_8$, there is more mass-dependent behavior due to variations in $\Omega_m$, where we see a significant enhancement (suppression) in black hole growth with increasing (decreasing) $\Omega_m$ for low-mass halos in SIMBA and high-mass halos in ASTRID. In terms of the SFHs, this is consistent with behaviour that shows earlier peaks and faster quenching with increasing $\Omega_m$ or $\sigma_8$.

\subsubsection{SMBHs and stellar feedback} 

\ch{As seen in the middle two panels of Figure \ref{fig:bh_content},} higher SN wind speed (\snb{}) across the three models generally reduces black hole masses across all halo masses, likely by reducing the gas supply available for accretion. Increasing the energy per unit SFR at fixed wind speed (\sna{} in TNG and ASTRID) reduces the black hole mass, while changing the mass loading factor (\sna{} in SIMBA) conversely increases the black hole mass. However, this makes sense because $\dot{E} \propto \eta v_w^2$, so increasing the mass loading factor at fixed wind speeds has the same effect as decreasing the energy per unit SFR. While the variations in black hole mass with varying stellar feedback are mostly independent of halo mass, the small dependence that some cases show reveals possible interactions between the stellar feedback and black hole growth. In TNG, the halo mass at which the average black hole to halo mass ratio peaks shifts with changing \sna{}, with higher values driving this peak to higher masses, possibly signaling a change in the point at which AGN feedback transitions from a primarily kinetic to thermal mode above a critical accretion rate. In contrast to TNG, SIMBA and ASTRID exhibit a greater variation in black hole mass when varying \snb{} compared to \sna{}. ASTRID exhibits the most dramatic response to scaling the wind speed, suggesting that its black hole growth model is more strongly coupled to the surrounding gas properties that are directly affected by stellar feedback. However, it is important to note that when comparing trends in black-hole mass as a function of stellar mass (see Oh et al., \textit{in prep.}), the trends are often inverted. This is particularly true for TNG and ASTRID at low stellar masses, where varying feedback affects both the star formation and the amount of gas available for the black holes to grow. 

\subsubsection{SMBHs and AGN feedback} 

Perhaps counterintuitively, varying the strength of AGN feedback does not necessarily affect the final mass of the black holes in the three models (\ch{as seen in the right two panels of Figure \ref{fig:bh_content}}), depending on how the three models implement black hole \ch{accretion and feedback}. In particular, varying the stochasticity of AGN feedback in TNG (\agnb{}), the strength of kinetic feedback per unit accretion in ASTRID (\agna{}), and the jet speed in SIMBA (\agnb{}) at low halo masses do not strongly impact the final black hole mass. 
TNG's \ch{relative black hole masses} show a relatively modest dependence on its AGN feedback parameters \ch{that scale kinetic feedback, suggesting effective self-regulation through thermal AGN feedback or through interactions with SN feedback}. In contrast, SIMBA's three-phase AGN feedback (radiative mode, momentum-driven winds, and jet mode) shows stronger variations with AGN parameters, particularly evident in the large suppression of black hole growth with increased momentum flux (\agna{}). This reflects SIMBA's distinct implementation where different feedback modes operate simultaneously rather than switching between modes. At low masses, varying the jet speed does not significantly affect the black hole mass, consistent with SIMBA's known efficient transport of matter to large distances by AGN feedback, which tends to blow gas out of the subhalo at low masses  \citep{2023ApJ...945L..17T, 2023MNRAS.526.2441B, 2024MNRAS.529.4896G}.  At high halo masses, SIMBA also shows a slight increase in black hole masses for lower jet speeds, where fewer baryons are blown out of the subhalo and can thus provide fuel for the black hole. \ch{ASTRID shows sensitivity to scaling the thermal AGN feedback (\agnb{}), with variations spanning nearly an order of magnitude in $M_{\rm BH}/M_h$ at fixed halo mass where black hole growth is strongly suppressed with higher thermal feedback per unit accretion. 
This strong dependence likely arises because \agnb{} in ASTRID directly scales the thermal feedback normalization, which is the primary channel through which SMBH growth self-regulation occurs in Bondi-based models \citep{2018MNRAS.479.4056W}. In contrast, in TNG the thermal feedback normalization remains fixed, while varying \agnb{} affects only the kinetic mode. This fundamental difference in parameter choices highlights how model-dependent implementations can lead to qualitatively different behaviors despite similar underlying physical processes.}

\subsection{Stellar-to-gas mass ratios and feedback} \label{sec:result_galstate_fgas}

\begin{figure*}
    \centering
    \includegraphics[width=0.99\textwidth]{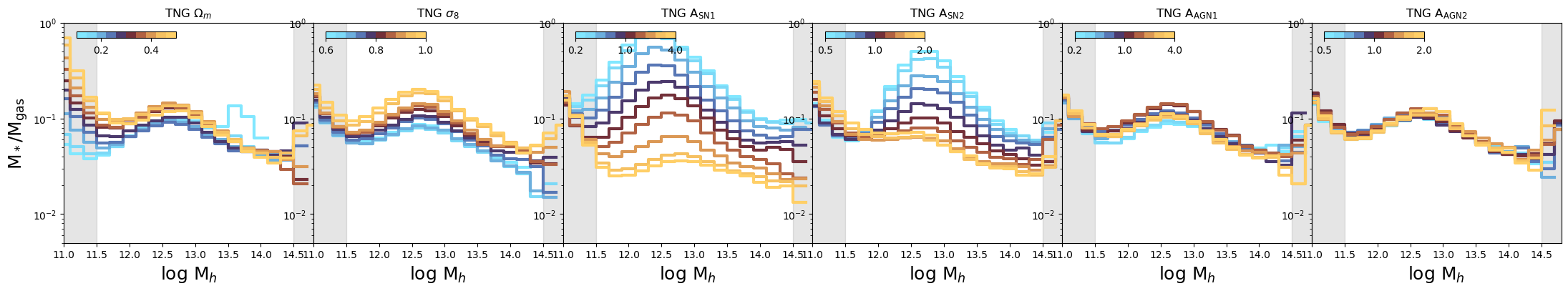}
    \includegraphics[width=0.99\textwidth]{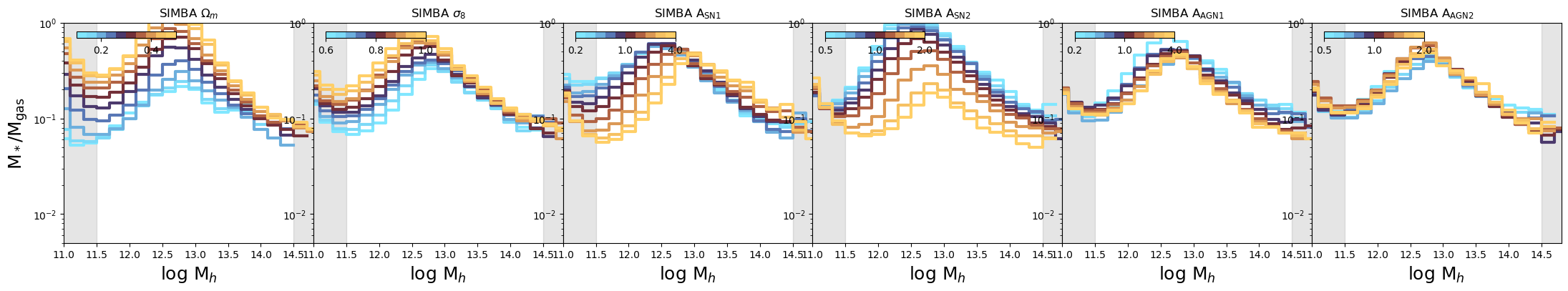}
    \includegraphics[width=0.99\textwidth]{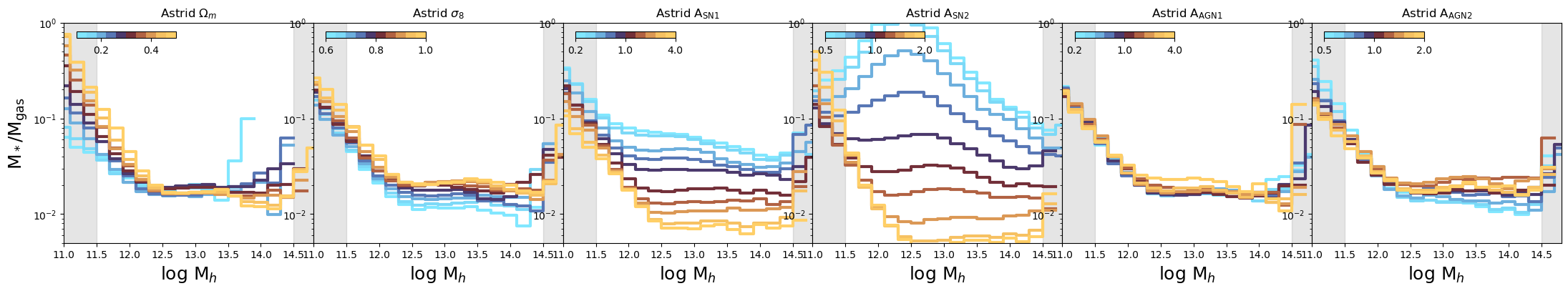}
    \caption{Similar to Figure \ref{fig:baryon_content}, but for the \textbf{ratio of stellar to gas mass} across the three simulations while varying feedback. The change in the stellar to gas mass ratio compared to the baryon fraction traces the star formation efficiency compared to the overall amount of baryons in the ISM+CGM of the halo. 
    }
    \label{fig:gas_content}
\end{figure*}

In this section, we examine the stellar-to-gas mass ratios as a function of halo mass across the three models while varying cosmology and feedback. 
This ratio quantifies the relative balance between the stellar and gas components within the subhalo at $z\sim 0$, and traces the net effect of adding gas (through accretion and recycling), removing gas (through feedback), and converting gas into stars over the galaxy's history. While this metric alone cannot disentangle the efficiency of star formation from gas removal through feedback and other processes (which is better traced by the baryon fraction discussed in Section \ref{sec:result_galstate_fbaryon}), it provides complementary information about the final state of baryonic matter that remains bound to the system.
Our main results are summarized in Figure \ref{fig:gas_content}. Overall, the ratio of stars to gas is $\sim 10\%$ and peaks around $M_h\sim 10^{12.5-13}M_\odot$ in TNG and SIMBA, while it is generally lower in ASTRID and falls with increasing halo mass, flattening around $M_h\sim 10^{12.5}M_\odot$. We save this quantity for the end because it is affected by both the total baryonic content in the subhalo (which correlates with the total available fuel for forming stars but also the amount of baryon cycling) as well as the mass of the black hole (and therefore the extent to which star formation is regulated by AGN feedback). 

\subsubsection{Stellar-to-gas mass ratios and cosmology} 

The response of stellar-to-gas mass ratios to cosmological parameters, \ch{as seen in the left two panels of Figure \ref{fig:gas_content}}, is consistent with the baryon fractions and black hole masses, showing a general increase in the amount of stars with an increase in both $\Omega_m$ and $\sigma_8$, with the effects generally being more pronounced at low halo masses. 

\subsubsection{Stellar-to-gas mass ratios and stellar feedback} 

\ch{The effects of galactic winds on the subhalo stellar-to-gas mass ratio is shown in the middle two panels of Figure \ref{fig:gas_content}.} In TNG, increasing the scaling for the energy per unit SFR (\sna{}) reduces the stellar-to-gas ratio across all masses, keeping more baryons in the ISM and CGM. It also shows a slight shift in the peak star formation efficiency to higher halo masses that mirrors the shift in the peak black hole mass. At masses of $M_h\sim 10^{12.5-13}M_\odot$, the amount of stars per unit gas mass can change by over an order of magnitude, showing the enormous effect it has on the efficiency with which gas can be consumed to form stars. ASTRID shows similar behavior, albeit around a significantly different base trend. Increasing the scaling for the mass loading (\sna{}) in SIMBA leads to lower stellar-to-gas mass ratios at low halo masses, with the trend inverting above masses of $M_h\sim 10^{13}M_\odot$, primarily due to a shift in the peak of the $M_*/M_{\rm gas}$ relation toward higher masses with increasing \sna{}. 
Scaling the wind speed (\snb{}) across all three models results in reduced stellar-to-gas mass ratios, \ch{with ASTRID showing the strongest effects of varying the wind speed, dramatically reducing the the efficiency of star formation at halo masses ($M_h > 10^{12.5}$M$_\odot$).}

\subsubsection{Stellar-to-gas mass ratios and AGN feedback}

Though they significantly affect the baryon fraction and black hole masses, scaling the strength of AGN feedback does not strongly affect the stellar-to-gas mass ratios across the three models, \ch{as seen in the right two panels of Figure \ref{fig:gas_content}}. The strongest effects are from changing the momentum flux (\agna{}) in SIMBA, which causes a slight suppression in the $M_*/M_{\rm gas}$ ratio, and the thermal feedback in ASTRID (\agnb{}), which suppresses the growth of SMBHs and leads to a slight rise in the $M_*/M_{\rm gas}$ ratio. 

\subsection{Putting it all together}

The effects on the baryon fraction, black hole growth, and stellar-to-gas mass ratios with changes in the CAMELS parameters are useful to identify the distinct channels by which galaxies grow in the three models. Cosmological parameters ($\Omega_m$ and $\sigma_8$) generally affect all three quantities in concert, primarily by altering both the timing and rate of structure formation. Higher values of $\sigma_8$ lead to earlier collapse of density perturbations and more rapid structure formation, resulting in galaxies that form earlier and have more time to form stars and grow larger black holes. This is reflected in the systematically higher black hole masses and stellar-to-gas ratios seen at fixed halo mass. Similarly, higher values of $\Omega_m$ increase the overall matter density, leading to deeper potential wells at fixed halo mass and more efficient accretion of material. This affects baryon fractions in two competing ways: while deeper potential wells can retain baryons more effectively against feedback, the earlier onset of star formation also leads to earlier \ch{BH growth and} stronger AGN feedback. The relative insensitivity to $\sigma_8$ at very high masses indicates that once halos become massive enough, internal feedback processes rather than cosmological parameters determine their evolution, consistent with our understanding of hierarchical structure formation.

In contrast to cosmological parameters, scaling the feedback parameters shows a wide range of distinct behaviors across the three models in terms of both sensitivity and mass-dependence. The effects of varying feedback are most pronounced at intermediate halo masses (regime ii) where a combination of the feedback processes are responsible for determining galaxy properties. In physical terms, the stellar feedback scalings are meant to control how efficiently stellar energy couples to the surrounding gas (\sna{}) and how fast the resulting winds propagate (\snb{}), which together determine the resulting effect on the ISM, and thus on subsequent star formation. Higher wind energies or speeds can launch gas to greater distances from the galaxy, but the ultimate outcome depends critically on the host halo's mass: in lower-mass halos ($M_h < 10^{12}M_\odot$), the shallow potential wells allow this material to be permanently ejected, reducing both baryon fractions and star formation efficiency. However, in more massive halos, the ejected material often remains bound in a hot CGM, leading to the counterintuitive result of higher baryon fractions but lower stellar-to-gas ratios (e.g. \snb{} in all three models). This redistribution of gas also has important consequences for black hole growth --- stronger stellar feedback typically results in lower black hole masses by reducing the supply of gas available for accretion. \ch{This is complemented by analyses like \citet{2023AJ....166..228T}, which studies redshift dependent changes in the number of black holes and average accretion rates in CAMELS/SIMBA and CAMELS/TNG.} The mass-dependent transitions in these effects, particularly evident in how \sna{} in TNG or SIMBA and \snb{} in TNG and ASTRID can shift the boundaries between feedback regimes, demonstrate how stellar feedback is responsible for setting the characteristic mass scales in galaxy evolution where different physical processes dominate.

The distinct effects of more mass-loaded winds (where more mass is ejected at a given velocity, as controlled by \sna{} in SIMBA, which changes mass loading at fixed wind speed) versus more energy-loaded winds (where less mass is ejected at higher velocities, as seen from varying TNG and ASTRID's \sna{}, which changes the energy per unit SFR at fixed wind speed) also show differences in how wind properties affect galaxy evolution. For a fixed energy budget, mass-loaded winds carry more baryons but at lower velocities, meaning they are more easily retained by the halo's potential well. This leads to more efficient recycling of material, where gas falls back to the galaxy on relatively short timescales, allowing for extended periods of star formation but at lower rates. In contrast, energy-loaded winds carry less mass but at higher velocities, allowing them to escape to larger distances or even from the halo entirely. This is particularly evident in how TNG's \sna{} and SIMBA's \sna{} show opposite trends in black hole growth --- higher energy winds more effectively reduce the central gas density, inhibiting black hole accretion, while more mass-loaded winds can actually enhance it through recycling \ch{(especially at late times; \citet{2023AJ....166..228T})}. The differences also manifest in how these winds interact with the circumgalactic medium: energy-loaded winds are more effective at heating the surrounding gas and creating hot halos, while mass-loaded winds are \ch{known to be} more efficient at redistributing metals and enriching the CGM \citep{2023ApJ...949...21C, 2024arXiv241012909B}; characterizing these properties can thus can lead to better predictions for observable quantities such as short-timescale stochasticity and enrichment, apart from their long-timescale imprints on galaxy SFHs that is the focus of the current work. 

Finally, varying the AGN feedback scalings (\agna{} and \agnb{}) can be used to study how the different implementations of black hole feedback can regulate galaxy properties through distinct physical channels. \ch{In TNG's feedback implementation, \agna{} affects the energy per accretion event, while \agnb{} controls the burstiness by changing the threshold of energy accumulated before it is released as kinetic feedback. Since in Bondi-like prescriptions for BH accretion, the subsequent self-regulation of BH growth through feedback tends to predominantly occur through thermal feedback \citep{2018MNRAS.479.4056W}, these parameters tend to affect the baryon fraction but have little impact on the final black hole masses.}
Varying the strength of SIMBA's jet mode AGN feedback (\agnb{}) --- which can efficiently transport material to large distances from the galaxy --- primarily affects the baryon fractions, but not the stellar content or black hole masses, indicating that the feedback efficiently prevents gas re-accretion without necessarily shutting off black hole growth. On the other hand, changing the momentum flux (\agna{}) can strongly inhibit the growth of SMBHs, leading to more star formation at late times, particularly in higher-mass halos. ASTRID shows the strongest coupling between AGN feedback and galaxy properties, with \agnb{} (thermal feedback) strongly affecting both black hole growth and baryon content, suggesting less efficient self-regulation compared to the other models. Interestingly, across all three simulations, AGN feedback parameters have relatively modest effects on stellar-to-gas mass ratios compared to their dramatic effects on baryon fractions and black hole masses, indicating that while \ch{(kinetic)} AGN feedback is efficient at moving \ch{baryons} around, it may be less effective at directly controlling the efficiency of converting available gas into stars \ch{by displacing or disrupting reservoirs of cold, dense gas. However, these results should be interpreted with several important caveats: our parameter exploration is very model dependent and primarily focuses on kinetic AGN feedback (except in ASTRID), in contrast to the effects of thermal feedback that are thought to play a larger role in regulating black hole growth. In larger volumes with more massive halos, and with parameter variations targeting other aspects of BH growth and AGN feedback, the effects on star formation might be more pronounced.}

\section{Discussion} \label{sec:discussion}

\subsection{What Shapes the SFHs of Galaxies in CAMELS?}

The systematic analysis of the average SFHs (Section \ref{sec:result_sfh}) and corresponding galaxy properties (Section \ref{sec:result_galstate}) across the CAMELS multiverse help us understand the effects of (and interplay between) the different physical processes regulating galaxy growth across cosmic time. By studying the response of these observables to variations in both cosmological parameters and feedback implementations, we can identify several key mechanisms that are at play:

\begin{enumerate}
    \item \textbf{Cosmology and early growth:} The cosmological parameters ($\Omega_m$ and $\sigma_8$) primarily affect the early evolution of galaxies by setting the rate of structure formation and halo mass assembly. Higher values of $\sigma_8$ lead to earlier collapse of density perturbations, and thus more rapid gas accretion and star formation, particularly evident in the rising slopes ($\beta$) of SFHs. The matter density ($\Omega_m$) affects SFHs by changing the amount of baryonic matter relative to dark matter, and thus the depth of the potential well at a fixed halo number density. Higher $\Omega_m$ results in deeper potential wells, enabling galaxies to retain more baryons against feedback processes. This manifests as systematically earlier rises and higher peaks for the SFHs, but also results in earlier black hole growth which quenches the galaxies earlier. 
    \item \textbf{Stellar Feedback:} affects all aspects of the average SFH by directly affecting star formation (i.e. by changing the amount of baryons available in lower-mass halos or affecting the star formation efficiency) and through indirect effects (for example by inhibiting the growth of black holes and thus delaying the onset of quenching). These effects can be quite mass dependent, and are difficult to disentangle without also considering the effects on other galaxy properties like the baryon fraction and relative black hole mass. 
    \item \textbf{AGN feedback:} primarily affects the timescale and manner of quenching, and predominantly affects higher-mass halos that host proportionately larger SMBHs. Stronger AGN feedback, however, does not always lead to the intuitive result of quenching galaxies faster, since inhibiting the further growth of the black hole through strong feedback can instead lead to more prolonged star formation. Our results show that AGN feedback and stellar feedback can be strongly coupled even in massive halos, and trends in how AGN feedback affects galaxies should be studied in the context of a given \ch{model of stellar feedback and have differing effects at different values of the feedback parameters}. 
    \item \textbf{Baryon cycling:} The efficiency of baryon cycling is roughly imprinted on galaxy SFHs through the duration over which prolonged, bursty, inefficient star formation can be sustained in the subhalo. This is traced through the baryon fraction and stellar-to-gas-mass ratio, and can be seen in the correlation between the $f_{\rm baryon}$ and SFH width ($\tau$). This is also seen in the effect of changing the energy per unit SFR, which drives more material out of the ISM. For higher-mass halos, this material stays in the CGM, eventually re-accretes and contributes to star formation, \ch{which can lead} to an overall lower star formation rate spread over a longer period of time. 
\end{enumerate}

While the broad qualitative effects of feedback are similar across models, the detailed implementation of wind launching, recycling, and AGN feedback modes leads to distinct signatures in the resulting SFHs. These differences provide valuable constraints for future model development and highlight the importance of multi-wavelength observations that can probe both stellar and gas properties across cosmic time. In the following sections, we try to generalize our results and contextualize them given the current literature. In Section \ref{sec:disc_univmodel}, we attempt to derive a set of equations that can describe the qualitative behaviour of the SFHs across all three models, and in Section \ref{sec:disc_cgm_juggling} we examine the role of mass vs energy-loaded winds in lifting gas \ch{from the ISM} to the CGM. In Section \ref{sec:disc_Z_fduty} we look at alternate observational signatures like chemical enrichment, duty cycles for star formation and quenched fractions, and in Section \ref{sec:disc_inference} we invert the problem and instead measure the constraints on cosmology and feedback that we can obtain using a sample of observed SFHs. Finally, in Section \ref{sec:disc_literature} we study how consistent our results are with the existing literature, and identify areas for further study.

\subsubsection{A Fundamental Set of Equations Describing Galaxy Star Formation Histories} \label{sec:disc_univmodel}

Through the preceding sections, we can build intuition for how the average SFHs are affected by cosmology and the strength of different types of feedback in the three CAMELS models we consider. Given a scenario, e.g. varying \snb{} in SIMBA, we would likely be able to account for the parameter being varied, look up its effect on the physical properties of galaxies in that model and predict the shape of the resulting average SFH. However, given the differences in the numerical implementation of star formation, black hole seeding, growth and feedback, etc. across the three models, and the fact that the CAMELS stellar and AGN feedback parameters themselves can mean different things depending on the model, it is difficult to generalize the shape of the average SFH in a physically motivated, model agnostic manner. 

To get at this, we attempt to derive a set of empirical equations that can describe how the various portions of the SFH (the rising and falling slopes ($\alpha, \beta$), time of peak/SFH width ($\tau$), normalization ($\phi$) and lookback time at which star formation starts ($\eta$)) depend on the physical state of the galaxy as well as the CAMELS parameters. To do this, we use a combination of the generalize additive model (GAM) and symbolic regression approaches described in Sections \ref{sec:method_gam} and \ref{sec:method_pysr} combined with the physical insights from Sec. \ref{sec:result_galstate}. Briefly, we use the heavily overparametrized GAM to obtain a full nonlinear mapping between each SFH parameter ($\Theta_{SFH} \in \{ \alpha, \beta, \tau,\phi, \eta \}$) and the corresponding galaxy state variables and CAMELS parameters (described in Eqn. \ref{eqn:gam_params}). Using the loss from this as the best-case scenario (i.e., similar to a Bayes risk or Pareto efficiency factor), we then use terms that are important based on the symbolic regression analysis of each model (shown in Figure \ref{fig:pysr_feature_importance}) to construct a set of common equations that have the same form (but different coefficients) for all three models. When we add terms to the equation, we impose the following constraints:
\begin{itemize}
    \item Adding a term should significantly improve the loss across the three models (i.e. a loss-complexity tradeoff). If an equation shows similar improvement from adding a linear term to a non-linear one, the linear term is preferred.
    \item If an equation shows similar improvement from adding a \colorb{CAMELS feedback parameter} or a \colorc{galaxy state variable}, the latter is always preferred. We do this to be able to generalize our results while reducing the dependence of our model on the specific parameter choices across the three CAMELS models.
    \item If the coefficients of all three models (TNG, SIMBA, ASTRID) are similar for any particular term in an equation, the coefficients are replaced by that numerical value.
\end{itemize}

The resulting equations obtained by following this procedure are shown in Eqn.\eqref{eqn:general_sfhs} below, with the relevant coefficients given in Table \ref{tab:general_sfh_coeffs}:

\begin{figure*}
    \centering
    \includegraphics[width=0.99\textwidth]{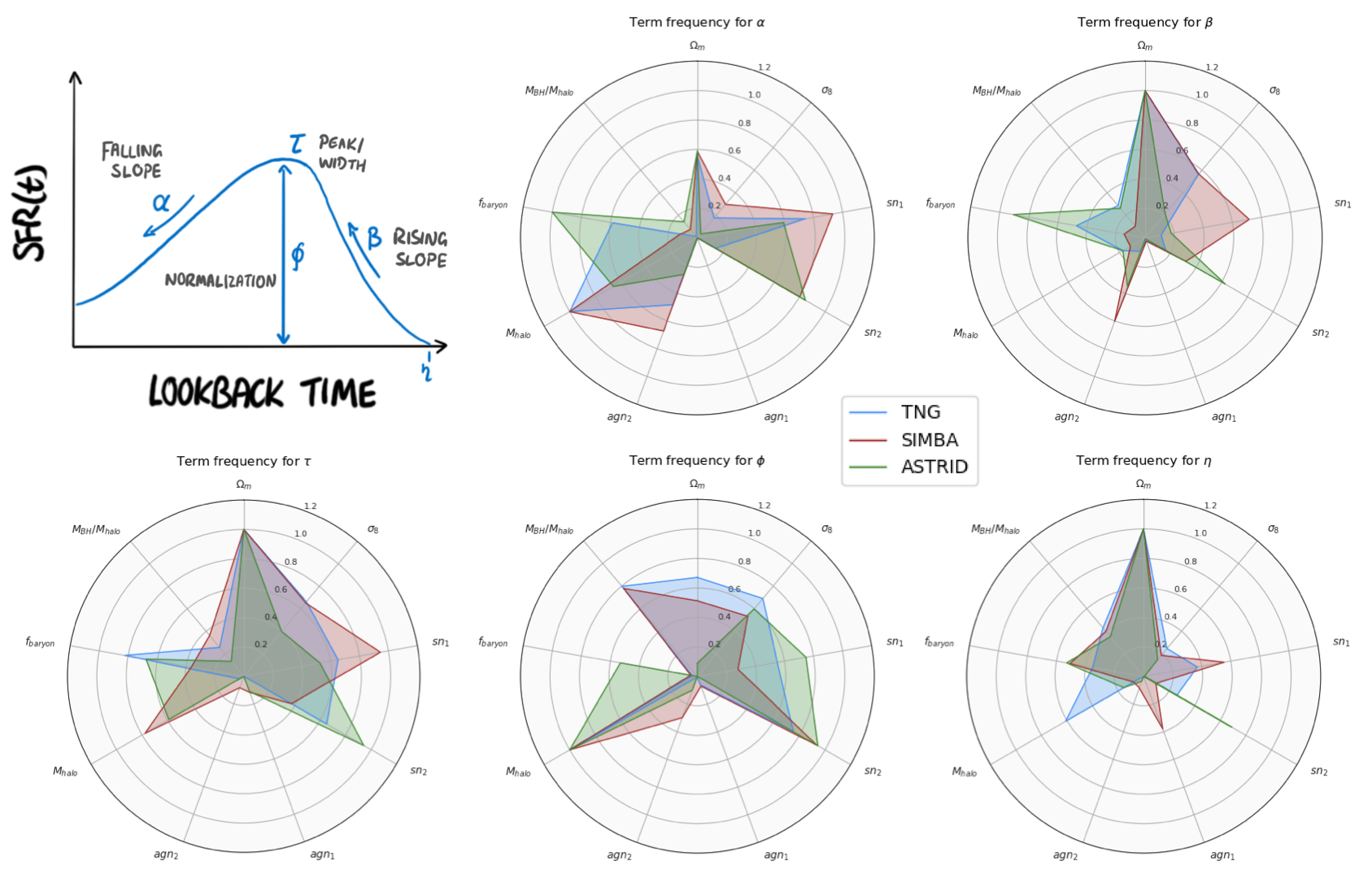}
    \caption{\ch{The frequency with which different terms occur} in the symbolic regression solutions to the SFH parameters across the three CAMELS models. While different from traditional metrics of feature importance (e.g. using saliency maps or random forests), \textbf{the occurrence of certain recurring terms (\ch{e.g. $M_{\rm halo}$ being an important feature for $\phi$ across all three models}) in the set of equations generated by symbolic regression aid in the construction of equations that generalize across the CAMELS models}.}
    \label{fig:pysr_feature_importance}
\end{figure*}

\begin{widetext}
\begin{equation} \label{eqn:general_sfhs}
    \begin{split}
    \mathrm{peak/width \rightarrow}~~ \tau &=  c_1 + \frac{c_2}{\colora{\Omega_m}}(1-0.55\colorc{M_{halo}} + 0.12\colorc{f_b} + c_3\colorc{M_{\rm BH}}) - 4\colora{\sigma_8} + c_4\colorb{sn_1} - c_5\colorb{sn_2}  \\
    \mathrm{falling~slope \rightarrow}~~ \alpha &= -c_1 + c_2\colorc{M_{halo}}(1+0.07\colorb{agn_2}+c_3\colorc{f_b}) - c_4\colorb{sn_1} - c_5\colorb{sn_2} - 2.5\colora{\Omega_m} \\
    \mathrm{rising~slope \rightarrow}~~ \beta &=  c_1 - c_2 \colora{\Omega_m}  - \frac{c_3(1 - c_4\colorc{f_b})}{\colora{\Omega_m}} \\
    \mathrm{normalization \rightarrow}~~ \phi &= -7.5 + 0.875\colorc{M_{halo}}(1 + c_1\colorc{M_{\rm BH}} - c_2\colorb{sn_1} - c_3\colorb{sn_2} + c_4\colora{\sigma_8})  \\
    \mathrm{(shift) \rightarrow}~~ \eta &=  12.5\colora{\Omega_m} - 3 
    \end{split} 
\end{equation}
\end{widetext}
where terms in \colora{green} denote the cosmology variables in CAMELS, \colorb{pink} denote the galactic wind and AGN feedback parameters in CAMELS, and \colorc{blue} denote galaxy state variables (${\rm M}_{\rm halo}$ is the log halo mass, ${\rm M}_{\rm BH}$ is the log of the relative BH mass to halo mass, and ${\rm f}_{b}$ is the ratio of the mass in baryons to the scaled dark matter mass in the subhalo). The SFH parameters are dimensionless ($\alpha, \beta$), in units of time (Gyr; $\tau, \eta$) or in units of SFR ($M_\odot/yr$; $\phi$).

\begin{table}[]
    \centering
    \begin{tabular}{c|ccccc}
     Equation & $c_1$ & $c_2$ & $c_3$ & $c_4$ & $c_5$ \\
    \hline
    $\tau_{\rm TNG}$ & \textbf{4.571} & \textbf{0.891} & \textbf{-0.173} & \textbf{0.930} & 3.236  \\
    $\tau_{\rm SIMBA}$ & 5.957 & 4.761 & 0.025 & 5.088 & \textbf{-2.085}  \\
    $\tau_{\rm ASTRID}$ & \textbf{3.795} & \textbf{1.958} & -0.035 & \textbf{0.498} & 4.710  \\
    \hline
    $\alpha_{\rm TNG}$ & 19.487 & 1.878 & -0.052 & 1.456 & \textbf{0.673}  \\
    $\alpha_{\rm SIMBA}$ & \textbf{10.530} & 1.208 & -0.066 & \textbf{-2.286} & 3.062  \\
    $\alpha_{\rm ASTRID}$ & \textbf{3.398} & \textbf{0.498} & \textbf{-0.826} & 0.907 & 3.564  \\
    \hline
    $\beta_{\rm TNG}$ & \textbf{7.128} & 7.635 & 0.559 & 0.412  & -  \\
    $\beta_{\rm SIMBA}$ & \textbf{4.340} & \textbf{3.043} & \textbf{0.289} & 0.675  & -  \\
    $\beta_{\rm ASTRID}$ & 10.475 & 11.667 & 0.885 & \textbf{0.360}  & -  \\
    \hline
    $\phi_{\rm TNG}$ & 0.021 & 0.052 & 0.077 & 0.091  & -  \\
    $\phi_{\rm SIMBA}$ & 0.015 & 0.001 & 0.061 & 0.065  & -  \\
    $\phi_{\rm ASTRID}$ & 0.017 & 0.053 & 0.175 & 0.032  & -  \\
    \hline
    $\eta_{\rm TNG}$  & -  & -  & -  & -  & -  \\
    $\eta_{\rm SIMBA}$  & -  & -  & -  & -  & -  \\
    $\eta_{\rm ASTRID}$  & -  & -  & -  & -  & -
    \end{tabular}
    \caption{Coefficients for the SFH shape parameters in our general set of equations \eqref{eqn:general_sfhs} for the three CAMELS models. Values in bold highlight large variations in the coefficients across the different models.}
    \label{tab:general_sfh_coeffs}
\end{table}

\begin{figure*}
    \centering
    \includegraphics[width=0.99\textwidth]{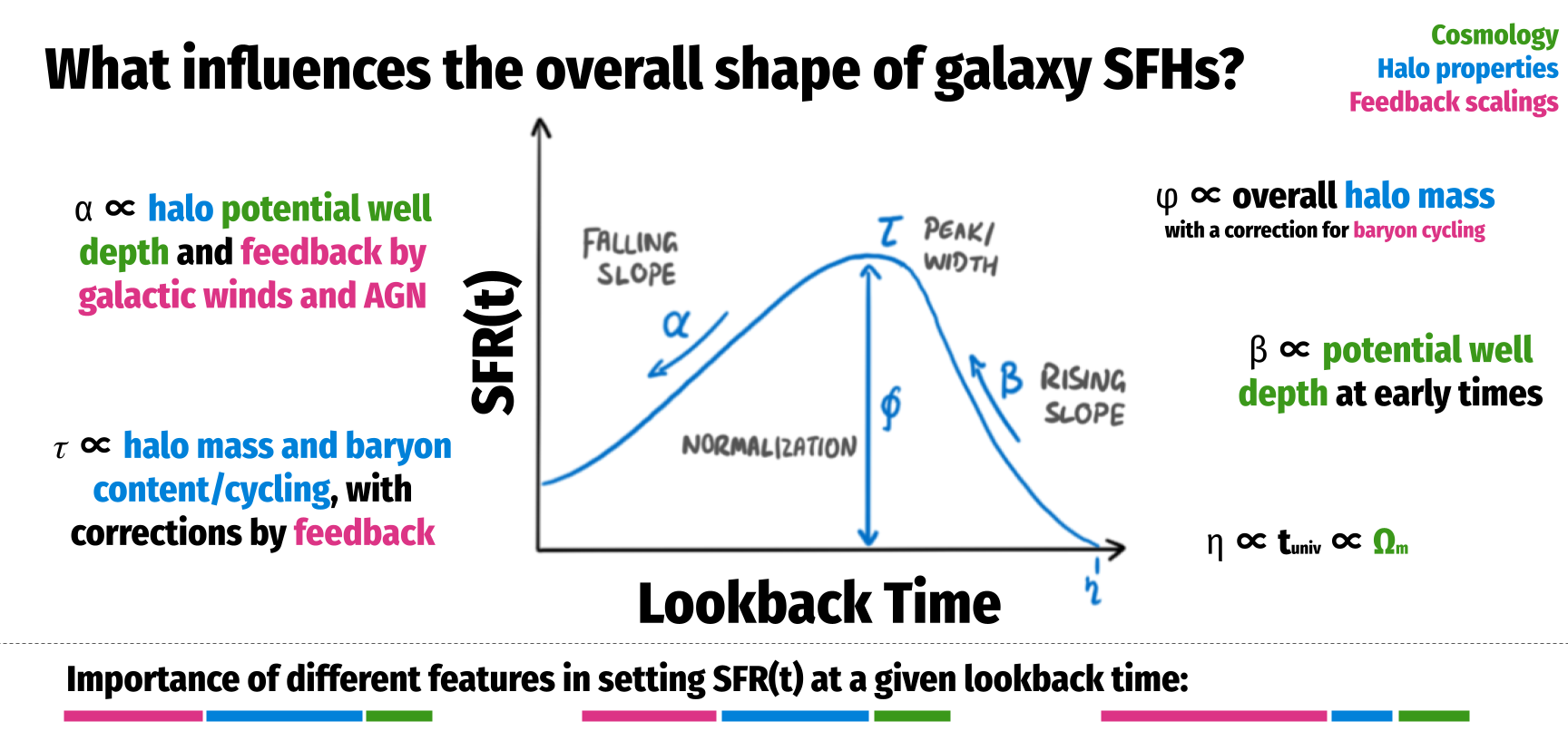}
    \includegraphics[width=0.99\textwidth]{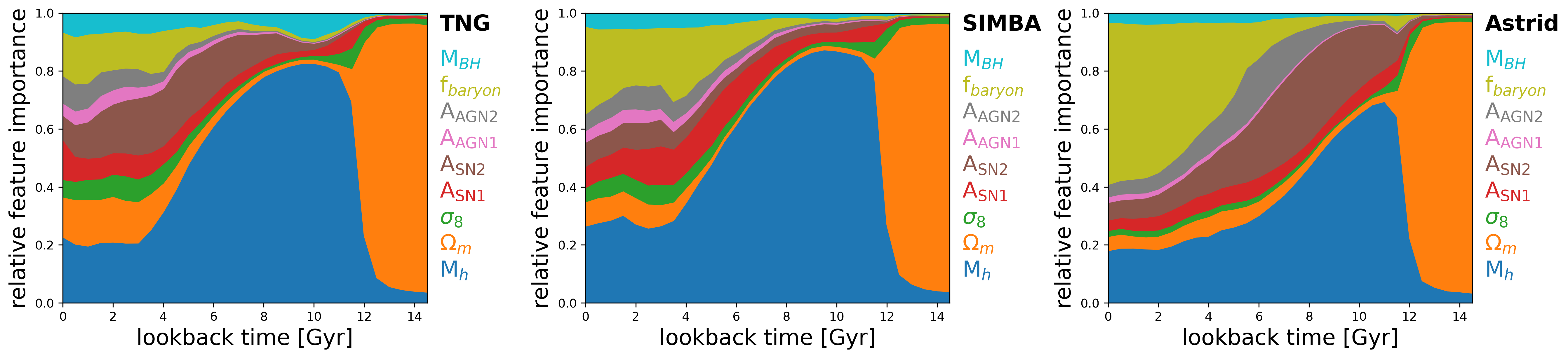}
    \caption{(Top) Summary of the various factors impacting average galaxy SFHs across the three CAMELS models. (Bottom) Feature importance of the cosmology, feedback scaling factors and galaxy properties as defined in Eqn. \eqref{eqn:general_sfhs} on the star formation rate as a function of lookback time. Cosmology dominates the average SFR of galaxies at early times, giving way to halo mass as they assemble. At late times, the SFR is a complex interplay of feedback and baryon cycling. This behavior generalizes across the three models, but the relative contributions differ.}
    \label{fig:sfh_results_summary}
\end{figure*}

Having derived a system of equations that captures the essential physics governing galaxy evolution across the three models, we now briefly examine the behavior of the various terms: The \textbf{shift parameter} ($\eta$), which determines when the star formation first started, depends purely on the matter density ($\Omega_m$). This makes sense, since the start of star formation occurs shortly after the big bang in all three models, and $\Omega_m$ is the only factor in our analysis that can change the age of the universe.

The \textbf{rising slope} ($\beta$) shows a remarkably simple dependence on cosmology, primarily scaling with $\Omega_m$ with a correction term corresponding to the width of the SFH, correlated with the baryon fraction. This suggests that early star formation is predominantly governed by the rate of structure formation, with feedback playing a secondary role through its effect on gas availability (though this could also be partially due to the slight degeneracy between $\tau$ and $\beta$ in setting the early time slope, as seen in Figure \ref{fig:sfh_param_explain}). 

To first order, the \textbf{overall normalization} ($\phi$) scales nearly linearly with halo mass but includes multiplicative corrections from the relative black hole mass and feedback parameters, capturing both the fundamental mass dependence of star formation on halo accretion and its regulation by feedback processes.

The \textbf{falling slope} ($\alpha$) is primarily determined by halo mass, with corrections from AGN feedback and baryon content indicating the role of these processes in quenching. We find that higher $\Omega_m$ universes show systematically steeper declining slopes, consistent with earlier structure formation and more rapid quenching. The much lower value for $c_1$ in ASTRID compared to the other models highlights the faster quenching due to more massive black holes. 
Counterintuitively, the inclusion of thermal AGN feedback as a variable in ASTRID contributes to $\alpha$ indirectly via the $f_b$ term rather than directly through the \agnb{} term, since its primary impacts are inhibiting the growth of the central black hole and prolonging baryon cycling. \ch{Due to variations in the wind speed and the thermal AGN feedback in ASTRID having a major impact on the baryon fraction, this has a much bigger impact on the falling slope (through the $c_3$ coefficient) compared to TNG and SIMBA.}
The differences in the strengths of coupling with \sna{} (through $c_4$) and \snb{} (through $c_5$) highlight how stellar feedback plays different roles in quenching galaxies across the three models. 

The parameter $\tau$, which governs the \textbf{time of the peak and (to an extent) the SFH width}, exhibits a complex dependence on a range of factors both cosmology and feedback, scaling inversely with $\Omega_m$ but modified by the halo mass, baryon fraction, and relative black hole mass. This reflects the competing effects of potential well depth and feedback processes in determining the rise and fall of star formation. The overall coupling through $c_2$ and $c_4$ is particularly strong in SIMBA, where the effect of varying these parameters can strongly delay the onset of star formation, pushing $\tau$ to later times. The sign for $c_5$ (the coupling to wind speed) is also inverted for SIMBA in comparison to TNG/ASTRID, likely due to the wind speed being varied at constant mass loading instead of E/SFR. 

The top row panel of Figure \ref{fig:sfh_results_summary} summarizes our conclusions from studying the general equations.
As an additional sanity check, we also perform an independent analysis of the importance of the various parameters on the distribution of SFRs at different lookback times. The bottom row of Figure \ref{fig:sfh_results_summary} shows the feature importance of each parameter (estimated using a random forest) on predicting the distribution of log SFR($t_i$) for a range of lookback times $t_i$ ranging from the big bang to present day. Consistent with our equations, we find that cosmology plays the largest role at early rimes, followed by halo mass (to first order) dictating how the galaxies assemble over time. Finally, a variety of feedback factors start to kick in after cosmic noon, and significantly influence the late time evolution and quenching behavior of galaxy SFHs. 
\ch{Interestingly, the \agna{}  parameter does not appear directly in our derived equations for most SFH parameters. This omission likely stems from a combination of factors: (1) In TNG and ASTRID, \agna{} corresponds to the energy normalization in the kinetic feedback mode, which primarily affects more massive galaxies that are underrepresented in our volume-limited sample; (2) in cases where it does make a difference (i.e. in SIMBA, where varying \agna{} affects both the black hole masses and accretion rate density), the effects of \agna{} are largely captured indirectly through its impact on the relative black hole mass ($M_{\rm BH}$) and baryon fraction ($f_b$) terms that appear in the equations; and (3) Our methodology prioritizes minimal models that generalize across all three simulation suites, potentially excluding parameters that have moderate effects in some models but weak effects in others. Rather than indicating that \agna{} is generally unimportant for galaxy evolution, this likely reflects the specific mass range and volume constraints of our analysis.}
It is interesting that the broad overall shape of the SFH is set by a combination of ${\rm M}_{\rm halo}$ and ${\rm f}_{b}$ at early and late times respectively, which together set the amount of gas available for star formation. This is also analogous to the cosmic star formation rate density (SFRD), which peaks around $z\sim 2-3$ and follows a similar trend.

The emergence of consistent functional forms across the three models implementing distinct sub-grid physics suggests these equations reflect robust physical mechanisms regulating galaxy evolution that go beyond specific numerical implementations. Across all models, the coefficients reveal a hierarchy in the importance of different physical processes: halo mass terms consistently have the largest coefficients, followed by cosmological parameters in the early-time equations, and finally feedback parameters at late times. This quantifies how the fundamental properties of dark matter halos set the stage for galaxy evolution, while feedback processes introduce secondary but significant modifications to the resulting star formation histories. Future work applying these equations to other CAMELS models such as the CAMELS-SAM \citep{2023ApJ...954...11P} or the SWIFT-EAGLE suite of CAMELS runs \citep{2024arXiv241113960L}, and perhaps even higher resolution simulations like FIRE \citep{2014MNRAS.445..581H, 2018MNRAS.480..800H} can help validate and refine these equations further. They also suggest specific observational tests that could help discriminate between models, such as examining the correlation between relative black hole masses and star formation history widths, where the models predict notably different $c_3$ coefficients in the $\tau$ equation.

\subsection{A physical reparametrization of SN feedback across CAMELS codes} \label{sec:disc_cgm_juggling}

\begin{figure*}
    \centering
    \includegraphics[width=\textwidth]{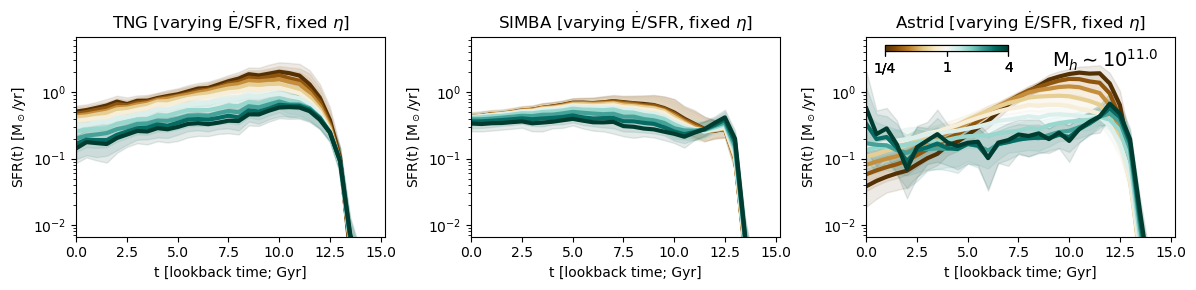}
    \includegraphics[width=\textwidth]{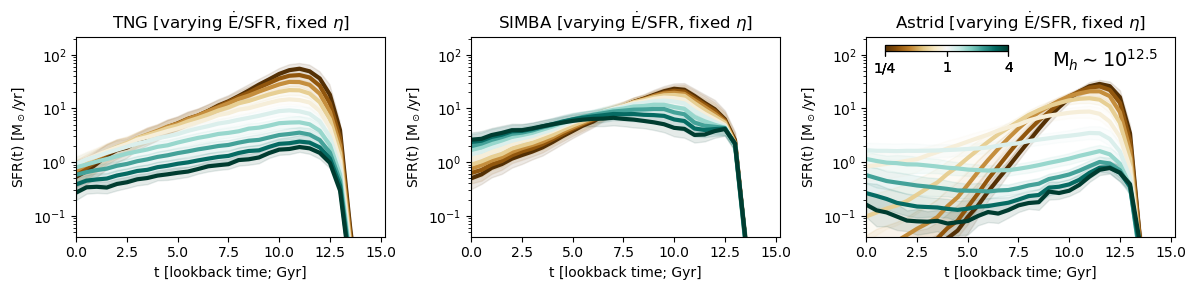}
    \includegraphics[width=\textwidth]{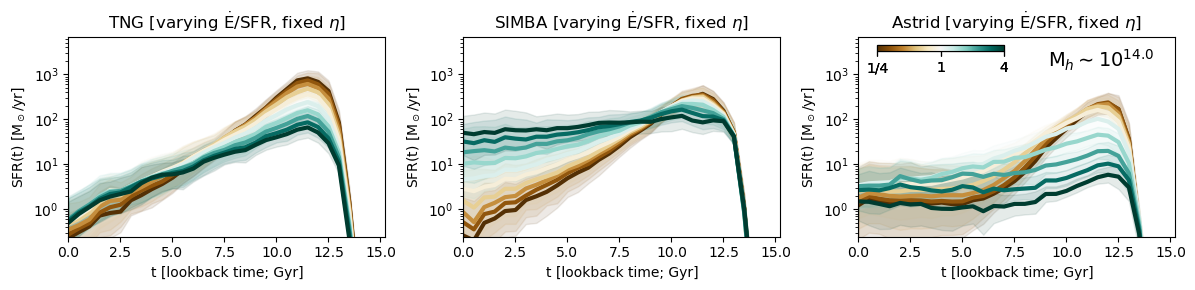}
    \caption{\ch{Changes in the SFHs for all three CAMELS models while varying \sna\ and \snb\ such that the energy per unit SFR is varied while keeping the mass loading factor fixed, at low (\textit{top}), intermediate (\textit{middle}) and high (\textit{bottom row}) halo masses (using the normalizing flow emulator developed in Section~\ref{sec:method_flows}). Note that (except for the high mass halos where BH feedback dominates), increasing the energy per units SFR (at fixed mass-loading $\eta$) leads to less star formation. }}
    \label{fig:fixedeta_mode_SFHs}
\end{figure*}

\begin{figure*}
    \centering
    \includegraphics[width=\textwidth]{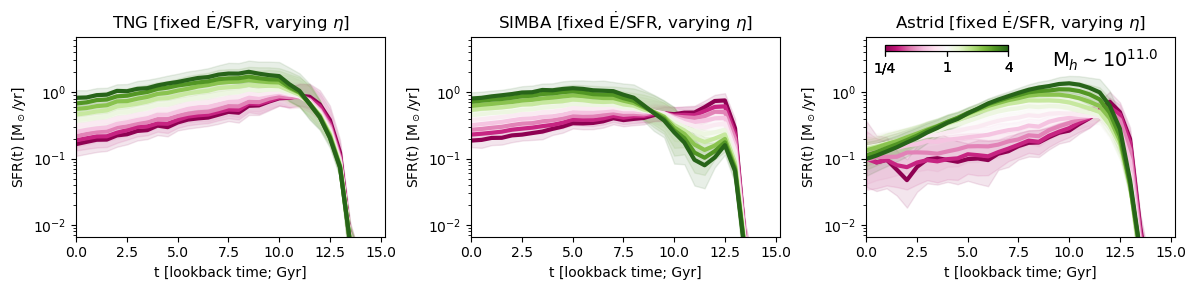}
    \includegraphics[width=\textwidth]{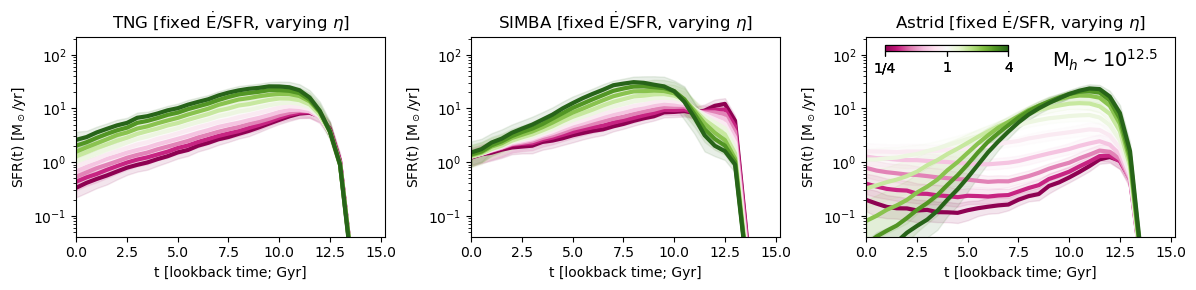}
    \includegraphics[width=\textwidth]{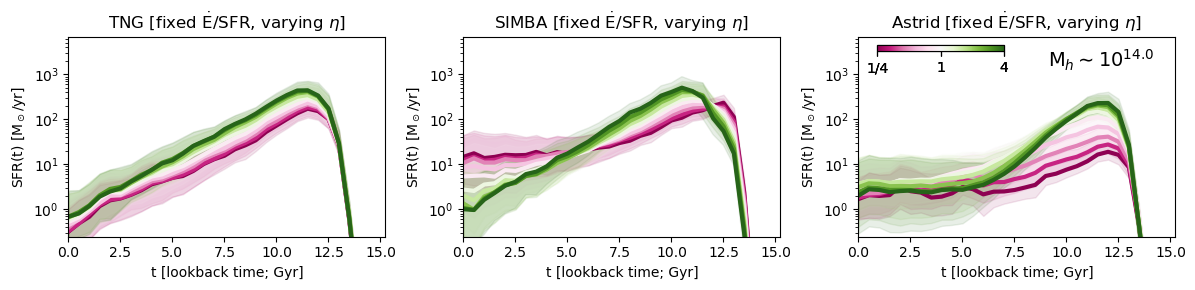}
    \caption{\ch{Same as Figure~\ref{fig:fixedeta_mode_SFHs} but now varying the mass loading $\eta$ while keeping the energy per star formation constant. Note that (except at very early times) increasing the mass loading of galactic winds actually leads to more star formation.}}
    \label{fig:fixedE_mode_SFHs}
\end{figure*}

\begin{figure*}
    \centering
    \includegraphics[width=\textwidth]{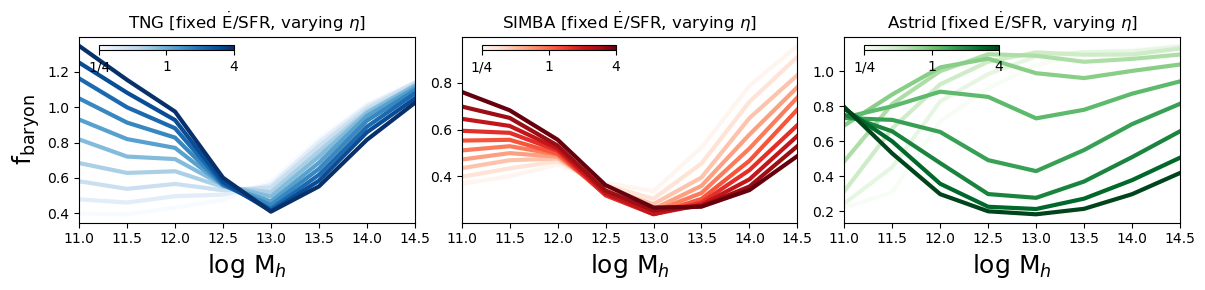}
    \includegraphics[width=\textwidth]{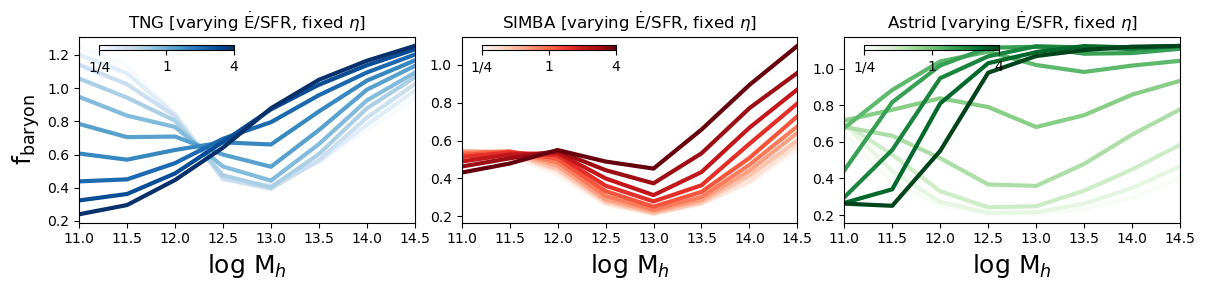}
            \includegraphics[width=\textwidth]{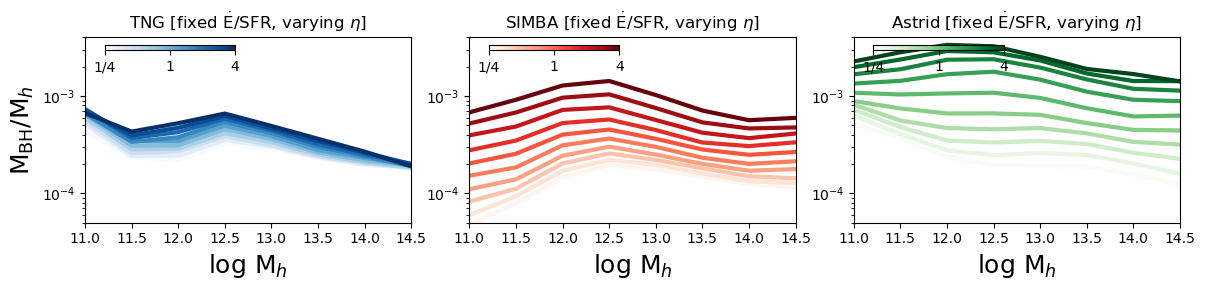}
            \includegraphics[width=\textwidth]{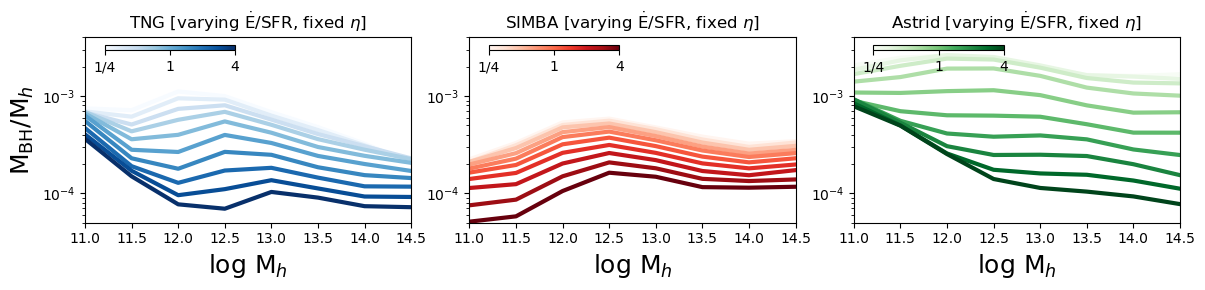}
            
    \caption{\ch{Variations in the baryon fraction (top panels) and black hole mass (bottom panels) for all three simulations (going from left to right: CAMELS/TNG, CAMELS/SIMBA, and CAMELS/ASTRID) as a function of halo mass. Within each set of six panels, the top column shows the result of varying the mass loading while keeping the energy per SFR constant while the bottom column is the opposite. }}
    \label{fig:varE_vareta_physquants}
\end{figure*}

\begin{figure*}
    \centering
    \includegraphics[width=\textwidth]{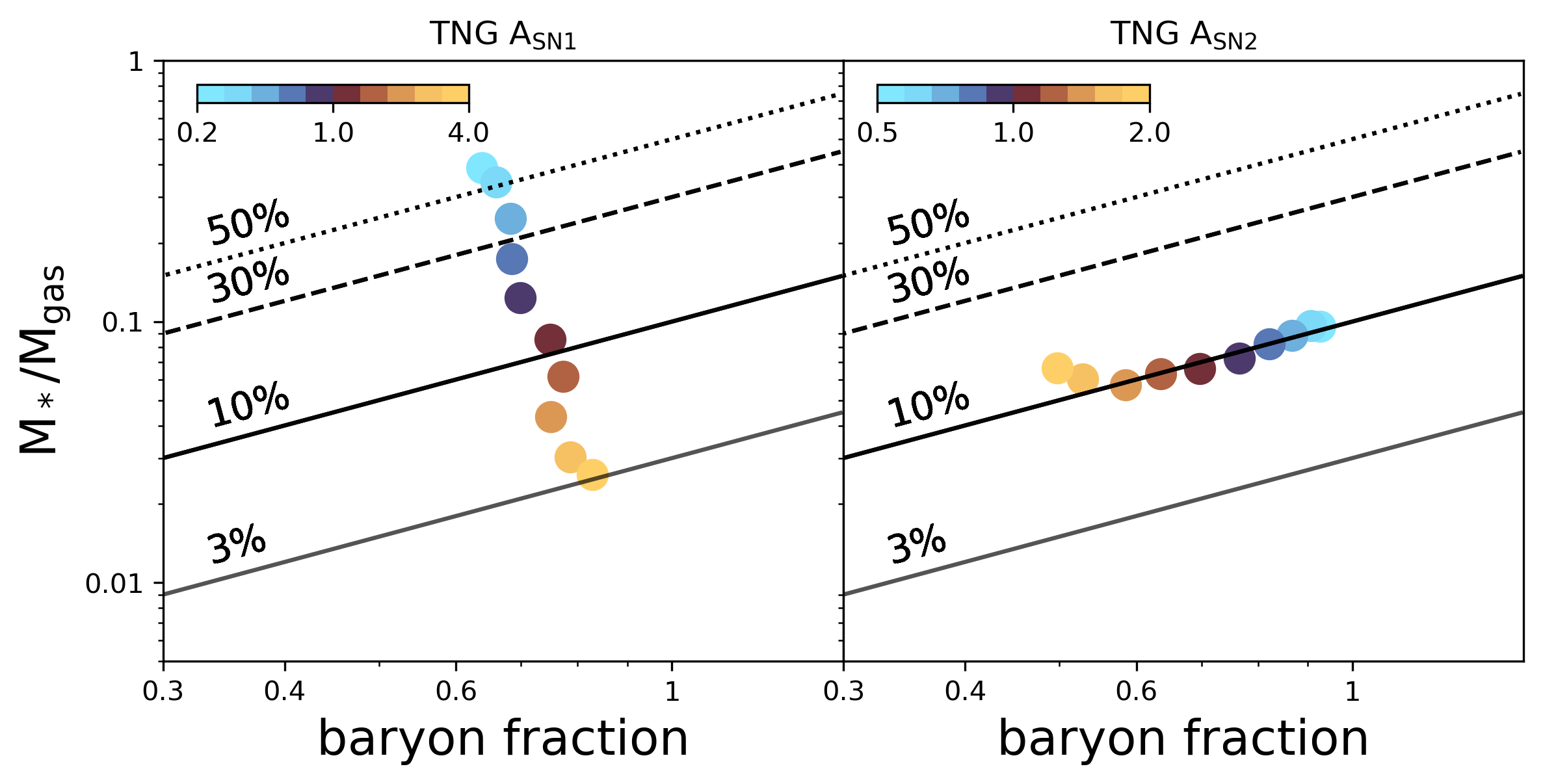}
    \caption{Varying the energy per unit SFR (\sna{}) and wind speed (\snb{}) in CAMELS/TNG have distinct effects on the baryon fraction and stellar-to-gas mass ratio (SGMR). This allows us to disentangle the effects of factors that change the gas content of halos from factors that change the efficiency of forming stars from that gas. The black lines show lines of constant baryon-fraction to stellar-to-gas mass ratio. While increasing \sna{} changes the star formation efficiency, changing \snb{} tends to affect the baryon content of the halo.}
    \label{fig:baryon_cycling_tng}
\end{figure*}

The interplay between stellar feedback strength and baryon cycling emerges as a key factor shaping galaxy evolution across the three CAMELS models we examine. In this section, we consider the `lifting' of gas into and out of the CGM due to feedback at early epochs in the context of \ch{winds that are more or less mass and energy-loaded} and how that influences the subsequent evolution of galaxies \citep{2022ApJ...933..133M, 2023ApJ...949...21C, 2023ApJ...956..118P, 2024ApJ...960...28V}. Since $\dot{E} \propto \eta v_w^2$SFR and $\dot{M} \propto \eta $SFR, we can reparameterize the two SN parameters in all CAMELS variations such that they are mass loading ($\eta$) and energy per unit SFR ($\dot{E}$/\rm{SFR}). For CAMELS/TNG and CAMELS/ASTRID, this is straightforward: 
\begin{equation}
\dot{E}/\rm{SFR} \propto  \rm{A_{SN1}} ;\qquad \eta \propto \rm{A_{SN1} A_{SN2}^{-2}}.
\end{equation}
For CAMELS/SIMBA, instead we have 
\begin{equation}
\dot{E}/\rm{SFR} \propto \rm{A_{SN1} A_{SN2}^{2}}; \qquad \eta \propto \rm{A_{SN1}}.
\end{equation}

\ch{Using the emulator developed in Section~\ref{sec:method_flows},} we show the resulting variations of these two new parameters in Figures~\ref{fig:fixedeta_mode_SFHs} and \ref{fig:fixedE_mode_SFHs}. \ch{These figures show all mass variations but the reparameterizations apply best to the low mass halos in which SN feedback dominates.} While the SFHs are certainly not identical across the simulation codes when varying these two physical parameters, they are more similar to the \sna{} and \snb{} variations in Figure~\ref{fig:sfh_1param_exploration} \ch{(compare in particular the \sna{} variations of TNG vs SIMBA)}. This similarity is evidence that this is a way of characterizing the SN feedback \ch{in a way that can be compared across all models}. The remaining differences are not surprising given that we have not attempted to match absolute parameter values of the new parameters; in addition, the other parts of the model (e.g. BH feedback) react differently, even if the SN feedback was identical.

\ch{Winds with higher energy loading (but fixed mass loading, top row of Figure~\ref{fig:fixedeta_mode_SFHs}) result in systematically less star formation, despite the fact that, by construction, the mass ejected out of the ISM is unchanged. We argue that this occurs because it is primarily energy that regulates mass inflow from the IGM to the CGM and from the CGM to the galaxy itself, consistent with analytic models of \citet{2023ApJ...949...21C} and \citet{2023ApJ...956..118P}.  }

\ch{We perform the opposite experiment in Figure~\ref{fig:fixedE_mode_SFHs}, keeping the energy per SFR fixed, while varying the mass loading factor. Again, focusing on the lower-mass halos, we see that increasing $\eta$ actually leads to \emph{higher} star formation rates. As explained in detail in \cite{2024ApJ...976..150V}, this arises because lower specific energy outflows are less able to lift gas out of the halo and tend to lead to higher halo densities and more radiative cooling. This all leads to higher rates of inflow onto the galaxy and therefore more star formation. }

\ch{Compared to increasing the energy per star formation rate, increasing the mass loading leads to relatively higher baryon fractions and higher black holes for low mass halos. This is shown in Figure \ref{fig:varE_vareta_physquants}, where we look at the halo baryon content and black hole mass as a function of halo mass. Focusing only on the low mass halos where SN feedback dominates, we see the results are consistent with the effects seen in the SFHs: increasing the energy per unit SFR lowers the gas density in the halo and decreases the baryon fraction, while increasing the mass loading factor leads to higher baryon fractions since the lower specific energy outflows cannot heat inflowing gas and prevent gas from entering the halo. A similar effect is seen for the black hole masses -- a higher mass loading factor boosts the black hole mass, while more energy per unit of star formation has the opposite effect. The high halo mass end is more complicated because of the importance of black hole feedback, but we argue the results are still consistent with an extension of the trends we have seen so far. Higher energy per SFR (at fixed mass loading) leads to more efficient regulation even in high-mass halos and so drives down the black hole mass since its regulation is not required at the same level. However, SN feedback is unable to lift mass out of (or prevent it from falling in to) the deeper potential wells of high mass halos, leading to the observed flip at Milky-Way masses and higher baryon fractions.}

The relative importance of energy and mass are further illustrated in Figure \ref{fig:baryon_cycling_tng}, where we look at the halo baryon content against the fraction of baryons that are converted into stars in TNG.
Consistent with the effects seen in the SFHs, increasing the energy per unit SFR suppresses the efficiency of star formation while modestly increasing the baryon fraction, while increasing the wind speed (specific energy content) \ch{prevents more material from entering the halo}, leading to lower baryon fractions and SFRs. This behavior is consistent with \ch{energy injection being the primary regulatory process} operating in TNG \citep[see also][]{2024ApJ...976..150V, 2024ApJ...976..151V}. 

Interestingly, we find that the choice between \ch{varying the mass- vs energy-loading of the stellar} winds creates distinct signatures in both the timing and efficiency of baryon cycling, consistent with current literature \citep{2020MNRAS.498.1668W, 2022MNRAS.516..883S, 2023MNRAS.524.5391A, 2023ApJ...956..118P, 2024ApJ...960..100S, 2024MNRAS.529..537K}, and might be a key component to understanding the mechanisms by which early galaxies form \citep{2022MNRAS.511.3895F}. Winds with higher mass-loading favor repeated cycles of gas ejection and re-accretion, spreading star formation over longer timescales but potentially at the cost of reduced peak star formation rates. On the other hand, while winds with higher energy-loading are more effective at regulating early star formation through gas removal, they may limit the gas reservoir available for late-time star formation, particularly in lower-mass systems. Complementary observations to the average long-timescale SFH shape, such as short-timescale SFR stochasticity \citep{2023ApJ...947...61S}, and chemical enrichment (tracing metal loading; \citealt{2023ApJ...949...21C}) can help further distinguish between the effects of ejective vs preventative feedback in different regimes, and are explored in the next section. 

\subsection{Inference of Cosmology and Feedback Parameters from Observations} \label{sec:disc_inference}

\begin{figure*}
    \centering
    \includegraphics[width=0.32\textwidth]{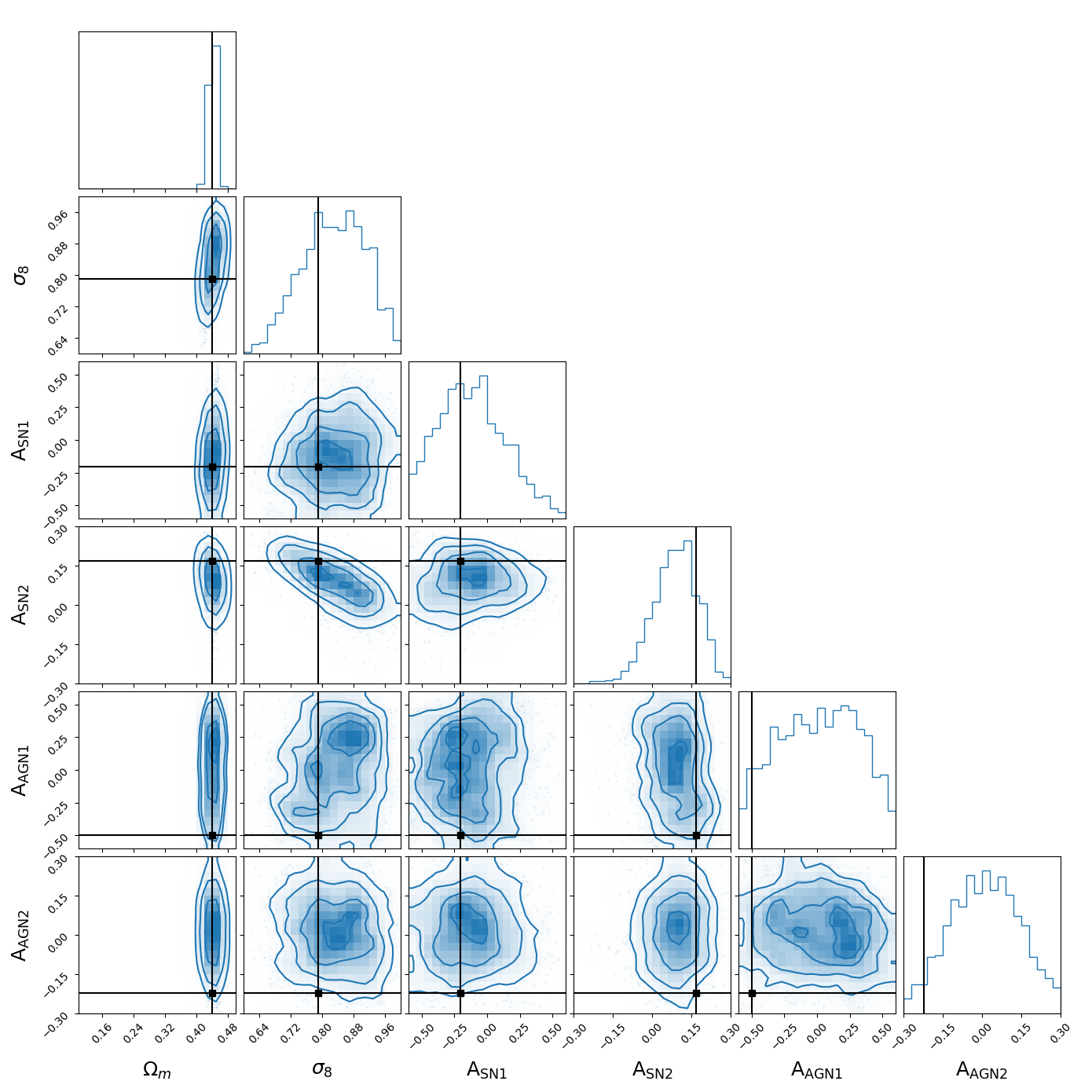}
    \includegraphics[width=0.32\textwidth]{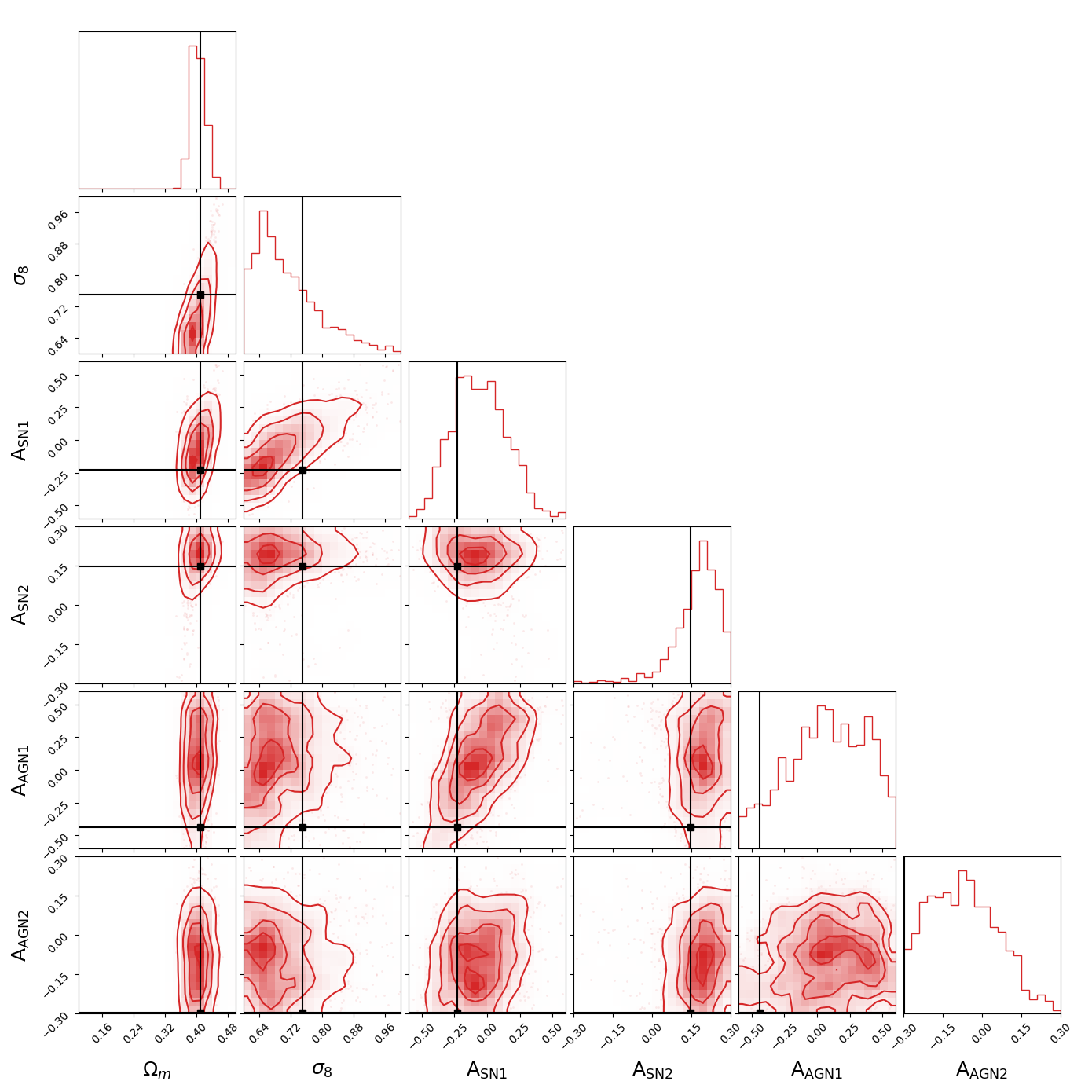}
    \includegraphics[width=0.32\textwidth]{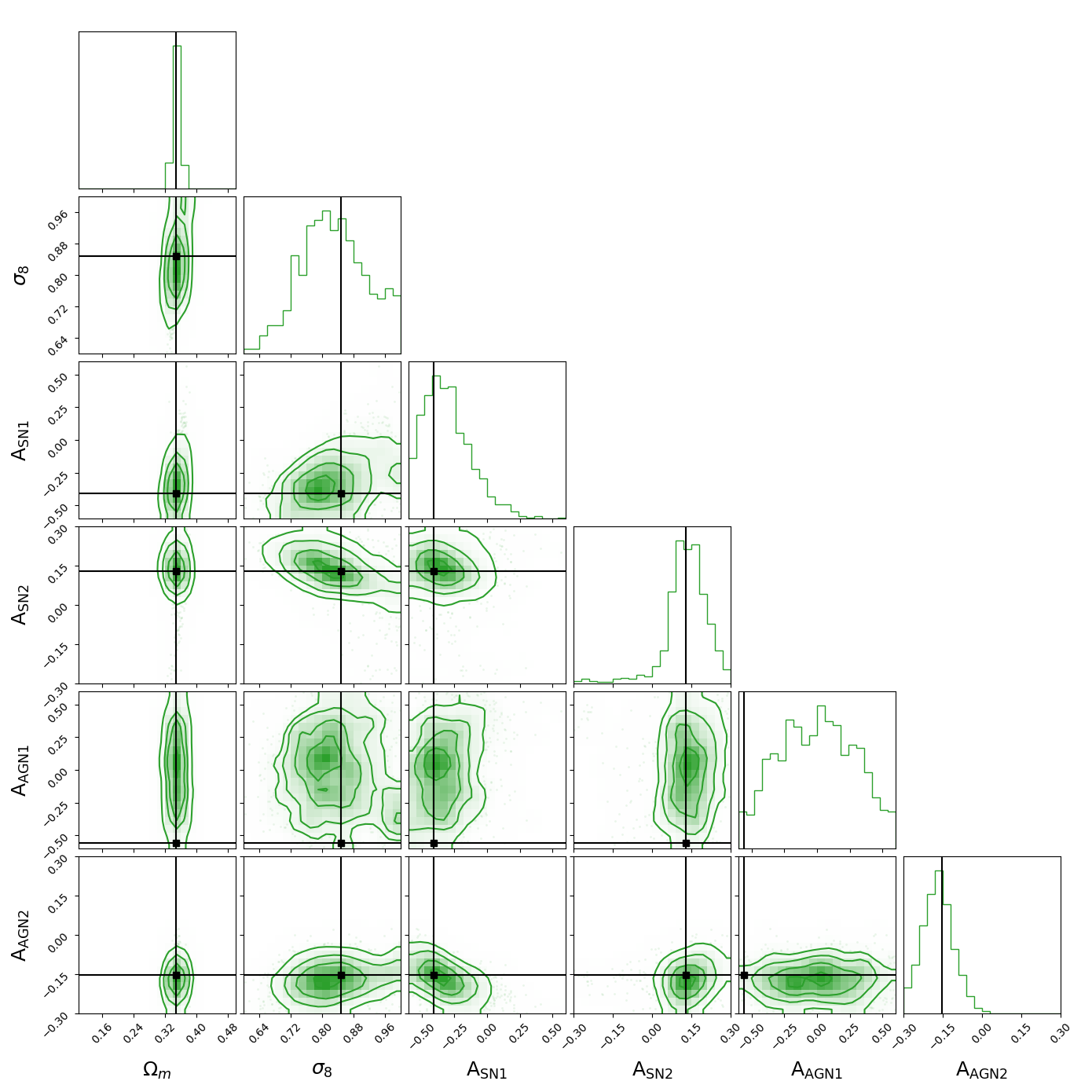}
    \includegraphics[width=\textwidth]{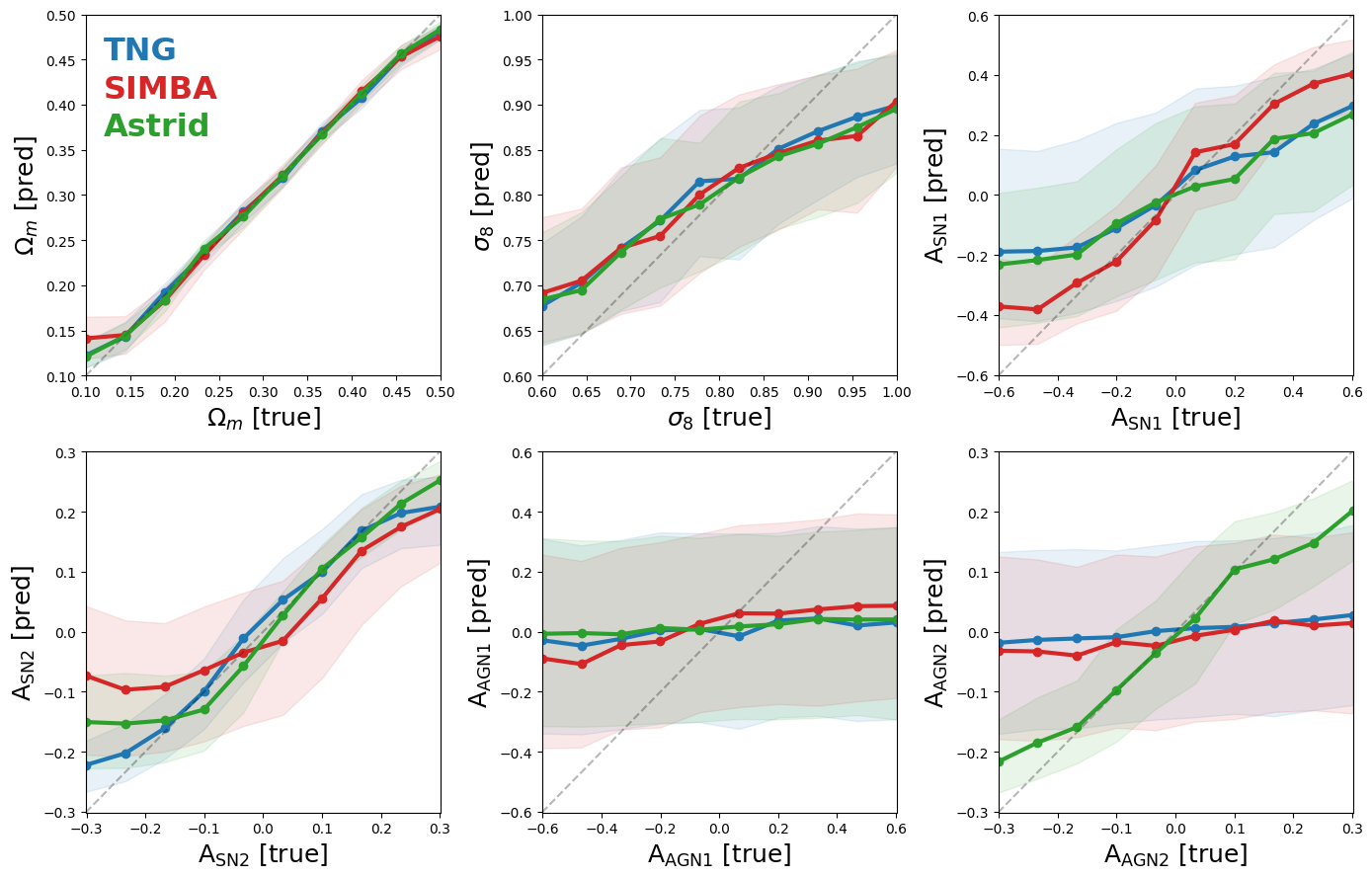}
    \caption{Inferring the CAMELS box parameters (both cosmology and feedback strengths) from a sample of 100 SFHs using SBI across the three models. Top: Corner plots showing the inferred posteriors for a single sample of SFHs corresponding to a realization of cosmology and feedback strength. Bottom: Average results across 1000 realizations across the full CAMELS parameter space. The cosmology is well constrained, as are the \sna{}, \snb{} parameters, while \agna{} and \agnb{} are not well constrained except for ASTRID. Since the feedback parameters do not always denote the same quantity across the models, caution needs to be exercised while interpreting absolute values. However, relative trends across galaxy populations will still provide robust estimates of how feedback changes across galaxy demographics. This represents a proof of concept, and application to real data will require forward modeling of noise and systematic effects.
    }
    \label{fig:reverse_flow_sbi}
\end{figure*}

In addition to understanding the impact of scaling feedback parameters on galaxy SFHs, it is also possible to invert the normalizing flow to instead infer the CAMELS parameters from a sample of galaxy SFHs \citep[see e.g.,][]{2024arXiv241113960L}. Figure \ref{fig:reverse_flow_sbi} shows a proof-of-concept for this procedure, using a simulation-based inference (SBI) framework to predict the cosmology and feedback parameters using the average SFHs of 100 galaxies with \massrange{9.5}{11.5} (similar to \ch{the sample considered in} the rest of this paper). We find comparable constraints on the cosmological parameters to other CAMELS papers \citep[see e.g., ][ and more]{2022ApJS..259...61V, 2022ApJ...933..133M, 2023ApJ...944...67J, 2024arXiv241204559L,2023MNRAS.525.1779P}, and slightly weaker constraints on stellar feedback strengths. AGN feedback remains largely unconstrained, except for \agnb{} in ASTRID, which strongly affects quenching and late-time star formation. This is in line with the behavior in Figure \ref{fig:sfh_1param_exploration}, where the SFHs over a wide stellar mass range are largely insensitive to AGN feedback. Comparing to the figures as a function of halo mass, it is certainly plausible that with a larger sample where we break down the inference over bins of stellar or halo mass or include additional enrichment information, it will be possible to obtain tighter constraints even on the AGN feedback parameters. \ch{This is largely meant to serve as a proof-of-concept demonstrating the viability of inferring feedback and cosmological parameters using the average SFHs of galaxies, and is not necessarily robust across models (i.e. when training on one model and applying to another, as in \citet{2023ApJ...952...69D}). In future work, we will construct a more general inference setup using the set of equations derived in Section \ref{sec:disc_univmodel}, using domain adaptation to account for the different coefficients across the models.} While \ch{the simple setup across a broad range in stellar mass described here} can only constrain the strength of feedback parameters that have an effect on the SFHs, doing so provides complementary information to other probes of feedback like the Lyman-$\alpha$ forest \citep{2023AJ....166..228T}, matter power spectrum \citep{2023MNRAS.526.5306D, 2024MNRAS.529.4896G}, structural properties \citep{2023MNRAS.525.6191B}, or galaxy color distributions \citep{2024arXiv241113960L}. It is possible that future studies could obtain joint constraints by combining multiple such probes across a range of environments, masses and epochs. 

\subsection{Feedback in Context} \label{sec:disc_literature}

Feedback mechanisms play a critical role in regulating star formation in galaxies. 
Here we briefly discuss the current literature on how feedback affects star formation and the baryon cycle in galaxies.

\subsubsection{Feedback Variations in the Literature}

\ch{Earlier studies have found that both stellar and AGN feedback suppress gas inflow into the halo at all redshifts, and promote baryon cycling at galaxy scales \citet{2015MNRAS.448...59N}, which qualitatively agree with the reduced baryon fractions and smaller $\tau$ we find at lower feedback strengths. 
Similar to our low \sna{} or \snb{} cases, the black hole masses in \citep{2018MNRAS.473.4077P} (sec 3.3-3.4) increase in the absence of galactic winds, consistent with our picture of higher stellar feedback inhibiting high star formation and black hole growth at early times in these halos. This also leads to generally higher gas fractions and longer quenching timescales from limiting the amount of gas blown out by kinetic AGN feedback, further highlighting the coupled nature of stellar and AGN feedback at late times. Consistent with this work, they also find lower normalizations and similar trends in gas fraction for the cases of stronger and faster winds, which correspond to the \sna{} and \snb{} terms in this work.
In a similar manner, Figure 10 of \citet{2015MNRAS.450.1937C} explores the effects of varying stellar feedback, finding that weaker feedback leads to more massive black holes, smaller sizes and lower star formation rates.}

\ch{On the AGN feedback side, \citet{2017MNRAS.465.3291W, 2020MNRAS.493.1888T} find that while the dominant channel for SMBH growth is the thermal mode, higher mass black holes self-regulate gas accretion through kinetic mode feedback and drive quenching. 
\citet{2020MNRAS.493.1888T} study how sSFR, stellar mass and black hole mass in massive galaxies are sensitive to the physics of black hole feedback in the TNG model, with quenching driven by the point where the energy from black hole kinetic winds (at low accretion rates) exceeds the gravitational binding energy of gas within the stellar radius, blowing gas out and preventing new stars from forming.}

\ch{It is also useful to compare this with studies of the baryon cycle in different models. 
\citealt{2022MNRAS.516..883S, 2024arXiv240407252S} study the effect of feedback on galaxy properties and the surrounding CGM in SIMBA, finding that the matter distribution for low-mass halos at early times ($z>2$) is influenced primarily by star-formation driven outflows, while higher mass halos past cosmic noon are affected most by AGN-driven jet feedback. Similar to Section \ref{sec:result_galstate_fbaryon}, \citealt{2022MNRAS.516..883S} find the baryon fraction for intermediate mass halos ($\sim 10^{12}-10^{13}$M$_\odot$) to be reduced due to evacuation by AGN jets.
\citealt{2017MNRAS.470.4698A} find that in-situ star formation from smooth accretion fuel the early growth of galaxies at all masses in FIRE, and a significant role is played by reaccreted gas ejected by galactic winds. They find a broad range of recycling times from $\sim 10$Myr to $\sim 1$Gyr, occurring at the scale radius of the halo. Median recycling times are $\sim 100-350$Myr, with some indication of longer times in dwarf galaxies relative to higher mass halos.
\citealt{2020MNRAS.498.1668W} find that baryons in EAGLE galaxies are more likely to come from smooth accretion than mergers compared to dark matter, with accretion at the halo scale affected by both stellar and AGN feedback and strongly dependent on subgrid physics. 
\citealt{2024arXiv240208408W} extend this analysis to the fiducial runs of three hydrodynamical models (EAGLE, SIMBA and IllustrisTNG), and inform our interpretations of the baryon fractions in Section \ref{sec:result_galstate_fbaryon}. Since the mass inflow and outflow rates correlate with the baryon fraction, we are able to use that quantity as a measure of the amount of baryon cycling and the effect of stellar and AGN feedback at low and high halo masses respectively. In addition, the authors find that galaxies in these models reach a given stellar mass and star formation rate at $z\sim 0$ in markedly different ways. One significant difference is that stellar feedback in low mass halos tends to drive material far beyond the halo in EAGLE and SIMBA, while in TNG it is stronger in the ISM (i.e. suppresses star formation more) but tends to blow material into the CGM from where it can get recycled. They also find that the scale of feedback driven outflows correlates with both the total baryon fraction within the halo and the prevention of cosmological inflow that can suppress star formation at late times.
The results in Section \ref{sec:disc_cgm_juggling} also qualitatively agree with the minimalist regulator model proposed in \citet{2024ApJ...976..150V, 2024ApJ...976..151V} in terms of the effects of increased mass- and energy-loading of supernova feedback on star formation and baryon fractions, though the addition of AGN feedback complicates the interpretation at high halo masses.}

\subsubsection{Other CAMELS Studies}

In addition to the SFH-feedback connection explored in this paper, there are numerous CAMELS papers that study the impact of varying feedback on other quantities, and the scope of what can be constrained using inference with various observations. These are often complementary to the approaches in this paper, and can be used in conjunction with the SFHs where available to obtain tighter constraints on the CAMELS parameters during inference.

\citealt{2021ApJ...915...71V, 2022ApJS..259...61V, 2023ApJ...944...67J, 2023mla..confE..21L} find that similar to the inference in Section \ref{sec:disc_inference}, we can predict cosmology, \sna{} and \snb{} using the cosmic star formation rate density, \ch{ stellar mass functions, or color distributions} in CAMELS/TNG, while the AGN parameters are \ch{poorly constrained}. They also highlight the possibility of using symbolic regression to derive analytic expressions for the SFRD that can be adapted to the SFHs in this work. However, given how ill-conditioned the equations tend to be, it can be difficult to obtain robust expressions.

Perhaps the closest analog to the current work, a companion paper by \cite{2024arXiv241113960L}, studies the inverse problem of constraining the CAMELS parameters using luminosity functions and distributions of galaxy colors, quantities that are directly affected by the SFHs and contain imprints of current and past star formation. That paper goes into detail about the care required in constructing detailed forward models for comparison with observations, and runs several tests for the amount of information contained in different observables, finding similar conclusions in terms of observational sensitivities to different parameter variations as in Section \ref{sec:disc_inference}.

\cite{2023AJ....166..228T} explore the effects of varying feedback on the low-redshift Lyman-$\alpha$ forest, finding that both AGN and stellar feedback play a role in shaping the thermal state of the IGM in SIMBA, primarily through the AGN jet feedback mode. The effects in TNG are more subtle, and primarily sensitive to stellar feedback, which indirectly influences AGN feedback by regulating the growth of the SMBH. Figures 6 and 7 in the paper explore the accretion rate density and number of black holes as a function of time. Comparing their curves with the black hole masses in Figure \ref{fig:baryon_content}, we find that the more massive black holes tend to have also formed earlier, which is also in line with the earlier \ch{times at which the SFHs peak} in those scenarios. \ch{While we do not vary the black hole seeding and accretion prescriptions in this paper, these are modeled in the extended 28-parameter CAMELS runs \citep{2023ApJ...959..136N} and will be explored further in future work. Figures 12-17 in \citet{2023ApJ...959..136N} show how varying factors like the seed mass and accretion behavior affect the overall stellar masses and cosmic SFRD in TNG and SIMBA.}

\citet{2024MNRAS.529.4896G} studies the impact of cosmological and feedback parameter variations on matter clustering, showing that the amount of spread of baryons relative to dark matter correlates roughly inversely with the baryon fractions shown in Figure \ref{fig:baryon_content}. 

Another set of model variations can be found in the FLAMINGOS suite of simulations \citep{2023MNRAS.526.6103K}, albeit at larger volumes with slightly lower overall resolution. The study uses an emulator trained on a sample of 32 smaller volume simulations, varying the strengths of stellar and AGN feedback to calibrate a larger model by learning the mapping between the subgrid parameters and corresponding variations in the stellar mass function and cluster gas fractions. As we consider more model variations (as in the much larger 28-dimensional parameter space covered in \citealt{2023ApJ...959..136N}) or scope for larger variations (e.g, using semi-analytic approaches \ch{such as those presented in} \citealt{2023ApJ...954...11P} in the CAMELS framework), these analyses can potentially be used for inference with a variety of observations for comparing and calibrating subgrid prescriptions for future simulations \citep[for e.g.,][]{2023arXiv231208277B, 2023MNRAS.526.5494M, 2024arXiv240312967E, 2024arXiv240217819L}.

\subsubsection{Higher Resolution Studies}
\label{sec:feedback_higher_res}


Higher resolution simulations can provide crucial insights into the detailed physics of feedback processes that complement our analysis of the CAMELS suite, better resolving the multiphase gas, structure formation and dynamics due to turbulence that are often approximated with a stiff equation of state in larger hydrodynamical simulations. While CAMELS enables exploration of parameter space at moderate resolution, zoom simulations and high-resolution runs can resolve the multi-phase structure of outflows and gas recycling that are responsible for the statistical trends we observe, and help estimate convergence and generalizability of our results. For example, the FIRE simulations \citep{2014MNRAS.445..581H, 2015MNRAS.454.2691M}, demonstrate that stellar feedback preferentially removes low-density material while creating channels through which outflows can escape. These simulations resolve the phase structure of the CGM at $\sim 100$pc scales, showing that outflows are inherently multi-phase, with cool clouds entrained in hot wind material - a complexity necessarily abstracted in the CAMELS subgrid models. The coupling between stellar and AGN feedback also emerges as particularly important in high-resolution studies, with \cite{2017MNRAS.470.4698A} finding that stellar feedback can effectively limit early black hole growth by disrupting gas inflows, aligning with our observation that stronger stellar feedback correlates with reduced black hole masses across all three CAMELS models. Furthermore, \cite{2024ApJ...964...54C} used parsec-scale resolution simulations to demonstrate that recycled stellar ejecta provides a crucial fuel source for central SMBHs, with a characteristic delay timescale of $\sim1.85$ Gyr between external gas accreted from mergers to energy the main galaxy and fuel SMBH growth. The ARTEMIS simulations \citep{2024arXiv240311692B} specifically investigated how feedback affects the assembly of Milky Way-mass galaxies, finding that both the strength and implementation of feedback significantly impact the morphology and structural properties of galaxies beyond just their star formation histories. This reinforces our finding that feedback effects manifest through multiple complementary observables, from metallicity scaling relations to duty cycles. These simulations provide a crucial testing ground for estimating the limitations of our current resolution, but also to validate and extend predictions such as our model in Eqn. \ref{eqn:general_sfhs} that generally describes the relation between the SFH shape and a galaxy's physical properties.


 


\subsubsection{Analytical Models for Star Formation in Galaxies}

While analytical models necessarily simplify the complex interplay of physical processes seen in simulations, they offer valuable insights into the fundamental scaling relations and physical dependencies that emerge from studies like ours. The classical ``bathtub'' or equilibrium models \citep{2010ApJ...718.1001B, 2012MNRAS.421...98D, 2014MNRAS.444.2071D} treat galaxies as gas reservoirs where star formation results from a balance between cosmological accretion and feedback-driven outflows. These models successfully reproduce the cosmic star formation rate density and broad galaxy scaling relations, but typically employ simplified prescriptions for feedback processes \citep[e.g., ][]{2013ApJ...768L..37T, 2018ApJ...868...92T}. Recent work by \cite{2020MNRAS.492.2418S} link the star formation rate to a balance between energy from stellar feedback and the depth of the halo potential well, with possible extensions to include the effects of AGN feedback, while \cite{2014arXiv1406.5191K} hypothesized that star formation histories (and resulting scaling relations) emerge from stochastic processes following central limit behavior, while \citet{2016MNRAS.457.2790T} found evidence for the SFR-M$_*$ scaling relation as a product of self-regulation of star formation. On the other hand, analytical models for black hole growth \citep{2017MNRAS.465...32B, 2020ApJ...897..102C} attempt to explain the co-evolution of SMBHs and their host galaxies, including how AGN feedback regulates galaxy growth and quenching, which in turn determines how the black hole can accrete gas and grow.

Our empirical equations describing SFH parameters (Eqn. \ref{eqn:general_sfhs}) contribute to this growing body of analytical descriptions that aim to capture the essential physics of galaxy evolution while maintaining interpretability, while extending to a range of models (TNG, SIMBA, ASTRID) and parameter spaces (folding in cosmology, and a variety of stellar and AGN feedback prescriptions). The double power-law parametrization we employ to describe the average SFHs of galaxies builds upon previous analytical frameworks that describe galaxy growth. For instance, the rising slope ($\beta$) encoding early star formation shows a robust correlation with the matter density ($\Omega_m$), reflecting fundamental connections between cosmic accretion and initial galaxy growth that have been explored in \ch{past models} 
\citep[e.g.][]{1974ApJ...187..425P}. The falling slope ($\alpha$) and characteristic timescale ($\tau$) parameters capture the complex interplay between feedback processes and baryon cycling that regulate late-time evolution, similar to ``bathtub'' or equilibrium models that treat galaxies as gas reservoirs with inflow and feedback-driven outflow term.

Our finding that the SFH shape parameters can be expressed through relatively simple combinations of physical quantities (halo mass, baryon fraction, and relative black hole mass) suggests that a large portion of the complex hydrodynamical effects can be captured by a reduced set of key variables. This aligns with work showing that galaxy evolution follows relatively simple scaling relations despite the underlying complexity of feedback processes \citep{2020MNRAS.498..430I}. However, the variation in coefficients across TNG, SIMBA, and ASTRID (Table \ref{tab:general_sfh_coeffs}) highlights how different implementations of sub-grid physics manifest in the emergent analytical relationships. By identifying the key physical quantities and relationships that govern SFH shapes, these models can help guide both the development of sub-grid prescriptions in simulations and the interpretation of observed galaxy properties in terms of underlying physical processes.


\section{Conclusions and Future Work} \label{sec:conclusions}

Using the CAMELS suite of cosmological simulations, we have systematically investigated how cosmology,  and stellar and AGN feedback shape the star formation histories of galaxies across cosmic time. By analyzing the effects of varying the CAMELS cosmology and feedback parameters on the average SFHs of galaxies in the TNG, SIMBA, and ASTRID models, we identify several fundamental aspects of how feedback regulates galaxy evolution:

\begin{enumerate}
    \item The coupling between stellar and AGN feedback plays a crucial role in determining SFH shapes beyond their individual effects. Strong interaction terms across all three models demonstrate how stellar feedback can regulate black hole growth, while AGN feedback modulates the efficiency of baryon cycling. These effects manifest distinctly across different mass scales, with stellar feedback dominating at early times and in low-mass halos, while AGN feedback becomes increasingly important for massive systems after cosmic noon.
    \item When we parametrize galaxy SFHs using a double power-law model, different parts of the SFH (rising and falling slopes, time of peak SFR and overall normalization) exhibit robust correlations with physical quantities including halo mass, baryon fraction, and black hole mass. The remarkable consistency of these correlations across different numerical implementations suggests they reflect fundamental physics of galaxy evolution rather than specific modeling choices. This provides a novel framework for using SFH shapes to constrain feedback processes in both simulations and observations.
    \item  Varying the parameters that affect the strengths of mass- and energy-loading of galactic winds create distinct signatures in both the timing and efficiency of baryon cycling which are consistent across all simulation codes. Winds with higher mass loading favor repeated cycles of gas ejection and re-accretion, leading to extended periods of moderate star formation. In contrast, winds with higher energy loadings more effectively regulate star formation through preventative feedback \ch{while also limiting} the gas reservoir available for late-time star formation, particularly in lower-mass systems.
\end{enumerate}

These results demonstrate that galaxy SFHs encode rich information about feedback processes that can complement traditional observational probes. Rich constraints on galaxy SFHs are already available from IFU spectroscopy at low redshifts \ch{from surveys including SDSS-IV MaNGA \citep{2015ApJ...798....7B}, CALIFA \citep{2012A&A...538A...8S} and SAMI \citep{2015MNRAS.447.2857B}}, with galaxy surveys like LEGA-C \citep{2016ApJS..223...29V} at intermediate redshifts, or JWST surveys like CEERS \citep{2025ApJ...983L...4F} and JADES \citep{2023arXiv230602465E} at even higher redshifts. Looking ahead, extending this analysis to observations, to higher redshifts, and validating the model's predictions using different datasets or high-resolution simulations all provide interesting avenues of study. 

The framework developed here provides a new way to assess and refine feedback implementations in cosmological simulations while offering physically motivated interpretations of observed galaxy properties. As both observational constraints on galaxy SFHs and computational capabilities continue to improve, this approach will become increasingly valuable for understanding the complex interplay between feedback processes that shape galaxy evolution.


%

\begin{acknowledgments}
This paper is the culmination of a long series of analysis, and KI would like to thank Eric Gawiser, Sandra Faber, Steve Finkelstein, Casey Papovich and Viraj Pandya for their sage advice and perspectives. \ch{We would also like to thank the anonymous referee for their insightful and constructive comments.} KI thanks the CAMELS collaboration for making their dataset public, Kathryn Johnston for coining the phrase 'the CAMELS Multiverse', and Noopur Gosavi for moral support while wrangling the figures down to a manageable level. This project started during the GALEVO-23 program: Building a Physical Understanding of Galaxy Evolution with Data-driven Astronomy. 
This research was supported in part by grant NSF PHY1748958 to the Kavli Institute for Theoretical Physics (KITP). 
This research was supported in part by the National Science Foundation under Grant No. NSF PHY-1748958.
Support for KI was provided by NASA through the NASA Hubble Fellowship grant HST-HF2-51508 awarded by the Space Telescope Science Institute, which is operated by the Association of Universities for Research in Astronomy, Inc., for NASA, under contract NAS5-26555.
\ch{TS is supported by NSF through grant AST-2510183 and by NASA through grants 22-ROMAN22-0055 and 22-ROMAN22-0013 and acknowledges the support of the NSF-Simons AI-Institute for the Sky (SkAI) via grants NSF AST-2421845 and Simons Foundation MPS-AI-00010513. }
GLB acknowledges support from the NSF (AST-2108470 and AST-2307419, ACCESS), a NASA TCAN award, and the Simons Foundation through the Learning the Universe Collaboration.
\ch{DAA acknowledges support from NSF grant AST-2108944 and CAREER award AST-2442788, NASA grant ATP23-0156, STScI JWST grants GO-01712.009-A, AR-04357.001-A, and AR-05366.005-A, an Alfred P. Sloan Research Fellowship, and Cottrell Scholar Award CS-CSA-2023-028 by the Research Corporation for Science Advancement.}
The Flatiron Institute is supported by the Simons Foundation. 
We thank the developers of open-source astronomical software that enabled this analysis, listed below.
\end{acknowledgments}

\vspace{5mm}


\software{numpy \citep{numpy}, 
scipy \citep{scipy}, 
astropy \citep{astropy1,astropy2,astropy3}, 
matplotlib \citep{matplotlib},
pytorch \citep{pytorch}, 
pathfinder \citep{pathfinder}, 
LTU-ILI \citep{2024arXiv240205137H}, 
pyGAM \citep{pygam}, 
pysr \citep{2024ascl.soft09018C}, 
sympy \citep{sympy}, 
scikit-learn \citep{sklearn}, and 
hickle \citep{hickle}.
      }



\appendix

\section{Comparison of the normalizing flows predictions against the 1P runs}
\label{app:1p_vs_nn}

Since we use the normalizing flow to sample SFHs in different parts of the feedback parameter space, it is important to verify that they produce accurate SFHs in the known portions of it. To this end, we use the 1P CAMELS boxes, which vary the feedback in each parameter while holding the remaining ones constant, to test the normalizing flow which has been trained on the LH dataset. In Appendix \ref{app:1p_vs_nn} we compare the SFHs for the fiducial normalizing flows (\massrange{9.5}{11.5}) against SFHs computed using the CAMELS 1P boxes for the three models (TNG, SIMBA and ASTRID). There is qualitative agreement between the 1P and normalizing flow SFHs, with the 1P `truth' consistent within the distribution of the sampled SFHs from the flow.

To check whether the SFHs from the normalizing flows are accurately able to predict the SFHs of galaxies in different regimes of the CAMELS parameter space, we test them against SFHs from the single parameter variation (1P) boxes of the various CAMELS runs. 

\begin{figure*}
    \centering
    \includegraphics[width=0.49\textwidth]{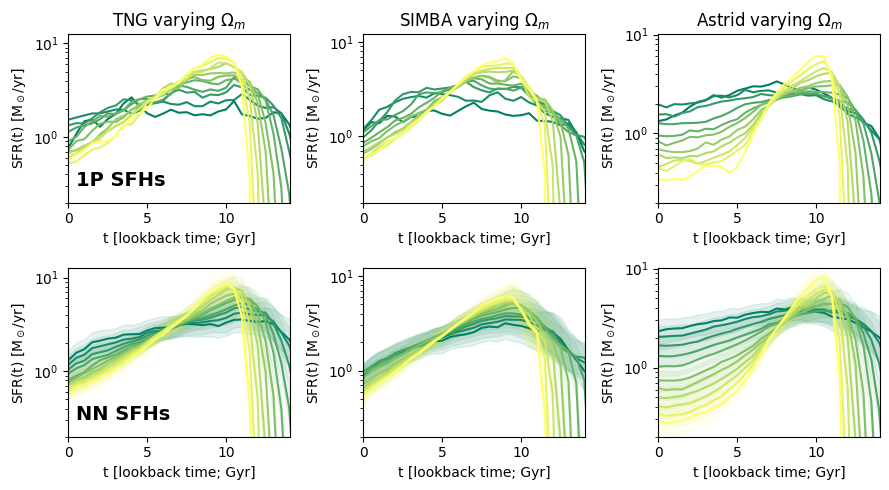}
    \includegraphics[width=0.49\textwidth]{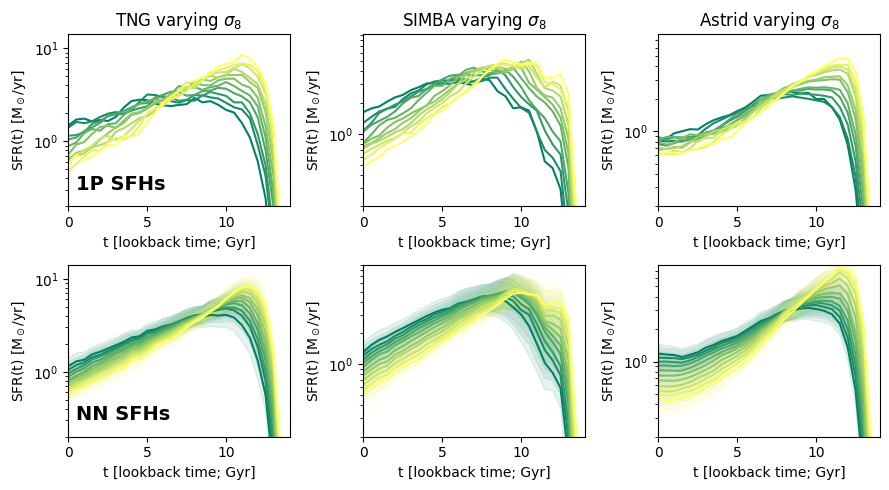}
    \includegraphics[width=0.49\textwidth]{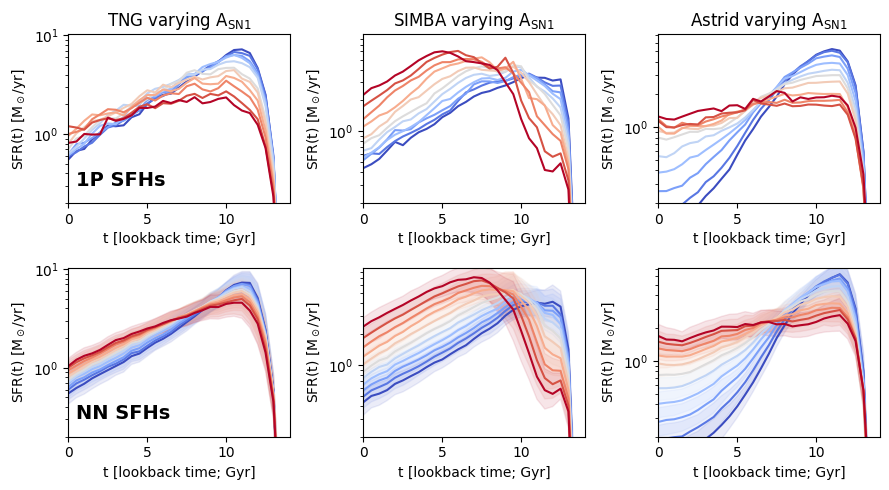}
    \includegraphics[width=0.49\textwidth]{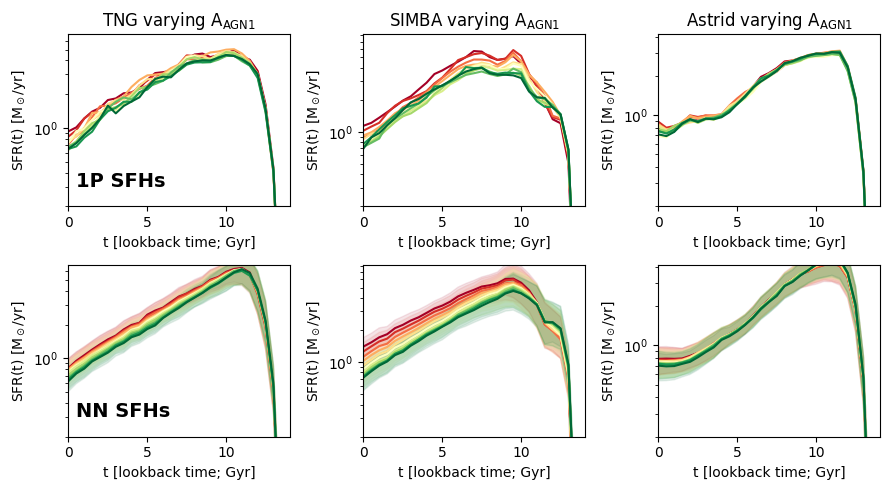}
    \includegraphics[width=0.49\textwidth]{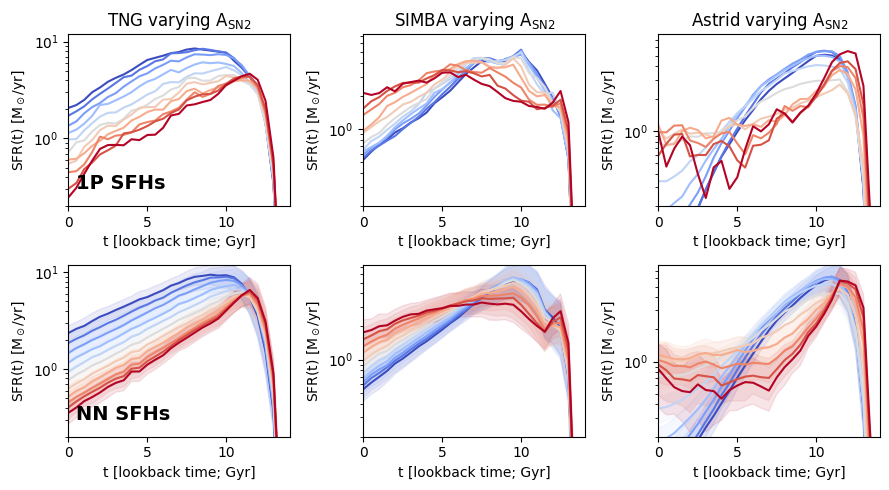}
    \includegraphics[width=0.49\textwidth]{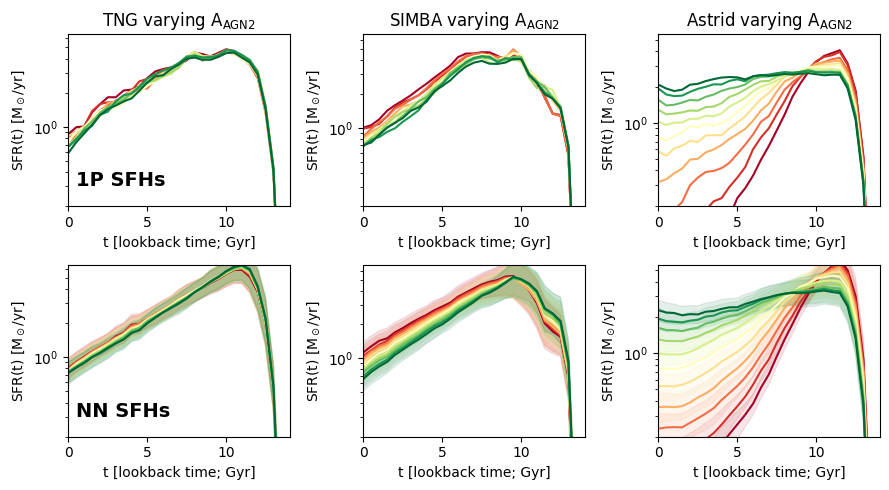}
    \caption{Validation plots comparing median (solid lines) and 16-84th percentiles (shaded regions) of the SFHs sampled from the trained normalizing flows (labeled NN) compared against SFHs from the CAMELS 1P dataset that vary a single parameter in the CAMELS box while keeping the others fixed at the fiducial value. The different panels show the 1P dataset for all six CAMELS box parameters. While the 1P datasets shows a single realisation of SFHs, the normalizing flow returns a distribution of SFHs corresponding to each set of parameters.}
    \label{fig:1p_vs_nn_validation}
\end{figure*}

We show the results of this analysis in Figure \ref{fig:1p_vs_nn_validation}. While the overall comparison is qualitatively similar, the 1P SFHs tend to look more stochastic than the normalizing flow predictions. This is because the predictions are a median of many realisations of what the SFHs can be, while the 1P sets are the SFHs corresponding to a single box, and cosmic variance can be quite significant for the small 25 Mpc boxes. 

\section{A full list of interaction terms that affect galaxy SFHs}
\label{app:all_interactions}

\begin{figure}
    \centering
    \includegraphics[width=0.32\linewidth]{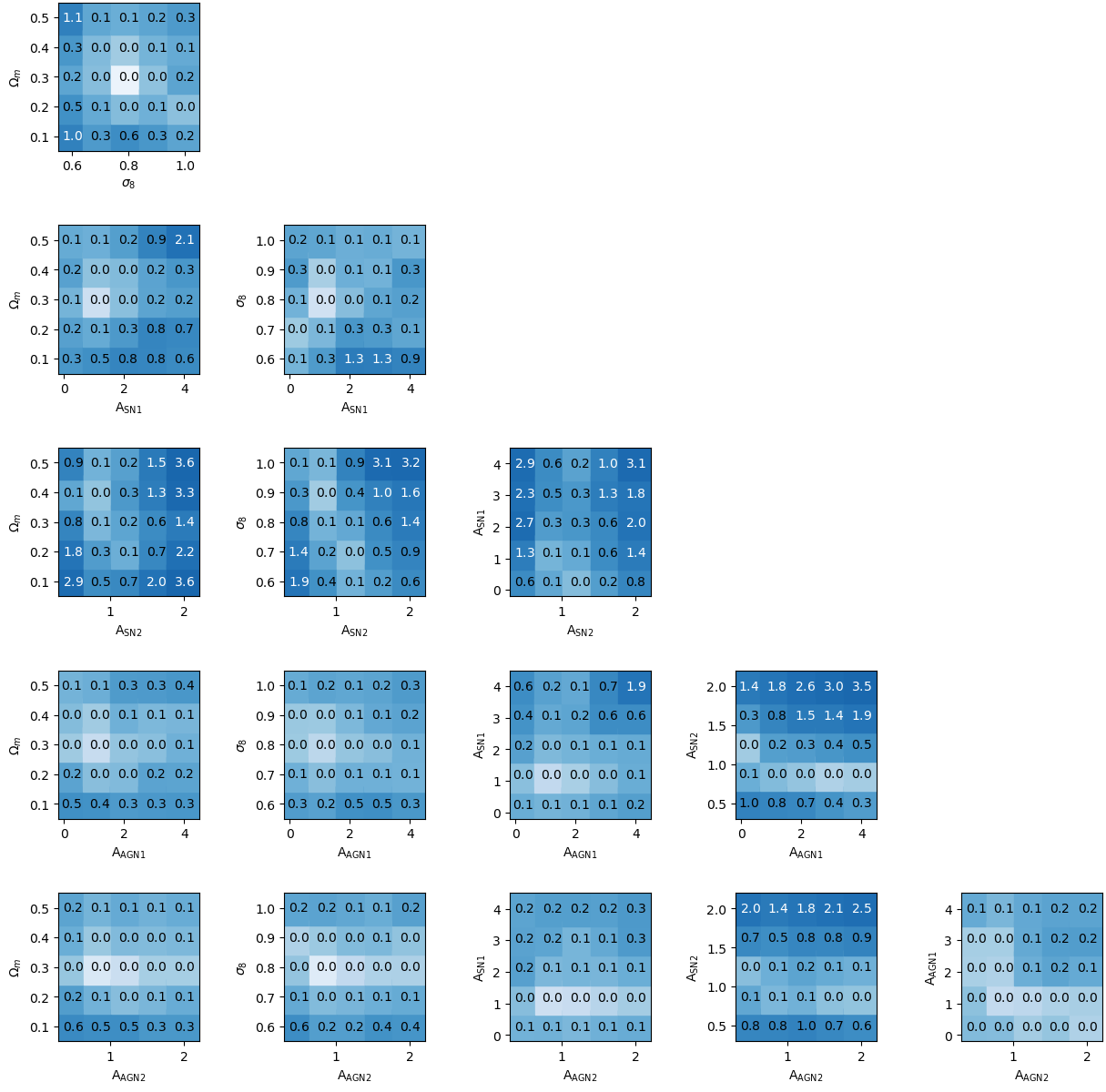}
    \includegraphics[width=0.32\linewidth]{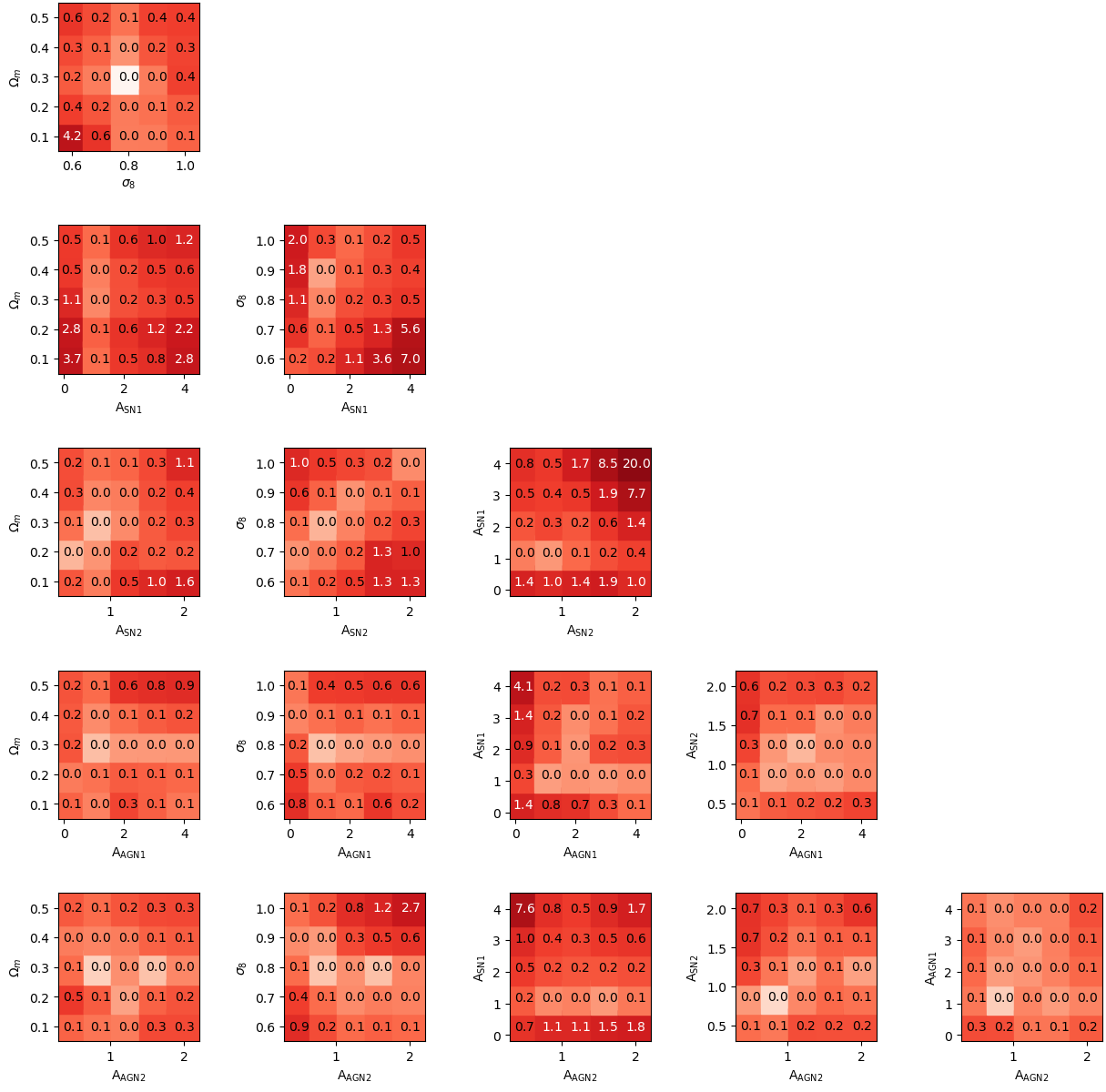}
    \includegraphics[width=0.32\linewidth]{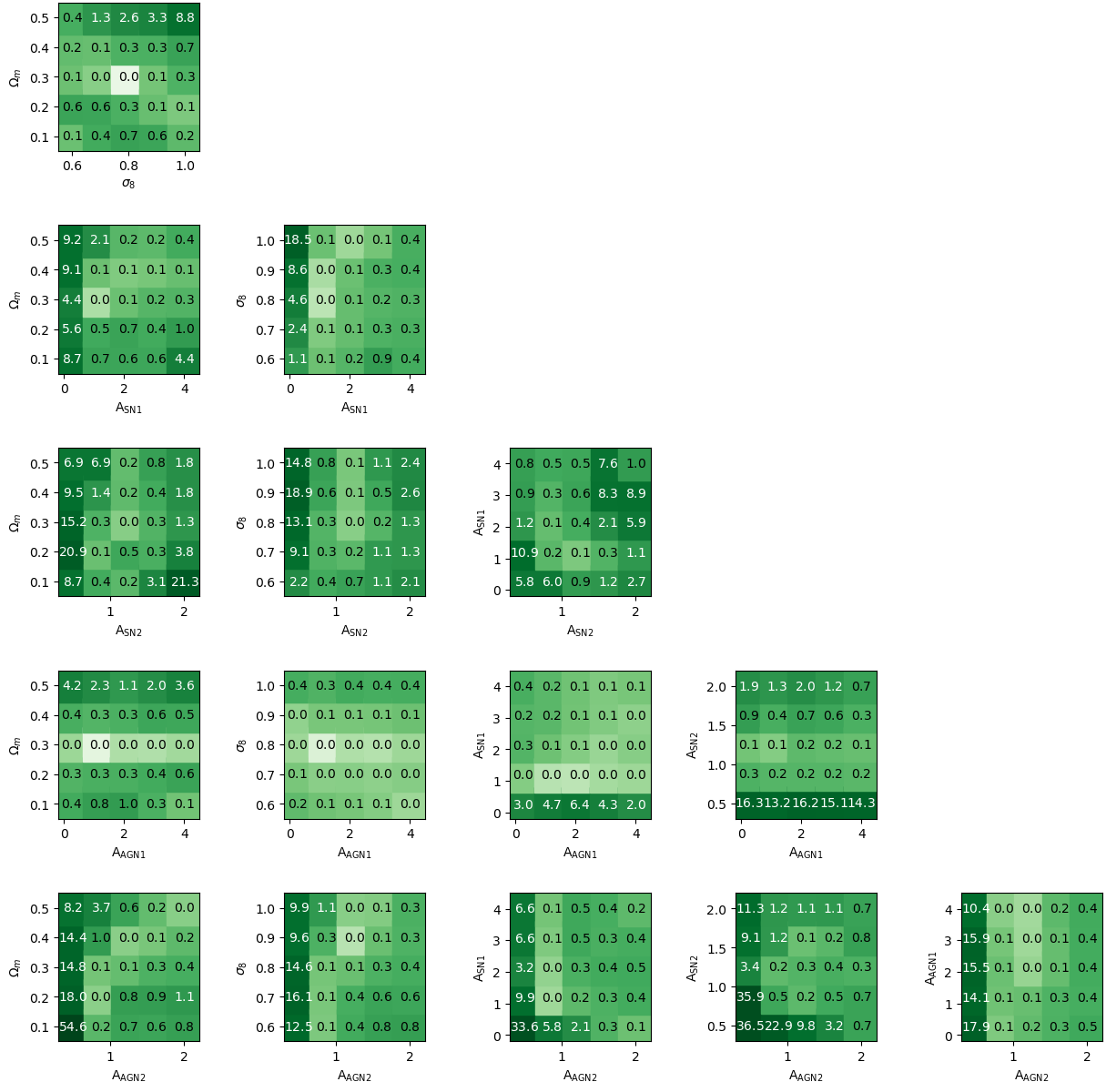}
    \caption{The distribution of $\chi^2_{SFH}$ values across the CAMELS parameter space when comparing the SFHs varying 2 parameters to the linear combination of the corresponding 1 parameter variations for the three CAMELS models: TNG (blue), SIMBA (red) and ASTRID (green). Values in each cell show the $\chi^2_{SFH}$ value, which also correspond the intensity of the colormaps.}
    \label{fig:all_interactions}
\end{figure}

Building on the analysis described in Section \ref{sec:result_sfh_interactions}, we present in Figure \ref{fig:all_interactions} a set of plots that show the regimes in which interactions between two parameters significantly change the resulting average SFHs of galaxies across the three CAMELS models. Using the same procedure as described in the section, we compute a $\chi^2_{SFH}$ by comparing SFHs from the normalizing flows sampled while varying 2 parameters to the linear combination of the normalizing flows sampled using the corresponding 1 parameter variations. While some regions show a common 2P interaction across the three CAMELS models (e.g. \sna{} $\times$ \snb{}), others are more model specific (e.g. \agnb{} $\times$ \snb{} in ASTRID).

\section{Baryon fraction vs stellar-to-gas-mass ratio for all CAMELS models and parameters}
\label{app:all_sfeplots}

Similar to Figure \ref{fig:baryon_cycling_tng}, here we show the trends in the baryon fraction and stellar-to-gas mass ratio (SGMR) while varying the CAMELS parameters for all three models, for halos $10^{11.5}<M_h<10^{12.5}$M$_\odot$. Given the sensitivity of the curves to model and halo mass choices, a full treatment of galaxies in this space is left for future study.

\begin{figure*}
    \centering
    \includegraphics[width=\textwidth]{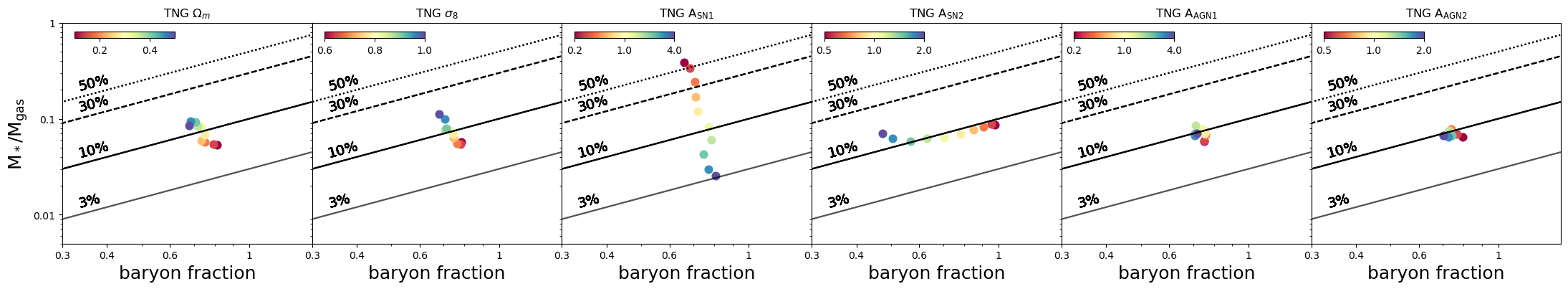}
    \includegraphics[width=\textwidth]{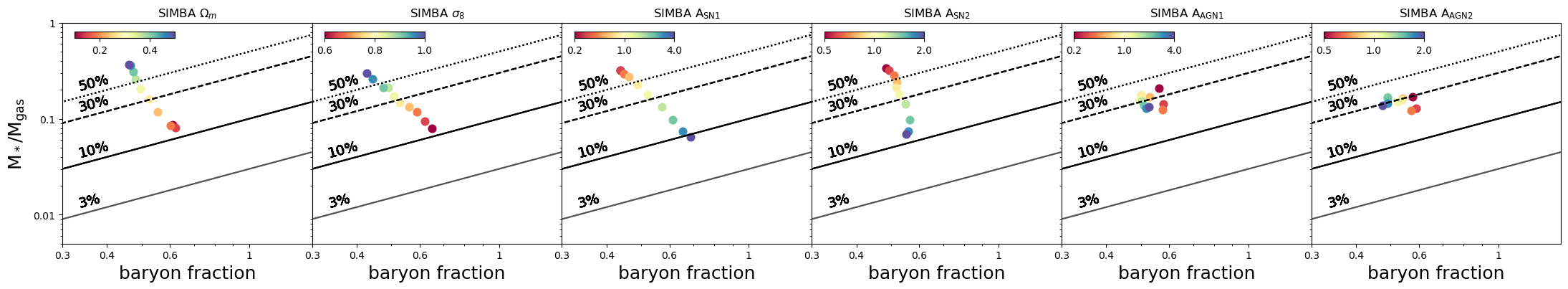}
    \includegraphics[width=\textwidth]{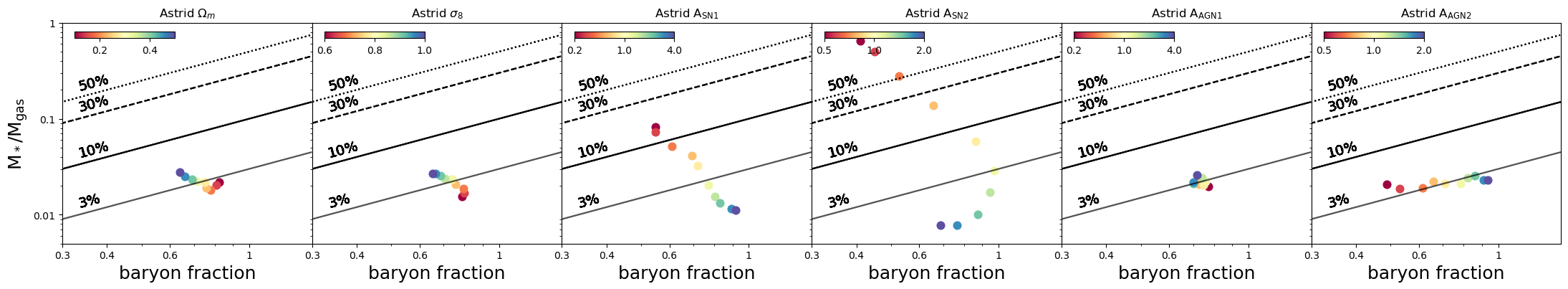}
    \caption{Similar to Figure \ref{fig:baryon_cycling_tng}, the plots here show the effects on the baryon fraction and stellar-to-gas mass ratio (SGMR) while varying the CAMELS parameters for all three models, for halos $10^{11.5}<M_h<10^{12.5}$M$_\odot$. The black lines show lines of constant baryon-fraction to stellar-to-gas mass ratio.}
    \label{fig:baryon_cycling_tng_extended}
\end{figure*}

\section{Chemical Enrichment and Duty Cycles: Complementary Windows into Feedback} \label{sec:disc_Z_fduty}

\begin{figure*}[ht]
    \centering
    \includegraphics[width=0.99\textwidth]{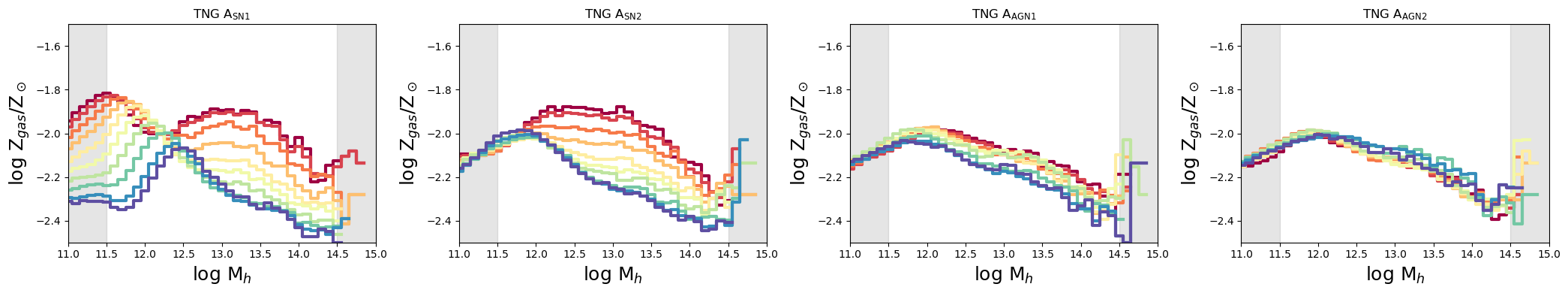}
    \includegraphics[width=0.99\textwidth]{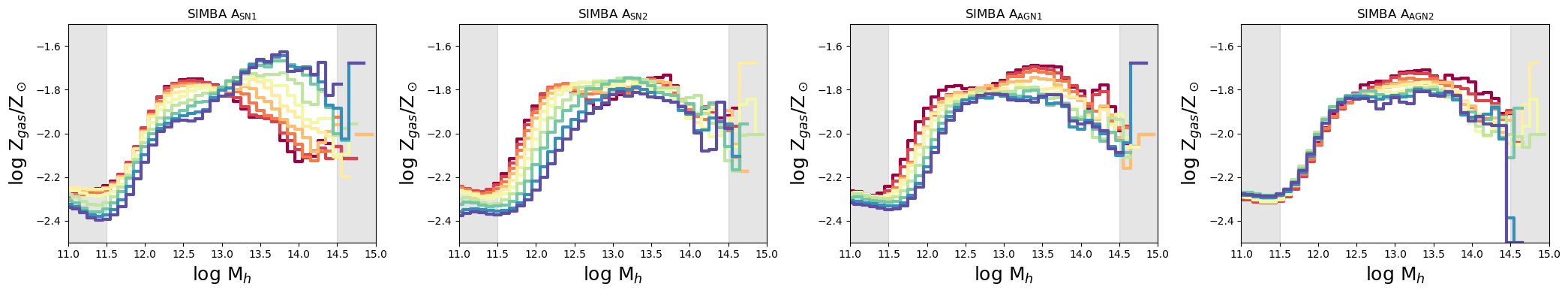}
    \includegraphics[width=0.99\textwidth]{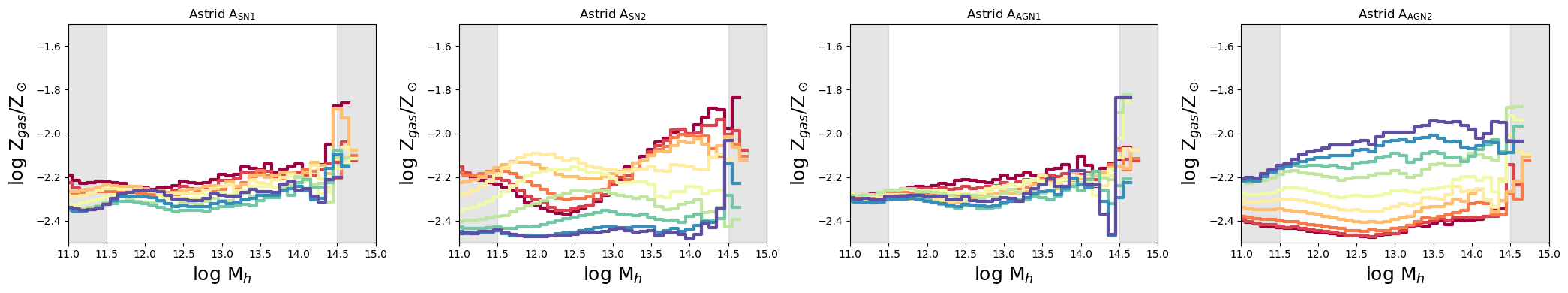}
    \caption{The impact of varying cosmology and feedback strengths on the gas phase metallicity of galaxies as a function of halo mass at $z\sim 0$ in the three models.}
    \label{fig:gasmet}
\end{figure*}

\begin{figure*}
    \centering
    \includegraphics[width=0.99\textwidth]{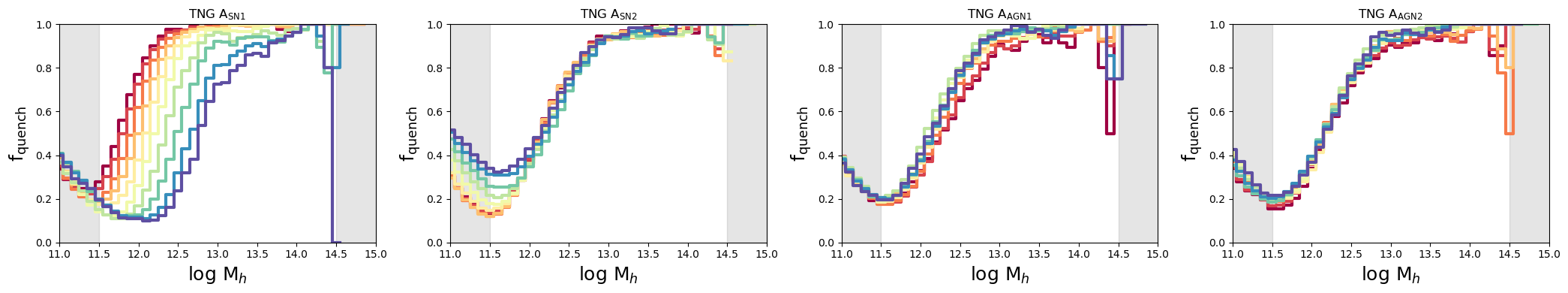}
    \includegraphics[width=0.99\textwidth]{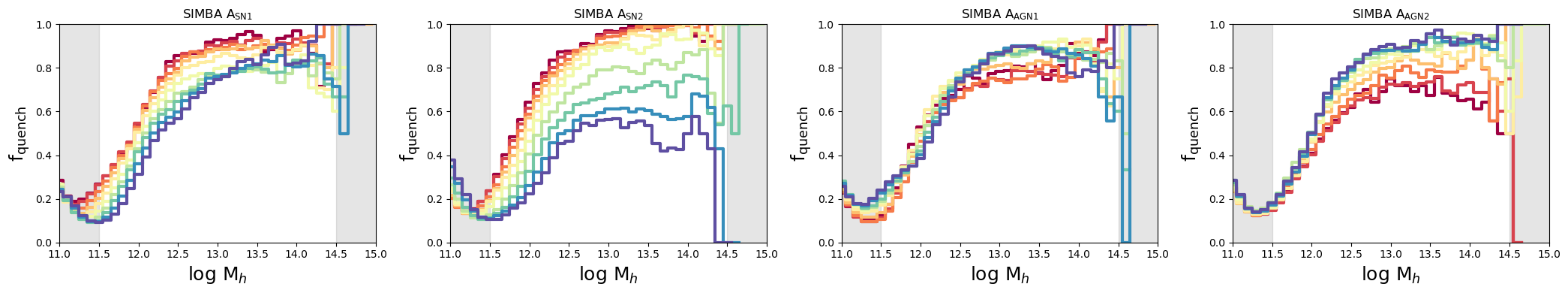}
    \includegraphics[width=0.99\textwidth]{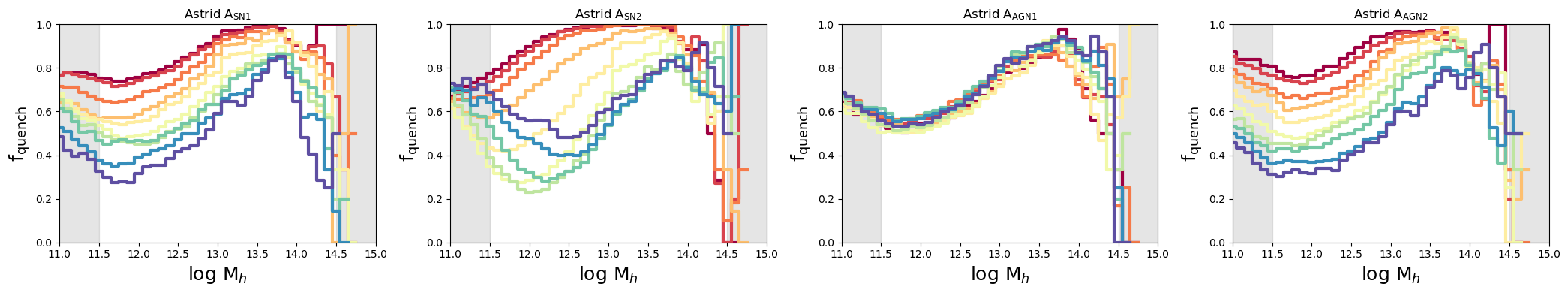}
    \caption{Similar to Figure \ref{fig:gasmet}, but for the fraction of quenched galaxies at $z\sim 0$.}
    \label{fig:quenched_fractions}
\end{figure*}

\begin{figure*}
    \centering
    \includegraphics[width=0.99\textwidth]{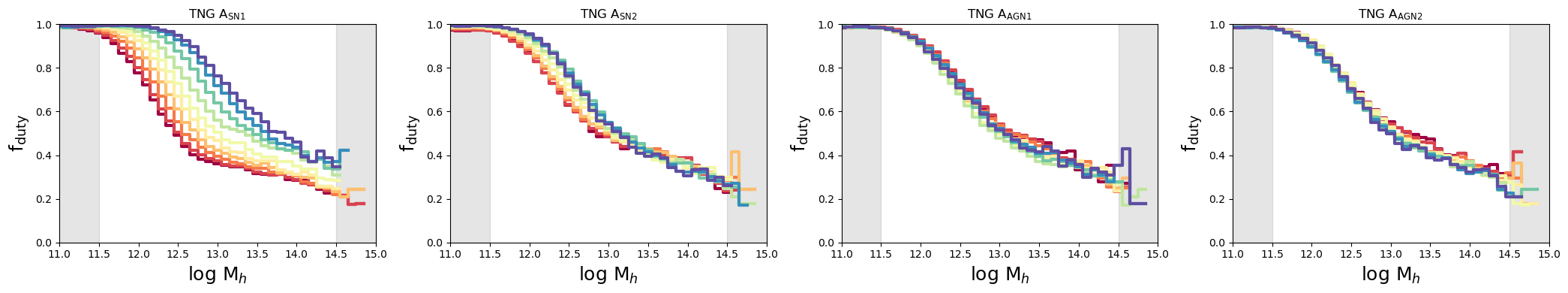}
    \includegraphics[width=0.99\textwidth]{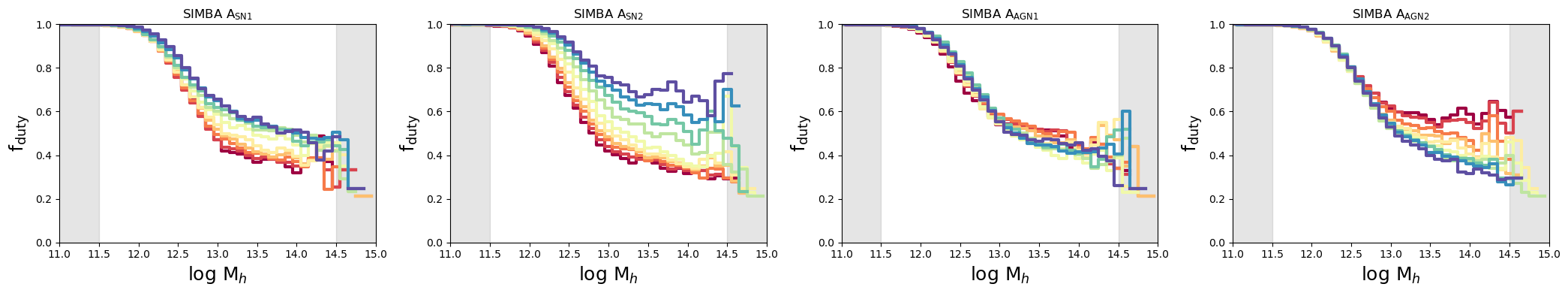}
    \includegraphics[width=0.99\textwidth]{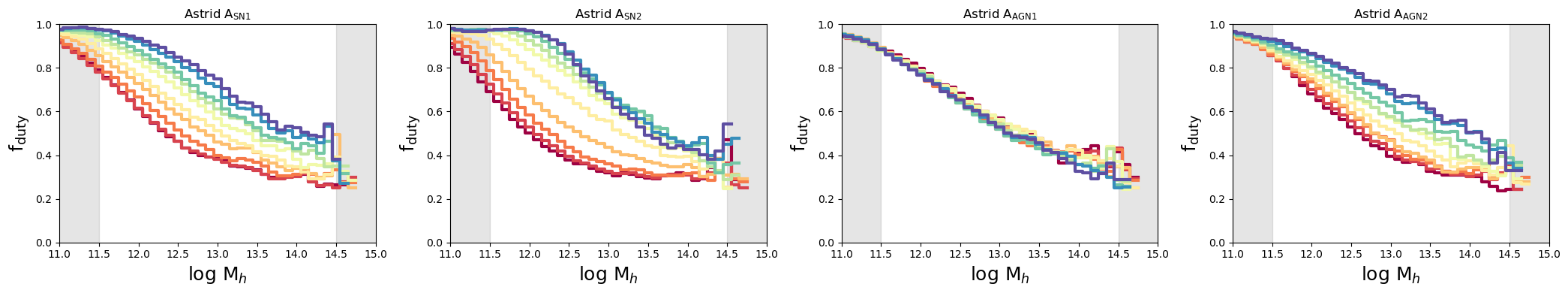}
    \caption{Similar to Figure \ref{fig:gasmet}, but for the mean duty cycle of the SFHs (i.e. how much time an SFH spends actively forming stars).}
    \label{fig:duty_cycle}
\end{figure*}

Varying the underlying cosmology and feedback strength leave distinct imprints not only on quantities like SFHs, baryon fractions and black hole mass, but also in the chemical enrichment patterns and alternate SFH based metrics like short timescale stochasticity and quenching behavior. Figures \ref{fig:gasmet}, \ref{fig:quenched_fractions}, and \ref{fig:duty_cycle} provide complementary views of how feedback shapes the gas phase metallicity, quenched fractions, and duty cycles of star formation.\textbf{This is especially useful in scenarios where high SNR data or tight constraints are not available on a few galaxy SFHs, instead of which large survey-level constraints are available for quantities like the quenched fraction and SFR duty cycles.}

Gas-phase metallicity distributions (Figure \ref{fig:gasmet}) show systematic variations with feedback strength that can help disambiguate different channels of baryon cycling. In TNG, increasing \sna{} leads to lower metallicities at fixed halo mass, though the metallicity is particularly insensitive at intermediate halo masses ($M_h\sim 10^{12.5}$M$_\odot$) and needs to be examined in future work. SIMBA shows a qualitatively different response - higher mass-loading (\sna{}) produces lower metallicities in low-mass systems but higher metallicities in massive halos. ASTRID's metallicities as a function of halo mass are systematically shallower, suggesting more efficient mixing between the ISM and CGM compared to the other models. Increasing wind speeds across all three models leads to lower metallicities, which is also true for AGN feedback except for \agnb{} in ASTRID, where SMBH growth is suppressed by increased thermal feedback, leading to more prolonged star formation and thus enrichment of the ISM. 

The fraction of quenched galaxies (Figure \ref{fig:quenched_fractions}) provides a complementary metric for how feedback affects quenching across different mass scales. All three models show a characteristic increase in quenched fraction with halo mass, but the transition scale and sharpness vary systematically with feedback strength. While scaling the feedback in most cases shows expected trends (higher quenched fractions in cases where star formation is suppressed), they can provide valuable statistical constraints on feedback strength from large galaxy surveys.

Similar to the quenched fractions, the duty cycles of galaxies - defined as the fraction of a galaxy's lifetime that it spends above a detectable threshold of star formation - provide a statistical or population level metric that can be used to constrain feedback. Here, we define the duty cycle similar to \citet{2011MNRAS.412.1123P} as the fraction of its lifetime that a galaxy spends actively forming stars, defined as $sSFR(t) > 0.2/t_{univ}(z)$.  Figure \ref{fig:duty_cycle} finds that the trends are similar to the quenched fractions in that they are largely monotonic but show significant mass dependence, and lower quenched fractions generally correspond to higher duty cycles.

\bibliography{sfh_feedback}{}
\bibliographystyle{aasjournal}

\end{document}